\pdfoutput=1

\documentclass[11pt]{article}
\usepackage[utf8]{inputenc}
\usepackage[T1]{fontenc}
\usepackage[english]{babel}

\usepackage[totalwidth=480pt, totalheight=680pt]{geometry}

\usepackage{xcolor}
\colorlet{mygreen}{green!50!black}
\colorlet{myblue}{green!20!blue}
\colorlet{myred}{red}

\usepackage{cite}
\usepackage{mathptmx}
\usepackage{amsmath,amsfonts,amsthm,amsbsy,amssymb}
\usepackage[hidelinks,%
						colorlinks=true,%
						linkcolor=myblue,%
						citecolor=mygreen,%
						urlcolor=myblue,%
						bookmarksnumbered=true]{hyperref}
\usepackage{mathtools}
\usepackage{nicefrac}
\usepackage{mathrsfs}
\usepackage{braket}
\usepackage{bbm}
\numberwithin{equation}{section}
\usepackage{lmodern}
\usepackage[bottom]{footmisc}

\usepackage{tikz}
\usetikzlibrary{shapes}
\usepackage{floatrow}
\usepackage[labelfont=bf, font={small}]{caption}

\def\beq{\begin{equation}}
\def\eeq{\end{equation}}
\def\bea{\begin{eqnarray}}
\def\eea{\end{eqnarray}}
\def\bit{\begin{itemize}}
\def\eit{\end{itemize}}

\def\we{\wedge}

\def\vp{\varphi}

\def\d{\delta}

\def\vp{\varphi}

\def\rd{{\rm d}}

\def\rL{{{\rm L}}}
\def\rV{{{\rm V}}}

\def\cL{{\cal{L}}}

\def\cJ{{\cal J}}
\def\cI{{\cal I}}

\def\cV{{\cal V}}
\def\cW{{\cal W}}

\def\mg{{\mathfrak{g}}}
\def\md{{\mathfrak{d}}}
\def\mt{{\mathfrak{t}}}
\def\mT{{\mathfrak{T}}}
\def\msl{{\mathfrak{sl}}}
\def\mL{{\mathfrak{L}}}
\def\mF{{\mathfrak{F}}}
\def\mH{{\mathfrak{H}}}
\def\mP{{\mathfrak{P}}}

\def\mW{{\mathfrak{W}}}

\def\ba{{\bf{a}}}

\def\bk{{\bf{k}}}
\def\bs{{\bf{s}}}
\def\br{{\bf{r}}}
\def\bs{{\bf{s}}}

\def\brr{{\bf{r}}_{-1}}
\def\bP{{\bf{P}}}

\def\bQ{{\bf{Q}}}
\def\bX{{\bf{X}}}
\def\bL{{\mathbf{\Lambda}}}
\def\bla{{\boldsymbol{\lambda}}}
\def\bd{{\boldsymbol{\delta}}}

\def\bx{{\boldsymbol{\xi}}}
\def\brho{{\boldsymbol{\rho}}}

\def\bE{{\mathbf{1}}}
\def\bD{{\mathbf{3}}}
\def\bF{{\mathbf{5}}}

\def\Ae{{^{[1]\!}A}}
\def\Al{{^{[\ell]\!}A}}
\def\Bl{{^{[\ell]\!}B}}
\def\Az{{^{[2]\!\!}A}}
\def\Bz{{^{[2]\!}B}}
\def\Ad{{^{[3]\!}A}}
\def\Bd{{^{[3]\!}B}}
\def\Av{{^{[4]\!}A}}
\def\Bv{{^{[4]\!}B}}

\def\Aff{{A_1^{(1)}}}
\def\Vir{{\rm Vir}}
\def\Ch{{\rm Ch}\,}

\def\E10{{{\rm E}_{10}}}

\newcommand{\nn}{\nonumber}

\def\ft{{\textstyle{\frac12}}}
\def\DD{{ {}_*^*}}

\newtheorem{theorem}{Theorem}
\newtheorem{definition}{Definition}

\begin{document}
\begin{titlepage}
\setcounter{page}{0}
\begin{center}
\vspace*{1.5cm}
\textbf{\LARGE A string-like realization\\[3mm]
 of hyperbolic Kac-Moody algebras}\\[2ex]
\vspace{2cm}
\textsc{\Large Saverio Capolongo$^1$, Axel Kleinschmidt$^{1,2}$,}\\[1ex]
\textsc{\Large Hannes Malcha$^1$ and Hermann Nicolai$^1$}\\
\vspace{1cm}
${}^1$\textit{Max Planck Institute for Gravitational Physics (Albert Einstein Institute)}\\
\textit{D-14476 Potsdam, Germany}\\[2ex]
${}^2$\textit{University of Vienna, Faculty of Physics, Boltzmanngasse 5, 1090 Vienna, Austria}\\
\vspace{2cm}
\textbf{Abstract}
\end{center}
\small{
\noindent
We propose a new approach to studying hyperbolic Kac-Moody algebras,
focussing on the rank-3 algebra $\mF$ first investigated by Feingold and Frenkel.
Our approach is based on the concrete realization of this Lie algebra in terms of a 
Hilbert space of transverse and longitudinal physical string states, which are 
expressed in a basis using DDF operators. When decomposed
under its affine subalgebra $\Aff$, the algebra $\mF$ decomposes into an infinite sum
of affine representation spaces of $\Aff$ for all levels $\ell\in\mathbb{Z}$. For $|\ell| >1$
there appear in addition coset Virasoro representations for all minimal models 
of central charge $c<1$, but the different level-$\ell$ sectors of $\mF$ do not form 
proper representations of these because they are incompletely realized in $\mF$. 
To get around this problem we propose to nevertheless exploit the coset Virasoro 
algebra for each level by identifying for each level a (for $|\ell|\geq 3$ infinite)
set of `Virasoro ground states' that are not necessarily elements of $\mF$ (in which 
case we refer to them as `virtual'), but from which the level-$\ell$ sectors 
of $\mF$ can be fully generated by the joint action of affine and 
coset Virasoro raising operators. We conjecture (and present partial evidence) 
that the Virasoro ground states for $|\ell|\geq 3$ in turn can be generated from
a {\em finite} set of `maximal ground states' by the additional action 
of the `spectator' coset Virasoro raising operators present for all levels $|\ell| > 2$. 
Our results hint at an intriguing but so far elusive secret behind Einstein's theory of 
gravity, with possibly important implications for quantum cosmology.
}
\vspace{\fill}
\end{titlepage}

\tableofcontents

\newpage
\section{Introduction}

\begin{figure}[H]
\centering
\begin{picture}(100,30)
\thicklines
\put(10,15){\line(1,0){40}}
\put(50,20){\line(1,0){40}}
\put(50,10){\line(1,0){40}}
\put(55,15){\line(1,1){10}}
\put(55,15){\line(1,-1){10}}
\put(85,15){\line(-1,1){10}}
\put(85,15){\line(-1,-1){10}}
\put(10,15){\circle*{10}}
\put(50,15){\circle*{10}}
\put(90,15){\circle*{10}}
\put(3,-2){$-1$}
\put(47,-2){$0$}
\put(88,-2){$1$}
\end{picture}
\caption{\label{fig:F}\textit{Dynkin diagram of $\mF$ with labelling of nodes.}}
\end{figure}
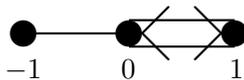

Our central object of interest in this paper is the hyperbolic Kac-Moody algebra (KMA)
$\mF \equiv \mg(A)$ associated with the indefinite Cartan matrix of rank three
\beq\label{Aij}
(A_{ij}) \,=\, 
\begin{pmatrix} 2 &-1 & 0 \\
-1 & 2 & -2\\
0 & -2 & 2\\
\end{pmatrix} \, , 
\eeq 
whose study was pioneered by Feingold and Frenkel~\cite{FF}
(see~\cite{Kac} for a general introduction to the theory of KMAs).
The associated Dynkin diagram with our labelling of roots is shown in Fig.~\ref{fig:F}.
The associated generators $\{e_i,f_i,h_i\}$ 
for $i \in \{-1,0,1\}$ obey the commutation relations
(`Chevalley-Serre presentation')
\begin{align}\label{CSerre}
\begin{aligned}
&[h_i, h_j] \,=\, 0 \, , \quad \quad \quad [e_i,f_j] \,=\, \delta_{ij} h_i\\
&[h_i, e_j] \,=\, A_{ij} \, e_j \, , \quad \, [h_i,f_j] \,=\, - A_{ij} \, f_j \, , \\
& {\rm ad}(e_i)^{1-A_{ij}}(e_j) \,=\, {\rm ad}(f_i)^{1-A_{ij}}(f_j) \,=\, 0\, .
\end{aligned}
\end{align}
With the $\msl(2)$ building blocks consisting of the triples $\{e_i, f_i, h_i\}$ 
the Lie algebra $\mg(A)$ is then defined to be the Cartan subalgebra (CSA) spanned by the 
$h_i$ plus the two free Lie algebras over the generators $e_i$ and $f_i$, 
respectively, modulo the above relations~\cite{Kac}.

The KMA based on the Cartan matrix~\eqref{Aij}, which is interchangeably designated 
as $\mF$ or $H\Aff$, or $AE_3$ or $A_1^{++}$ (the latter two designations
being preferred in the physics literature), is the simplest hyperbolic KMA
with a null root, and thus admits a distinguished affine subalgebra
$\Aff \equiv A_1^+$. Although not much is known about $\mF$, the following
facts have been established~\cite{FF}. The `germ' of the algebra $\mF$ resides in the 
beginnings of its graded decomposition w.r.t. its distinguished affine 
subalgebra for levels $|\ell|\leq 1$
\beq\label{VgV}
\overline\rV \;\oplus \; \mF^{(0)} \,\oplus \; \rV \, ,
\eeq
where at the center we have the affine subalgebra $\mF^{(0)} \equiv \Aff\subset\mF$.
At level one, $\rV \equiv \mF^{(1)} = L(\bL_0+2\bd)$ 
is the basic representation, 
while $\overline\rV\equiv \mF^{(-1)}$ is the conjugate representation
(see the following section 
for our definitions, conventions, and nomenclature). The algebra $\mF$ can 
then be generated by multiply commuting $\rV$ and $\overline\rV$. This task is, however,
complicated enormously by the need to divide out ideals generated by the Serre relations
({\em i.e.} the last line in~\eqref{CSerre}).
At level 2, this is still relatively simple, and we have~\cite{FF}
\beq\label{mg2} 
\mF^{(2)} 
\, \cong \, \rV \wedge \rV \big/ \cJ_2 \,\equiv\, 
\mF^{(1)} \wedge \mF^{(1)} \big/ \cJ_2 \, ,
\eeq
where $\cJ_2$ is the ideal generated by the Serre relation
involving the over-extended root with index $-1$ and $\cJ_2$ carries an action of the affine
algebra $\mF^{(0)}$.\footnote{An analogous decomposition for the maximal rank 
hyperbolic KMA $\E10$ is given in~\cite{KMW}.}
The above formula~\eqref{VgV} is only the beginning of an
infinite string of vector subspaces $\mF^{(\ell)}$ extending in both directions with 
$\ell\in\mathbb{Z}$,
see~\eqref{mgA} below, where each subspace $\mF^{(\ell)}$ consists of an infinite sum
of affine representation spaces for $|\ell| > 1$.
Consequently, the main obstacle towards a more `global' understanding 
of $\mF$ is that the procedure of dividing out Serre relations gets more 
and more cumbersome with higher levels already for levels $\ell=3$ 
and $\ell =4$~\cite{Kang1,Kang2}. For those levels, the complications 
are also evident from the formulas in~\cite{BB} and our explicit 
results for $\ell=3$ and $\ell =4$. 

As for products of affine representations, it has long been known that~\cite{KW1}
\beq\label{L0L0}
L(\bL_0+2\bd) \wedge L(\bL_0+2\bd) \,=\, \Vir(\ft,\ft) \otimes L(2\bL_1 + 3\bd) \, , 
\eeq
where $\Vir(\ft,\ft)$
is the minimal representation of the coset Virasoro
algebra with central charge $c=\frac12$ and $h = \frac12$; we recall that such a 
coset Virasoro algebra always accompanies the product of affine 
representations~\cite{GKO}.\footnote{Our conventions for tensor products of this 
type and the treatment of shifts by the null root $\bd$ are explained below in 
section~\ref{sec:MG}.} However, in the algebra $\mF$ the nice product structure 
of the r.h.s. is lost because one has to remove the top state associated with the Serre
relation, and thus one whole affine representation $L(2\bL_1 + 3\bd)$, so that 
by~\eqref{mg2} the level-2 sector of $\mF$ has the vector space structure~\cite{FF}
\beq\label{Level2}
\mF^{(2)} 
\,=\, \Big( \Vir(\ft,\ft) \ominus \mathbb{R} v_0 \Big) \otimes L(2\bL_1+3\bd) \, ,
\eeq
where $v_0$ is the vacuum state of the $\Vir(\ft,\ft)$ representation. 
Taking out the subspace $\mathbb{R} v_0$ leaves a `hole' in the 
coset Virasoro representation space $\Vir(\ft,\ft)$, as a result of which
the level-2 sector of the KMA is {\em not} a representation of the coset Virasoro
algebra anymore. Indeed, as we will show explicitly, the Virasoro algebra is no
longer obeyed on the truncated representation space, a statement which
extends to all levels of the Lie algebra $\mF$.
We will exhibit a structure similar to~\eqref{Level2} also for higher levels, where
similar `holes' will appear in the relevant coset Virasoro representations.

As a consequence, there is no `easy' way to construct the algebra by simply
multiplying affine representations as in~\eqref{L0L0}, and to obtain the Lie 
algebra elements of a given level-$\ell$ sector by application of the affine 
and coset Virasoro raising generators to a given set of ground states that
belong to $\mF$. In order to circumvent this difficulty, one main new tool 
employed in this work is to fill the `holes' by introducing `virtual states' which
belong to the relevant tensor products (corresponding to the l.h.s. of~\eqref{L0L0}), 
but vanish as elements of the Lie algebra, in this way restoring the full coset 
Virasoro representation.

At least for levels $|\ell|\leq 2$ this trick enables us to generate the whole 
level-$\ell$ sector by acting with the coset Virasoro algebra and the
affine algebra on a finite set of states that we will refer to as `maximal ground states'.
For levels $\ell > 2$ we encounter a vector space structure similar to~\eqref{Level2}
but with a `pile-up' of coset Virasoro representations stemming from calculations similar
to~\eqref{L0L0} (and more generally,~\eqref{TP}). This pile-up generates infinitely many 
copies of the finitely many maximal ground states. We call these copies `Virasoro ground
states'. The application of only affine and coset Virasoro raising generators does not allow
us to generate these additional Virasoro ground states from the maximal ground states.
Hence for level $\ell = 3$ we propose yet another set of operators that does exactly this.
We conjecture that there exists a generalization of this operator for all $\ell > 3$. Together
with the affine and coset Virasoro raising operators these operators would allow us to
generate any level-$\ell$ sector $\mF^{(\ell)}$ of $\mF$ from the finite set of maximal
ground states (which are essentially in one-to-one correspondence with the
allowable weights at level $\ell$ (\ref{Lambdaell})). An interactive visualization of
the associated root systems is presented in~\cite{VisualLie}.

A second new tool we rely on is the vertex operator formalism in the specific version
developed in~\cite{GN,GN1}, which builds on the seminal work of~\cite{Borcherds,IF,FLM}.
In this formalism, the Lie algebra is realized as a subspace of a certain Hilbert space of
physical string states, such that the elements of the Lie algebra are explicitly 
given in terms of DDF states built on certain tachyonic ground states, rather than
in terms of multi-commutators (the Del Giudice, Di Vecchia, Fubini (DDF)
formalism~\cite{DDF} is a well known and convenient tool to generate physical states
in string theory). A key feature first pointed out in~\cite{GN} is that for all levels 
$|\ell|>1$, there also appear {\em longitudinal} DDF states in the algebra, in addition
to the transversal DDF states familiar from the critical string.
One main advantage of the vertex operator algebra formalism is that we do not 
have to worry about Jacobi identities and the Serre relations
as these are automatically taken care of with the definition~\eqref{StateOperator}.
That is, unlike in~\cite{FF,KMW,BB} there is no need to take out affine 
representations `by hand', subtracting sub-representations
and compensating for over-subtractions.
Here, we will give explicit expressions for the maximal ground states 
for $\ell \leq 4$ in terms of the DDF basis.
In this way, we seek
to develop a perspective on hyperbolic KMAs different from the one 
usually taken in the mathematics literature, with the aim of gaining a more 
`global' understanding of its structure, as well as a more concrete realization of 
the algebra itself (as opposed to merely counting root multiplicities).

The Virasoro ground states are here determined by imposing the conditions
\eqref{VirasoroGS} on a given ansatz in terms of DDF states. With increasing level
this method becomes more and more unwieldy (for instance, at level
$\ell =4$ the `deepest' such state is so far inaccessible by our methods).
Therefore it would be desirable to determine the maximal ground states
by independent and more efficient means. If this can be done, we would 
have an efficient tool to explore higher levels. 
On top of unbounded pile-up of coset Virasoro representations described
above, there is the added
difficulty that in the final product for general level $\ell$
certain subspaces of affine representations
must be taken out, in analogy with~\eqref{Level2}. The real complication
is therefore not so much with products of affine representations 
but with the `holes' in the coset Virasoro representations, which become
more and more difficult to deal with as the level is increased. This proliferation
of complications is reminiscent of the fractal structure of a Mandelbrot set, 
although we know of no Lie algebra analog of the self-similarity features.

To conclude this introduction we wish to underline the potential relevance of the 
KMA $\mF$ for physics. We claim that this algebra hides a deeply buried secret about
Einstein's theory! To explain this point, observe that from the Cartan matrix~\eqref{Aij} 
we see that $\mF$ possesses two distinguished rank-two subalgebras, both of which
appear in the dimensional reduction of Einstein's theory to lower dimensions. Namely,
the upper 2-by-2 submatrix corresponding to a $\msl(3)$ subalgebra is associated 
with the Matzner-Misner SL(3) (actually GL(3)) group obtained by reducing Einstein
gravity from four to one dimension. 
On the other hand, the lower 2-by-2 submatrix is associated with an $A_1^{(1)}$ affine 
symmetry, which is just the Lie algebra underlying the Geroch group of general 
relativity, obtained by reducing Einstein's theory to two dimensions~\cite{Julia,BM}.
The lower-most diagonal entry corresponds to the Ehlers $\msl(2)$ symmetry 
obtained by dualizing the Kaluza-Klein vector in three dimensions.
The Geroch algebra and the Matzner-Misner $\msl(3)$ intersect in the middle entry 
corresponding to the Matzner-Misner SL(2), which likewise has been known for a long 
time in general relativity. 

All this suggests that one might try to find a concrete physical realization of $\mF$ by 
simply combining the Matzner-Misner SL(3) and the Geroch symmetry~\cite{Nic0}. 
However, it turns out that a simple dimensional reduction to one dimension cannot 
accomplish this because to realize the Geroch group, we need {\em two} coordinates 
for the duality transformations (\!\!\cite{Nic0} tried to circumvent this problem
by means of a null reduction, but again finds that the bulk of $\mF$ is realized 
only trivially; see also~\cite{Penna:2021apa} for a recent related investigation). 
The conclusion is that we cannot find a non-trivial realization
of $\mF$ by sticking with Einstein's theory and standard notions of 
space-time based field theory, but need an extension from which 
standard general relativity `emerges' only in a specific limit. Hints of such a theory 
have emerged from the study of cosmological billiards~\cite{DHN}. In particular, the
celebrated BKL analysis~\cite{BKL} of cosmological singularities
can be rephrased in terms of a cosmological billiard that takes place 
in the Weyl chamber of the Weyl group of $\mF$ (it is a main result of~\cite{FF}
that the even part of the Weyl group for $\mF$ is the modular group PSL$_2(\mathbb{Z})$).

In view of the compelling links with Einstein gravity on the one hand~\cite{DHN} and the 
horrendous complexity of $\mF$ on the other, one may also ask about 
possible implications for Big Bang cosmology. From a physics perspective,
the pile-up of truncated Virasoro modules with increasing level 
may indicate that more and more degrees of freedom 
`open up' in the approach towards the cosmological singularity.
It is for this reason that ~\cite{Nic} conjectured the emergence of 
a mathematically well-defined notion of non-computability towards the
singularity which may thwart attempts at mathematically understanding the 
beginning of time, unless a more `global' description of $\mF$ can be found.
At the very least this shows that the restriction to finitely many degrees 
of freedom that underlies most investigations in quantum cosmology
({\em e.g.} by means of a mini-superspace approximation, where
keeping only diagonal metric degrees of freedom would correspond 
to restricting $\mF$ to its CSA) may be far too na\"ive to 
understand the quantum origin of our universe.

\section{Basic facts about \texorpdfstring{$\mF$}{F}}\label{sec:2}

For~\eqref{Aij} we denote the simple roots by $\{\brr, \br_0, \br_1\}$, such that
their inner products yield 
\beq
A_{ij} \,=\, \br_i\cdot \br_j 
\eeq
with the over-extended root $\brr$.
The affine null root is $\bd = \br_0 + \br_1$, so that $\brr\cdot\bd = -1$.
The one-dimensional Ehlers $\msl(2)$ root sublattice of the 
$\mF$ root lattice is simply $\mathbb{Z}\br_1$, 
consisting of the elements $\pm n\br_1$ of length $2n^2$.
There are two important regular rank-two subalgebras, namely $A_2\equiv \msl(3)$ with 
simple roots $\{\brr, \br_0\}$, and the affine $A_1^{(1)}$ with simple roots $\{ \br_0 ,\br_1\}$.

The three simple roots of $\mF$ are associated with the generators $\{e_i,f_i,h_i\}$ for 
$i \in \{-1,0,1\}$ which satisfy~\eqref{CSerre}. With the root lattice 
$Q = \mathbb{Z}_{-1} \br_{-1} \oplus \mathbb{Z}_{0} \br_{0} \oplus \mathbb{Z}_{1} \br_{1}$
we write the roots as 
\beq
(a_{-1},a_0,a_1) \,\equiv \, \br \,=\, a_{-1} \brr + a_0 \br_0 + a_1 \br_1 \, .
\eeq
We can alternatively parametrize them in terms of the null root as
\beq\label{root1}
\br\,=\, a_{-1} \brr + k_{-1} \bd + a_1 \br_1 \equiv (a_{-1} ,k_{-1}, a_1 + k_{-1}) \, .
\eeq
For this combination to be a root we must have 
\beq
\br^2 \,=\, 2a_{-1} (a_{-1} - k_{-1}) + 2a_1^2 \leq 2 \,.
\eeq 
The level $\ell$ of a root $\br$ is defined by 
\beq
\ell\,\coloneqq\, \br\cdot\bd \,=\, - a_{-1} 
\eeq
and thus counts the number of occurrences of $e_{-1}$ or $f_{-1}$ in a multi-commutator.
Here we adopt the conventions of~\cite{KMW,GN1}, so {\em positive level}
is associated with {\em negative roots}.

\begin{figure}[H]
\vspace{.5cm}
\begin{tikzpicture}
\draw [fill] (-2,2) circle (.1);
\draw [fill] (0,2) circle (.1);
\draw [fill] (2,2) circle (.1);
\draw [fill] (-2,0) circle (.1);
\draw [fill] (0,0) circle (.1);
\draw [fill] (2,0) circle (.1);
\draw [fill] (-2,-2) circle (.1);
\draw [fill] (0,-2) circle (.1);
\draw [fill] (2,-2) circle (.1);
\draw[thick,->] (0.4,0) -- (1.6,0);
\draw[thick,->] (-0.4,0) -- (-1.6,0);
\draw[thick,->] (0.3,-0.3) -- (1.7,-1.7);
\draw[thick,->] (-0.3,0.3) -- (-1.7,1.7);
\node at (-1,0.3) {$f_1$};
\node at (1,-1.5) {$f_0$};
\node at (1,0.3) {$e_1$};
\node at (-1,1.5) {$e_0$};
\end{tikzpicture}
\vspace{.5cm}
\caption{The action of the Chevalley-Serre generators $e_i$ and $f_i$ depicted in
the affine root sublattice $Q^\prime$.}
\label{fig:RootLatticeCS}
\end{figure}
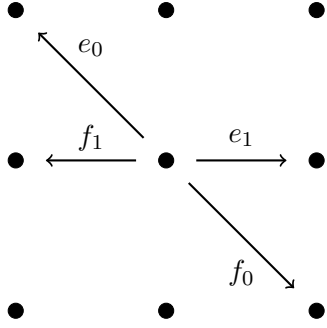

The affine algebra $\Aff\subset\mF$ identified above 
will play a central role in the remainder, as we will focus exclusively on 
the decomposition of $\mF$ w.r.t. this affine subalgebra, as in~\cite{FF}. 
On the affine root sublattice 
$Q^\prime \,=\, \mathbb{Z}_{0} \br_{0} \oplus \mathbb{Z}_{1} \br_{1}$,
the Chevalley-Serre generators
induce the transformations shown in Fig.~\ref{fig:RootLatticeCS}.
The generators $h_0$ and $h_1$ do not induce transformations in the root lattice but
give rise to eigenvalue equations. 
A more explicit description of $\Aff$ is afforded by
\beq
\Aff
\,=\, {\rm span}_{\mathbb{R}} \,\big\{ E_m, F_m,H_m, K, \md \ | m\in \mathbb{Z} \big\} \, , 
\eeq
with the standard commutation relations
\begin{align}\label{AffineCom}
\begin{aligned}
&[H_m, E_n] \,=\, 2 E_{m+n} \, , 
&&[H_m, F_n]\,=\,- 2 F_{m+n} \, , \\ 
&[E_m, F_n]\,=\,H_{m+n} + mK \delta_{m+n,0} \, , 
&&[H_m, H_n]\,=\,2mK \delta_{m+n,0} \, , \\
&[E_m, E_n]\,=\,[F_m , F_n ]\,=\,0 \, 
\end{aligned}
\end{align}
and where 
\begin{align}
\begin{aligned}
E_0 &= e_1\,,& \quad F_0 &= f_1\,,&\quad H_0 &= h_1\,, \\
F_{1} &= e_0\,,& \quad E_{-1}&= f_0 \,,&\quad K-H_0 &= h_0\,.
\end{aligned}
\end{align}
Furthermore, we define
\beq
K\,=\, h_0 + h_1\;\; , \qquad \md=h_{-1}+ 2h_0+ 2h_1 \, ,
\eeq
where $K$ commutes with all affine generators; its eigenvalue on an
arbitrary element of $\mF$ is the level $\ell$ associated to that element and for this reason
we will call $K$ the `level counting operator'. $\md$ is the `depth counting operator' with 
\beq
\big[\md \,,\, T_m \big] \,=\, - mT_m \quad 
\text{for all} \quad
T_m \in \big\{E_m, F_m, H_m \big\} 
\eeq 
and records the coefficient of $-\bd$ for any given root~\eqref{root1} of $\mF$.
Thus the depth {\em increases by one}, if the root is shifted by $-\bd$ in 
our conventions.\footnote{The Cartan generator $\md$ is related to the standard 
derivation $d=h_{-1}+h_0+h_1$ by $\md=d+K$.}
Together with $h_1$, the operators $K$ and $\md$ span the CSA of~$\mF$.

After these preparations the algebra can 
be decomposed into eigenspaces of $K$
\beq\label{mgA}
\mF \,=\, \bigoplus_{\ell\in\mathbb{Z}} \, \mF^{(\ell)} \, . 
\eeq
We will refer to this decomposition as the `level decomposition'
of $\mF$. The level-$\ell$ subspace $\mF^{(\ell)}$, on which $K=\ell$,
is thus the linear span of all
multi-commutators $[f_{i_1}, \ldots , [f_{i_{n-1}}, f_{i_n}]\ldots]$ with
$\ell$ generators $f_{-1}$. Negative levels are similarly associated
with multi-commutators $[e_{i_1}, \ldots , [e_{i_{n-1}}, e_{i_n}]\ldots]$
with $|\ell|$ generators $e_{-1}$ in the multi-commutator.
Hence, each subspace $\mF^{(\ell)}$ decomposes into (generally
infinitely many) irreducible representations of $\Aff$.
$K$ is the central charge which commutes with all elements of
$\Aff$, but not of $\mF$. Furthermore,
the subspace $\mF^{(-\ell)}$ is conjugate to $\mF^{(\ell)}$ and thus does not 
need to be studied separately. Below, we will therefore restrict attention 
to positive levels, {\em i.e.} {\em highest weight} representations of $\Aff$,
hence multi-commutators of $\{ f_i\}$.

The following theorem is of central importance (for readers' convenience we include a 
short proof of this Theorem in Appendix~\ref{app:pf}, see also Theorem~1 in~\cite{GN}).

\begin{theorem}[Feingold-Frenkel~\cite{FF}]\label{th:FF}~\\
Any level-$\ell$ element of $\mF$ can be obtained as a 
linear combination of commutators of level-one and level-$(\ell\!-\!1)$ elements,
that is
\beq
\mF^{(\ell)} \,=\, \left[ \mF^{(1)} , \, \mF^{(\ell -\! 1)} \right] \, .
\eeq 
\end{theorem}
Thus, one can proceed to explore the algebra level by level, moving up
in level by one step at a time. The commutator of two elements thus
provides a map from the tensor product $\mF^{(1)}\otimes \mF^{(\ell -1)}$
to $\mF^{(\ell)}$ which we denote by $\cI^{(\ell)}$, so that
\beq\label{Iell}
\cI^{(\ell)} \;:\; \mF^{(1)} \;\otimes \; \mF^{(\ell -1)} \;\rightarrow \; \mF^{(\ell)} \quad , \qquad
\cI^{(\ell)} \big( u \; \otimes \; v \big) \,\coloneqq\, [u,v] \qquad
\big( u\in \mF^{(1)}\,,\, v\in \mF^{(\ell-1)} \big)\,.
\eeq
The map $\cI^{(\ell)}$ is surjective by Theorem~\ref{th:FF}, but has a non-trivial kernel,
as a result of which we have the vector space isomorphism
\beq\label{FKer}
\mF^{(\ell)} \;\cong \, \big( \mF^{(1)} \otimes \mF^{(\ell-1)} \big) \,\big/ \, 
{\rm Ker} \, \cI^{(\ell)}\,.
\eeq
While the tensor product of affine representations in the numerator can be evaluated
by standard techniques, at least in principle, the main difficulty is in determining 
the kernels for all $\cI^{(\ell)}$, which in particular include affine representations 
associated with the Serre relations. A main novelty of the present approach is that,
once we have $\mF^{(1)}$ and $\mF^{(\ell -1)}$ in terms of DDF states, 
the commutator of any given pair of states can be directly evaluated 
by means of the universal formula~\eqref{StateOperator}, which at least
in principle gives all elements of $\mF^{(\ell)}$, again in terms of DDF states,
and in this way also furnishes information about the `structure constants' of $\mF$.
Let us stress that the quotient in~\eqref{FKer} must be distinguished from the division 
of the free Lie algebra by the `Serre ideal' $\oplus_{|\ell|\geq 2} \cJ_\ell$ that is 
employed in more standard approaches, {\em cf.}~\cite{BB} and section~3 
in~\cite{KMW}. Note also that our $\cI^{(\ell)}$ is not the same as $I_\ell$ in~\cite{BB},
the latter being defined for the free Lie algebra.

We will also need the (hyperbolic) fundamental weights $\{\bL_j\}$, which are defined by
\beq\label{Lambda1}
\br_i \cdot \bL_j \,=\, \d_{ij} \, .
\eeq
For the algebra $\mF$ they are given by 
\beq\label{Lambda2}
\bL_{-1} \,=\, - \bd\,\,, \quad
\bL_{0} \,=\, - \brr - 2\bd\,\,, \quad
\bL_{1} \,=\, - \brr - 2\bd + \frac12 \br_1 \, .
\eeq
From~\eqref{Lambda2} it follows immediately that the highest affine weights which can 
appear at level $\ell$ are 
\beq\label{Lambdaell}
\bL \,=\, p_0\bL_0 + 2p_1 \bL_1 + p_{-1} \bL_{-1} \quad
\text{with} \quad p_0 + 2p_1 \,=\, \ell \quad \text{and} \quad 
p_0\,,\,p_1 \in \mathbb{Z}_{\geq 0} \, , \ p_{-1} \in \mathbb{C} 
\eeq
($2p_1$ because $\br_1$ can only appear with integer coefficients 
as $\bL$ must be an element of the root lattice of $\mF$). For given $\ell$, the
number $p_{-1}$ will always be an integer;
in fact, for level $|\ell|\geq 2$ representations $L(\bL)$ with infinitely 
many different values $p_{-1}$ will occur for given $p_0$ and $p_1$. In the analysis
of $\mF$ we will encounter almost all irreducible highest-weight representations of the 
affine algebra for $\bL$ of the form~\eqref{Lambdaell}. They are uniquely characterized 
by providing the highest weight from~\eqref{Lambdaell}
and denoted by $L(\bL)$. 
The coefficients $p_0$ and $p_1$ are constrained to be non-negative integers, 
but the coefficient $p_{-1}$ of $\bL_{-1}=-\bd$ is arbitrary and the corresponding 
representation spaces differ only by the $\md$ eigenvalues of the highest weight vectors. 

For each level, there is an associated Sugawara realization of the Virasoro algebra, 
with generators ${}^{[\ell]}\!\cL_m^\text{sug}$ $(m\in\mathbb{Z})$
\beq\label{Sugawara}
{}^{[\ell]}\!\cL_m^\text{sug} \,\coloneqq\, \frac{1}{2(\ell + 2)} \sum_{k \in \mathbb{Z}} 
\DD \left[ \frac{1}{2} H_k H_{m-k} + E_k F_{m-k} + F_k E_{m-k} \right] \DD \, .
\eeq
The normal ordering for the affine generators in~\eqref{Sugawara} is defined by 
(with a sum $A$ over the $\mathfrak{sl}(2)$ generators paired by the Killing form)
\beq\label{TT}
\DD \, T^A_m T^A_n \,\DD \,\coloneqq\, 
\begin{cases}
T^A_m T^A_n & m<0 \, , \\
T^A_n T^A_m & m \geq 0 \, . 
\end{cases} 
\eeq
On each $\mF^{(\ell)}$ we have
\beq\label{SugTmCom}
\big[{}^{[\ell]}\!\cL_m^\text{sug} , T_n \big] \,=\, - n T_{m+n}
\eeq
for any affine generator $T_n \in \{E_n, F_n, H_n\}$
acting on the level-$\ell$ representations in $\mF^{(\ell)}$. 
The corresponding (Sugawara) central charge at level-$\ell$ is~\cite{GO}
\beq
c_\ell^\text{sug} \,=\, \frac{3\ell}{\ell +2} \, .
\eeq
Even more important for us is the fact that for each level there is a {\bf coset
Virasoro algebra} with generators ${}^{[\ell]}\!\mL_m^\text{coset}$ which commutes
with the affine generators~\cite{GKO}. The action of ${}^{[\ell]}\!\mL_m^\text{coset}$ does 
not affect the affine representation but shifts the associated affine weight diagrams by
$m\bd$. The key point here is that the action ${}^{[\ell]}\!\mL_m^\text{coset}$ is defined
only on the tensor product $\mF^{(1)}\otimes \mF^{(\ell -1)}$, but not directly on 
$\mF^{(\ell)}$ where its implementation would lead to inconsistencies, for which 
we will give some examples below. Equivalently, there is no consistent action of the 
coset Virasoro algebra on the kernel ${\rm Ker} \,\cI^{(\ell)}$.
In general, we therefore do not have a proper representation of the level-$\ell$ 
coset Virasoro algebras on the level-$\ell$ sectors of the Lie algebra.

More specifically, consider a level-one element $u \in \mF^{(1)}$ and 
a level-$(\ell\!-\!1)$ element $v \in \mF^{(\ell-1)}$ (for $\ell >1$) and 
their tensor product $w = u \otimes v$. The action of the affine generators 
$T_m \in \{E_m, F_m, H_m\}$ on $w$ obeys the usual distributive law
\beq\label{TmTp}
T_m w \,\equiv\, T_m \left( u \otimes v \right) \,=\, (T_m u) \otimes v + u \otimes (T_m v) 
\eeq
and remains valid in this form if the tensor product is replaced by a 
commutator. The action of the coset Virasoro element on tensor products is defined in
terms of the Sugawara generators~\eqref{Sugawara} by
\beq\label{mLm}
{}^{[\ell]}\!\mL_m^\text{coset} w \,\equiv\, {}^{[\ell]}\!\mL_m^\text{coset} 
( u \otimes v )
\,\coloneqq\, \Big({}^{[1]}\!\cL_m^\text{sug} u \Big) \otimes v \,+\, 
u \otimes \left( {}^{[\ell -1]}\!\cL_m^\text{sug} v \right) \, \,-\, 
{}^{[\ell]}\!\cL_m^\text{sug} w \, .
\eeq
In general, level-$\ell$ elements are sums of such tensor products, 
in which case this formula applies summand by summand. 
When one replaces the tensor product by a Lie algebra commutator $[u,v]$
inconsistencies arise whenever this commutator vanishes although 
the tensor product does not. {\em For this reason formula~\eqref{mLm} 
must not be used with Lie algebra commutators}, as this
will lead to inconsistencies but only to elements of the
tensor product $\mF^{(1)}\otimes \mF^{(\ell-1)}$. Likewise
applying this formula to a Lie algebra element that can be reached in 
two different ways by lower level commutators will
lead to contradictory results. Below we will 
exhibit explicit examples of this phenomenon, and show how the
coset Virasoro operator on the commutator fails already on level 2.

From the above theorem, it follows that the coset Virasoro central 
charge~\cite{GKO} associated with $\mF^{(1)}\otimes \mF^{(\ell -1)}$,
and hence also with $\mF^{(\ell)}$, is
\beq\label{CentralCharge}
c_\ell^\text{coset} \,=\, 1 + \frac{3(\ell-1)}{\ell+1} - \frac{3\ell}{\ell+2}
\,=\, 1 - \frac6{(\ell+1)(\ell+2)} \, .
\eeq
Consequently, all minimal Virasoro representations will occur in the
analysis of $\mF$. Each level $\mF^{(\ell)}$ will thus decompose into 
sums of products of certain level-$\ell$ representations of the affine algebra 
and the associated truncated representations of the level-$\ell$ coset Virasoro algebra, 
furthermore adorned by an increasing tail of products of lower level coset Virasoro
characters. The fact that the central charge~\eqref{CentralCharge} is bounded from above 
by $1$ is not generally true, and in fact violated for higher rank hyperbolic algebras 
such as $E_{10}$. For the minimal series the allowed ${}^{[\ell]}\!\mL_0^\text{coset}$
eigenvalues at level $\ell$ are then contained in the following list~\cite{GKO}
(see also~\cite{DiFrancesco})
\beq\label{hrs}
h_{r,s}^{(\ell)} \,=\, \frac{[(\ell+2)r - (\ell+1)s]^2 -1}{4(\ell+1)(\ell+2)} \, . \qquad
(r=1,...,\ell \,; \ s=1,...,r)
\eeq
These are the values that can be assumed by the virtual ground states, but
are shifted by integers in the coset Virasoro descendant states. There is
no such restriction on the eigenvalues $h$ to a discrete set for $c>1$ 
\cite{DiFrancesco}. In fact, in the ultimate analysis of $\mF$ there will appear such 
representations galore once one tries to simplify the `porous' coset 
Virasoro representations by reducing products.

\section{DDF construction}

Following~\cite{GN,GN1} our main tool to analyze the algebra
is to represent it in terms of a certain subspace of a Hilbert space of
physical string states. More specifically, we will be dealing with a
subcritical compactified bosonic string whose target space-time dimension is
$d=3$, equal to the rank of $\mF$, and whose momenta lie on
the Lorentzian root lattice $Q$ which is the $\mathbb{Z}$-linear span of the 
three simple roots $\{\brr, \br_0,\br_1\}$. Because $d=3 < 26$ there will
also appear longitudinal states in addition to the transversal states~\cite{GN}, and
these will show up for all levels $|\ell|\geq 2$. For the details of this construction
we refer to~\cite{GN}, and here only summarize some salient points.

\subsection{Lie algebra of physical states}
As usual, the string Fock space that we will associate to the Lie algebra
comes equipped with elementary Virasoro operators 
\beq\label{Lm}
\rL_m \,\coloneqq\, \frac12 \sum_{n\in\mathbb{Z}} 
\boldsymbol{:} \alpha^\mu_{m-n} \alpha_{n\mu} \boldsymbol{:}
\eeq
with the usual string oscillators $\alpha_m^\mu$ for $\mu = 0,1,2$ and 
$m \in \mathbb{Z}$. We define the space $\mP_n$ by
\beq
\vp\,\in\,\mP_n \quad \Leftrightarrow \quad \rL_m\vp \,=\, 0 \quad (m\geq 1) \quad \text{and} 
\quad (\rL_0 - n)\vp\,=\,0 
\eeq
and physical states belong to $\mP_n$ for $n=1$. As shown in the theory of vertex
operator algebras~\cite{Borcherds,FLM} (see also~\cite{GN}
for an introduction) the following quotient space is then a Lie algebra
\beq\label{mH}
\mH \,\coloneqq\, \mP_1 / \rL_{-1}\mP_0 \, ,
\eeq
where, as explained at length in these references, the commutator between any two 
elements $\vp, \psi \in \mH$ is defined via the state-operator correspondence 
through the formula
\beq\label{StateOperator}
[\vp\,,\, \psi ] \,\coloneqq\,
\oint\frac{\mathrm{d}z}{2\pi i} \ \cV(\vp;z) \, \psi 
\eeq
and where $\cV(\vp;z)$ is the vertex operator associated to the state $\vp$.
As shown in~\cite{Borcherds,FLM} this definition
satisfies all the requisite properties of a Lie bracket, to wit, antisymmetry and the 
Jacobi identity, {\em modulo} elements of $\rL_{-1}\mP_0$. This is the reason 
for restricting $\mP_1$ to the quotient~\eqref{mH}. The actual evaluation of
\eqref{StateOperator} becomes more laborious with the excitation level, since for
each state $\vp$ one first has to work out the associated vertex operator
by use of standard formulas, and then re-express the result of the calculation
in terms of DDF operators of the appropriate level. For more detailed 
explanations and simple examples see~\cite{GN}.

The affine subalgebra $\mF^{(0)}$ is a (tiny!) subspace of $\mH$. Adopting physicists'
bra and ket notation, its Chevalley-Serre generators are associated with the following 
states in $\mP_1$
\beq\label{CSGenerators}
e_i \,\coloneqq\, \ket{\br_i} \, , \quad 
f_i \,\coloneqq\, - \ket{- \br_i} \, , \quad 
h_i \,\coloneqq\, \br_{i \mu} \alpha^\mu_{-1} \ket{0} 
\qquad (i=0,1)
\eeq
so $e_i$ and $f_i$ correspond to tachyonic states, while $h_i$
correspond to photon states~\cite{GN}. Using~\eqref{StateOperator} it can 
be easily checked that the definitions~\eqref{CSGenerators} 
are such that~\eqref{CSerre} is satisfied. More generally, for the affine 
raising and lowering operators we have the operator-state 
correspondence
\beq\label{EFH}
E_m \,\widehat{=}\, \ket{\br_1 + m\bd} \, , \quad
F_m \,\widehat{=}\, - \ket{- \br_1 + m\bd} \, , \quad
H_m \,\widehat{=}\, \br_{1\mu} \alpha^\mu_{-1} \ket{m\bd} \, .
\eeq
With $\br_1 = \bd - \br_0$ we can equivalently write these states
in a Matzner-Misner type basis with $\br_1$ replaced by $\br_0$. The affine 
raising and lowering operators are thus all associated with tachyonic or photonic states. 
{\em The simple formulas~\eqref{CSGenerators} and~\eqref{EFH} contain all that is
required for the description of $\Aff$ as a subspace of $\mH\,$.}\vspace*{0.5cm}

\begin{figure}[H]
\vspace{.5cm}
\begin{tikzpicture}
\draw [fill] (-2,2) circle (.1);
\draw [fill] (0,2) circle (.1);
\draw [fill] (2,2) circle (.1);
\draw [fill] (-2,0) circle (.1);
\draw [fill] (0,0) circle (.1);
\draw [fill] (2,0) circle (.1);
\draw [fill] (-2,-2) circle (.1);
\draw [fill] (0,-2) circle (.1);
\draw [fill] (2,-2) circle (.1);
\draw[thick,->] (0.4,0) -- (1.6,0);
\draw[thick,->] (-0.4,0) -- (-1.6,0);
\draw[thick,->] (0,0.4) -- (0,1.6);
\draw[thick,->] (0,-0.4) -- (0,-1.6);
\draw[thick,->] (0.3,-0.3) -- (1.7,-1.7);
\draw[thick,->] (0.3,0.3) -- (1.7,1.7);
\draw[thick,->] (-0.3,0.3) -- (-1.7,1.7);
\draw[thick,->] (-0.3,-0.3) -- (-1.7,-1.7);
\node at (-1,0.2) {$F_0$};
\node at (-1,1.5) {$F_1$};
\node at (0.3,1) {$H_1$};
\node at (1,1.5) {$E_1$};
\node at (1,0.2) {$E_0$};
\node at (-1,-1.5) {$F_{-1}$};
\node at (0.5,-1.3) {$H_{-1}$};
\node at (1.6,-1) {$E_{-1}$};
\end{tikzpicture}
\vspace{.5cm}
\caption{The action of the affine generators $E_m$, $F_m$ and $H_m$ on the 
level-$\ell$ root sublattice $Q^\prime$. 
$H_0$ gives rise to an eigenvalue equation. The positive Chevalley generators are
identified as $e_1 \equiv E_0$ and $e_0 \equiv F_1$.} 
\label{fig:RootLatticeAffine}
\end{figure}
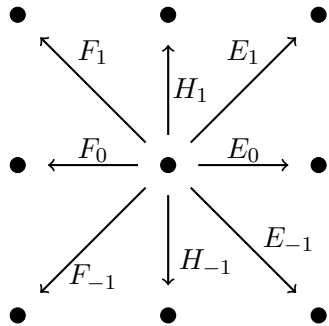

The affine generators induce the motions on the affine root sublattice (affine ladder
diagram) $Q^\prime$ shown in Fig.~\ref{fig:RootLatticeAffine}.

At levels $|\ell|\geq 1$ there will be many more states, in particular those 
corresponding to higher excited string states. We will furthermore
have many more tachyonic states. To describe them, we introduce
for any root $\br$ its DDF decomposition by 
\beq
\br \,=\, \ba - n \bk \qquad \text{with} \qquad \ba^2 \,=\, 2 \, , \quad \bk^2 \,=\, 0 
\quad \text{and} \quad \ba\cdot\bk \,=\, 1 \, .
\eeq
We refer to $\ba$ as the tachyonic momentum associated to the root $\br$.
In addition, we will need polarization vectors 
$\bx\equiv\bx(\ba)$ obeying
\beq
\bx(\ba)\cdot \ba \,=\, \bx\cdot\bk \,=\, 0
\eeq
and that we will normalize to 1.
For a level-$\ell$ root $\br = - \ell \brr + \cdots$, the null vector 
$\bk = \bk(\ell) \equiv \bk_\ell$ depends on the level, and we have
\beq
\bk_\ell \,=\, \frac1{\ell} \bd \, .
\eeq
Importantly, {\em for $\ell >1$, $\bk_\ell$ does not belong to the root lattice},
but is nevertheless a necessary ingredient in the DDF construction.
For level one we will usually drop the subscript, {\em i.e.} write $\bk\equiv \bk_1= \bd$.
At level $\ell$ the relevant tachyonic momenta are 
\beq\label{TachyonMomentum}
\ba_n^{(\ell)} \,=\, - \ell \brr - \left(\ell + \frac{n^2 -1}{\ell}\right)\bd + n\br_1 \, , 
\qquad (n\in \mathbb{Z})
\eeq
with associated tachyon states $\ket{\ba_n^{(\ell)}}$. For $\ell = 1$ all these momenta 
belong to the root lattice, while only for selected values of $n$ if $|\ell|>1$. 
The fact that they generally do not for $|\ell|>1$ is an important feature of our construction
of higher level states in $\mF$. On the space of tachyonic states we have
the positive definite scalar product\footnote{It is the fundamental discreteness 
of our construction that allows such a discrete scalar product, whereas in the
continuum we would have $\braket{ \ba | \ba' } = \delta(\ba - \ba')$.}
\beq\label{aman}
\braket{ \ba^{(\ell)}_m \,|\, \ba^{(\ell')}_n } \,=\, \delta_{mn} \delta^{\ell\ell'} \, , 
\eeq
which eventually can be extended to the full Hilbert space by means of the 
DDF algebra~\eqref{DDF}.
Denoting the `spin' of a state by half its $h_1$ eigenvalue, the spin range 
of the ($2n{+}1$)-dimensonal (Ehlers) $\msl(2)$ multiplet generated from
$\ket{\ba_n^{(\ell)}}$ is $ -n \leq s \leq n$. 
The depth of $ \ba^{(\ell)}_n$ is $\md= \ell + \frac{n^2-1}{\ell}$.

Furthermore, denoting the elementary Weyl reflections associated with 
$\br_0$ and $\br_1$, respectively, by $w_0$ and $w_1$, we have the following 
action of the affine Weyl group on $\ba^{(\ell)}_n$ for $k\in\mathbb{Z}$
\begin{align}\label{WeylTach}
\begin{aligned}
(w_1w_0)^k (\ba_n^{(\ell)}) 
\,&=\, \ba^{(\ell)}_{n-k\ell}\,, \\
w_0 (w_1w_0)^{k-1} (\ba_n^{(\ell)})
\,&=\, \ba^{(\ell)}_{-n+k\ell}\,.
\end{aligned}
\end{align}
We recall that the affine Weyl group is the semi-direct product
of an abelian translation group with a finite Weyl group~\cite{Kac}; in the present case 
$\mathbb{Z}\rtimes \mathbb{Z}_2$ and we take the translation group to be generated 
by $\mt \equiv w_1w_0$.
From~\eqref{WeylTach}, we see that the level-$\ell$ tachyonic momenta group into 
$\left[\frac{\ell}2\right] + 1$ distinct Weyl orbits. This is also the number 
of possible choices for $\bL_0$ and $\bL_1$ at level $\ell$ in~\eqref{Lambdaell}.
Hence for each admissible level-$\ell$ weight there is an appropriate tachyonic 
vacuum on which to build a DDF tower of string states, as explained below.

Physical states are built by acting on these tachyonic vacua with
DDF operators~\cite{DDF}. Unlike in non-compact string theory, for $\mF$ this is a 
{\em discrete} construction, in the sense that for any given level only a discrete subset
of the momentum space continuum is admitted, of which the root lattice is a subset.
This discrete set fills the continuum more and more densely as $\ell\rightarrow\infty$. 
 
\medskip
 
Because the string is subcritical, the DDF operators come in two varieties. 

\medskip
 
1)
The {\em transversal} level-$\ell$ DDF operators $\Al_m$ (with $m \in \mathbb{Z}$) 
are defined by
\beq\label{TransversalDDF}
\Al_m \,\coloneqq\, \oint\frac{dz}{2\pi i} \, \bx_\mu \bP^\mu(z) e^{im\bk_\ell\cdot \bX(z)}
\,=\, \oint\frac{dz}{2\pi i} \, \bx_\mu \bP^\mu(z) e^{(m/\ell)\bd\cdot \bX(z)}
\eeq
in terms of the usual string coordinate fields 
\beq
\bX^\mu (z) \,=\,q^\mu - ip^\mu \ln z + i \sum_{m\neq 0} \frac1{m} \alpha^\mu_m z^{-m} 
\, , \quad
\bP^\mu(z) \,=\, i\frac{d}{dz} \bX^\mu(z) \, , 
\eeq
where again $\mu = 0,1,2$. By construction, the DDF operators are physical, to wit,
\beq\label{DDFphys}
\big[ \rL_m\,, \Al_n \big] \,=\, 0 \, .
\eeq
The operator $\Al_{m}$ shifts the momentum of 
the state on which it acts by $ m\bk_\ell = (m/\ell) \bd$. The transversal polarization 
vector $\bx^\mu$ appearing in~\eqref{TransversalDDF} is given by
\beq\label{xin}
\bx\big( \ba_n^{(\ell)} \big) \,=\, \frac1{\sqrt{2}} \big( \br_1 -2n\bk_\ell \big) \, .
\eeq
Because for the non-zero modes in~\eqref{TransversalDDF} a shift of the 
polarization vector by a multiple of $\bd$ drops out as it is a total derivative,
the non-zero mode DDF operators at a given level are the same for all $m\neq 0$ 
in~\eqref{TachyonMomentum}. For simplicity, we will therefore set $n=0$ 
in~\eqref{xin} for {\em all} $A_m$ including the zero mode $A_0$. 
Unlike for higher rank ($d>3$) algebras there is only one transversal
polarization, whence the DDF oscillators carry no transverse indices 
for $A_1^{(1)}$. Because of the orthogonality properties and $\bd^2 =0$ no 
normal ordering is required in~\eqref{TransversalDDF}. 

On the level-$\ell$ subspace the DDF transversal operators obey the 
standard commutation relations 
\beq\label{DDF}
\big[ \Al_m \,, \Al_n \big]\,=\, m \delta_{m+n,0}
\eeq
because $m$ is the eigenvalue of the operator $(m/\ell)K$
on $\mg^{(\ell)}$. On the tachyonic vacua they obey
\beq\label{Alma}
\Al_m \ket{\ba^{(\ell)}_n} \,=\, 0 \qquad \text{for } m\geq 1\,.
\eeq
The key property of the DDF operators is that their application 
to any physical state creates another physical state by~\eqref{DDFphys}.
The same is true for all the `composite operators' 
(affine generators and Sugawara operators) 
that we will construct from the transversal DDF operators.
\vspace*{.5cm}

 2)
The {\em longitudinal} level-$\ell$ DDF operators $\Bl_m$ (with $m \in \mathbb{Z}$) 
are defined by~\cite{Brower:1972wj}
\beq
\Bl_m (\ba)\,\coloneqq\, \oint \frac{dz}{2\pi i} \ \boldsymbol{:} \left[ 
- \ba _\mu \bP^\mu(z)
+ \frac{m}{2} (\ba \cdot \bk_\ell) \frac{\mathrm{d}}{\mathrm{d}z} 
\log \left(\frac{\bk_{\ell \, \mu}}{\ba \cdot \bk_\ell} \right) 
\bP^\mu(z)\right] 
e^{i m \bk_\ell\cdot \bX(z)} \boldsymbol{:} \, , 
\eeq
where we use $\Bl_m$ rather than the more common notation
$\Al_m^-$ for easier notational distinguishability between transversal
and longitudinal DDF operators. Note also that we do {\em not} include the
usual contribution
quadratic in transversal DDF operators in this definition (this modification is
often included to make transversal and longitudinal DDF operators commute,
unlike~\eqref{BAcomm}). The argument $\ba$ is the tachyonic momentum of the state on 
which the longitudinal DDF operator $\Bl_m$ acts. The definition implies~\cite{GN}
\beq\label{BAcomm}
\big[\, \Bl_m(\ba)\,,\, \Al_n \big] \,=\, - n \, \Al_{m+n} + 
\frac{n}{\sqrt{2} \ell} \, (\ba \cdot \br_1) \, \delta_{m+n,0} \ \bd_\mu \alpha^\mu_0 \, .
\eeq
The longitudinal DDF operators are likewise physical, {\it viz.}
\beq
\big[ \rL_m\,,\Bl_n \big] \,=\, 0 \, .
\eeq
Finally, the longitudinal DDF operators obey a Virasoro algebra of their own~\cite{GN}
\beq
\big[ \Bl_m\,,\, \Bl_n \big] \,=\, (m-n) \, \Bl_{m+n} + 2 (m^3 - m) \delta_{m+n,0}
\eeq 
with central charge $24$. Let us emphasize that these operators are only well-defined 
on a subset of the full Hilbert space $\mH$, and only on states for which 
$\ba\cdot\bk \neq 0$.

In summary, the space of physical states at level $\ell$ is the linear span of states
\beq\label{PhysState}
\prod_{i=1}^M \Al_{-m_i} \prod_{j=1}^N \Bl_{- n_j} \, \ket{ \ba^{(\ell)}_n} \qquad
\text{for} \quad 
\begin{array}{l}
m_1 \ge \ldots \ge m_M \ge 1 \, , \\
n_1 \ge \ldots \ge n_N \ge 2 \, , 
\end{array} \quad
\quad 
\text{and}
\quad 
M, N \ge 0 \, .
\eeq
The restriction $2\leq n_j$ is due to the fact that $\Bl_{-1}$ creates null physical states
because
\beq \label{Lm1}
\big(\ldots\big) \ \Bl_{-1} \ket{ \ba^{(\ell)}_n} \,=\, 
\big(\ldots \big) \rL_{-1} \ket{ \ba^{(\ell)}_n - \bk_\ell } \,=\,
\rL_{-1} \big(\ldots \big) \ket{ \ba^{(\ell)}_n - \bk_\ell } \, , 
\eeq
where $(\ldots)$ stands for any combination of DDF operators (which all commute 
with $\rL_{-1}$). These states must be omitted. The total momentum of the 
state~\eqref{PhysState} is 
$\ba^{(\ell)}_n + (\sum_{i=1}^M m_i + \sum_{j=1}^N n_j) \bk_\ell$,
which in order to be an element of the root lattice $Q$ must thus satisfy the condition
\beq\label{rootmomentum}
\sum_{i=1}^M m_i + \sum_{j=1}^N n_j \,=\, \ell P- 1 \, . \qquad (P\in\mathbb{N}) 
\eeq
Importantly, {\em our Lie algebra formulas make sense only for states obeying this
condition, in the sense that all elements of $\mF$ are associated with momenta
subject to~\eqref{rootmomentum}.} While the `in-between' states with momenta not on 
the root lattice do exist as physical states, various
operations, and in particular the definition of the commutator, fail
for them, because appropriate cocycle factors cannot be consistently 
defined for fractional momenta (see Appendix~\ref{app:Cocycle}). 

However, even restricting to admissible DDF states with momenta on the 
root lattice, we will see that the Lie algebra $\mF$ contains only a subset of 
these states. $\mF\subset \mH$ is thus a proper subspace of $\mH$. This
can already be seen from the fact that the number of physical states~\eqref{PhysState}
for a given root $\br$ is $\vp^{-2}(1 -\br^2/2) - \vp^{-2}(-\br^2/2)$, whereas the known
root multiplicities of $\mF$ are bounded above by the smaller $\vp^{-1}(1-\br^2/2)$ (in
accord with Frenkel's conjecture~\cite{IF}), with the usual partition function~\eqref{phi}.
The main challenge is therefore to characterize how precisely $\mF$ is embedded in 
$\mH$. Below we will exhibit several examples where one can see which DDF states appear
in $\mF$.

\subsection{Level 1}
As a `warm-up' let us look at the level-one subspace $\mF^{(1)}$ where the longitudinal
DDF operators
do not yet appear, and where we have a fully explicit description of the
so-called basic representation of $\Aff$ in terms of the well-known Frenkel-Kac
construction~\cite{FK} (see also~\cite{Segal:1981ap} and the introductory 
reviews~\cite{GO0,GO}). We here rephrase these
results in DDF language, as explained in~\cite{GN}.
By~\eqref{DDF} the level-one transversal DDF operators are
\beq\label{DDF0}
A_m \, \equiv \, \Ae_m 
\,=\, \oint\frac{dz}{2\pi i} \ \bx_\mu \bP^\mu(z) e^{m \bd\cdot \bX(z)} \, , 
\eeq
with the tachyonic ground states $\ket{\ba_n}$ whose momenta are of the form 
\beq\label{ba}
\ba_n \, \equiv \, \ba_n^{(1)} \,=\, -\brr - n^2\bd + n\br_1 \, . \qquad (n \in \mathbb{Z}) 
\eeq
They belong to a single Weyl orbit in the root lattice of $\mF$.
With these definitions it is straightforward to write down the first few levels of the basic 
representation, and thus $\mF^{(1)}$, for depths $\md\leq 4$ in terms of transversal DDF
states
\begin{align}\label{DDF1}
\begin{aligned}
&\md\,=\,0: \quad\quad\quad\quad& \ket{\ba_0} \quad \\
&\md\,=\,1: & A_{-1} \ket{\ba_0} \, , \ \ket{\ba_{\pm 1}} \quad \\
&\md\,=\,2: &A_{-2} \ket{\ba_0} \, , \ A_{-1}^2\ket{\ba_0} \, , \
A_{-1} \ket{\ba_{\pm 1}} \quad \\
&\md\,=\,3: &A_{-3} \ket{\ba_0} \, , \
A_{-2} A_{-1} \ket{\ba_0} \, , \ A_{-1}^3 \ket{\ba_0} \,,\, 
A_{-2} \ket{\ba_{\pm1}} \,,\ A_{-1}^2 \ket{\ba_{\pm 1}} \quad \\
&\md\,=\,4:\quad & A_{-4} \ket{\ba_0} \, , \ A_{-3} A_{-1} \ket{\ba_0} \, , \
 A_{-2} A_{-2} \ket{\ba_0} \,,\ A_{-2}A_{-1}^2 \ket{\ba_0} \,,\ A_{-1}^4 \ket{\ba_0} \, , \ \, \\
& {} \quad &A_{-3} \ket{\ba_{\pm1}} \, , \
A_{-2} A_{-1} \ket{\ba_{\pm 1}} \, , \
A_{-1}^3 \ket{\ba_{\pm 1}} \,,\ \ket{\ba_{\pm 2}} \quad
\end{aligned}
\end{align}
(recall that the depth $\md$ is the coefficient of $(-\bd)$ for the root 
corresponding to the given DDF state in the form~\eqref{root1}).
Note that with our choice of polarization we have
\beq\label{A0}
A_0 \, \ket{\ba_n} \,=\, \sqrt{2} n \, \ket{\ba_n} \, .
\eeq
We refer to the uppermost state $|\ba_0\rangle \equiv \ket{- \brr} = \ket{\bL_0 + 2\bd}$
(corresponding to the Chevalley-Serre generator $f_{-1}$) as the `ground state'
of $\mF^{(1)}$. It is easily seen that the states~\eqref{DDF1} correspond to
the $\msl(2)$ representations 
\beq\label{SL2Rep}
\bE_0 \oplus \bD_1 \oplus(\bE\oplus\bD)_2 \oplus (\bE\oplus\bD\oplus\bD)_3 \oplus 
(\bE\oplus\bE\oplus \bD\oplus\bD\oplus\bF)_4 \, .
\eeq
This list can be easily continued. As a special case of~\eqref{aman} we have 
the scalar product of the tachyonic states 
\beq
\braket{ \ba_m | \ba_n } \,=\, \delta_{mn} \, , 
\eeq
which can be extended to the full basic representation because we can evaluate all scalar
products of DDF states by means of the commutation relations~\eqref{DDF}.

For a graphical representation of the level-one subspace 
$\mF^{(1)}$ see Figure~\ref{fig:L1RootLattice} in Appendix~\ref{app:Figures}.
The level-one sector is a single affine representation, hence all its states
can be reached by acting on the ground state with affine raising operators.
The DDF states at different depths can also be connected via the level-one Sugawara
 generators~\cite{GN}
\beq\label{Sug}
{}^{[1]}\!\cL_m^\text{sug} \,=\, \frac12 \sum_{n\in\mathbb{Z}} \
\DD A_n A_{m-n} \DD \quad \Rightarrow \quad
[\, {}^{[1]}\!\cL_m^\text{sug} , A_n ] \,=\, -n A_{m+n} \, 
\eeq
where the normal ordering is defined as in~\eqref{TT}. This simple quadratic formula works 
only for the basic representation, whereas at higher levels we encounter non-polynomial 
expressions in the DDF operators, see next section. With our choice of polarization the
 operator
\beq
{}^{[1]}\!\cL_0^\text{sug} \,=\, \frac12 A_0^2 + A_{-1}A_1 + A_{-2} A_2 + \ldots 
\eeq
counts the depth of a given DDF state in~\eqref{DDF1}, and can thus
be identified with the operator $\md$. The Sugawara operators 
provide a convenient tool, since ${}^{[\ell]}\!\cL_m^\text{sug}$ while moving up in depth,
does not change the $\msl(2)$ representation content of a given state. Hence one can 
directly build infinite `strings' of $\msl(2)$ singlets, triplets, {\em etc.} by straightforward 
application of Sugawara operators 
${}^{[\ell]}\!\cL_{-1}^\text{sug}\,,\, {}^{[\ell]}\!\cL_{-2}^\text{sug} \,,\dots$ 
To move between the different $\msl(2)$ representations contents we must use the 
affine generators defined below. It is also straightforward to re-express
the list~\eqref{DDF1} in terms of multi-commutators, see {\em e.g.}~\cite{GN}.

We can also rephrase these results in terms of characters. Recall that
$ \mF^{(1)} = L(\bL_0 + 2 \bd)$ and so $\Ch \, \mF^{(1)} = q^{-2} \, \Ch \, L(\bL_0) $
with $q = e^{-\bd}$, 
for which we have~\cite{FF} 
\beq\label{chLevel1}
\Ch \, \mF^{(1)}\,=\,\frac{q^{-1}}{\varphi(q)} \ \Theta_{\bL_0} \quad, \qquad
\Theta_{\bL_0} \equiv q^{-1} \sum_{n\in\mathbb{Z}} \exp \big( \mt^n (\bL_0) \big) \,.
\eeq
where $\mt \coloneqq w_1 w_0$ for the elementary affine translation.
See Appendix~\ref{app:character} for a definition of the Euler function $\varphi$ 
and the generalized theta function.

\section{General results for higher levels}

\subsection{Affine and Sugawara generators at arbitrary level}\label{sec:Generators}
In order to implement the action of the affine and the Sugawara generators 
for arbitrary levels, we here specialize the formulas of~\cite{GKN} and~\cite{GN1}
to the case of $\mF$ giving the affine step operators and the 
Sugawara operators explicitly in terms of transversal DDF operators
for each level-$\ell$ subspace. In particular, these formulas generalize the well-known
formula~\eqref{Sug} at level-one to arbitrary levels. The essential new feature
here (in comparison with the standard vertex operator construction) is that
the exponents appearing in these expressions are expressed not in terms
of standard string oscillators, but rather in terms of transversal DDF 
operators~\cite{GN,GN1};
in the form given they are thus different for each level, and valid only on
the respective level-$\ell$ subspace. The use of transversal DDF operators implies 
that both the affine and the Sugawara operators map physical states to physical 
states without changing the level, hence map elements of $\mF^{(\ell)}$ 
to other elements of $\mF^{(\ell)}$. 

We first give the formulas for the affine generators. The generators of the Heisenberg
subalgebra are identified with a (for $\ell >1$ proper) subset of the transversal
DDF operators
\beq\label{Hm}
{}^{[\ell]}\!H_m \,\coloneqq\, \oint\frac{dz}{2\pi i} \ \br_{1 \, \mu} \, 
\bP^\mu(z) e^{m \bd\cdot \bX(z)} \,=\, \sqrt{2} \ \Al_{\ell m} \, .
\eeq
So it is only at level one that we have a one-to-one correspondence between
the Heisenberg generators and the transversal DDF operators. This is the basic reason
for the simplicity of formula~\eqref{Sug} and its equivalence to a free
field energy momentum tensor. For the affine raising and lowering
operators we have the formulas
\begin{align}\label{EmFm}
\begin{aligned}
{}^{[\ell]}\!E_m &\,\coloneqq\, \oint \frac{\mathrm{d}z}{2\pi i} \ z^{\ell m} \ 
\exp\left( + \sum_{n > 0} \frac{ \sqrt{2}}{n} \ \Al_{-n} z^n \right) 
e^{ i \br_1 \mathbf{Q}} z^{\br_1 \mathbf{P}} 
\exp\left( - \sum_{n > 0} \frac{ \sqrt{2}}{n} \ \Al_{n} z^{-n} \right) c_{\br_1} \, , \\[2mm]
{}^{[\ell]}\!F_m &\,\coloneqq\, - \oint \frac{\mathrm{d}z}{2\pi i} \ z^{\ell m} \ 
\exp\left( - \sum_{n > 0} \frac{ \sqrt{2}}{n} \ \Al_{-n} z^n \right) 
e^{- i \br_1 \mathbf{Q}} z^{- \br_1 \mathbf{P}} 
\exp\left( + \sum_{n > 0} \frac{ \sqrt{2}}{n} \ \Al_{n} z^{-n} \right) c_{- \br_1} \, , 
\end{aligned}
\end{align}
where $c_{\pm \br_1}$ denotes the cocycle factor (see Appendix~\ref{app:Cocycle}),
and where we have written out the normal ordering. The extra label on the affine generators
is meant to indicate that these formulas are only valid on the corresponding
level-$\ell$ subspace $\mF^{(\ell)}$ (as always we drop the label $\ell$
whenever statements are valid for all levels).
As explained in~\cite{GN1} the zero mode contribution 
$\propto\br_1\cdot\bQ$ is {\em not} a shift in momentum space, but rather 
a Lorentz boost which rotates the tachyonic momenta~\eqref{TachyonMomentum}
belonging to different depths into one another, as appropriate. One can now check
that the operators~\eqref{Hm} and~\eqref{EmFm} do satisfy the commutation relations
\eqref{AffineCom} on the level-$\ell$ subspace.

All affine generators act on DDF states~\eqref{PhysState}. The result of such an action
can then be properly re-ordered by moving all annihilation operators to the right by means 
of~\eqref{DDF} and~\eqref{BAcomm}, until they annihilate the tachyonic ground state 
by virtue of~\eqref{Alma}, to obtain a new DDF state. $\Al_0$ commutes with all other 
(transversal and longitudinal) DDF operators.

In the sequel we will refer to the Ehlers $\msl(2)$ multiplet belonging to an 
affine highest weight state as an {\bf affine ground state} or
{\bf affine ground state multiplet}. The members $\psi$ of this $\msl(2)$ multiplet obey 
\beq\label{affg}
T_m \psi \,=\, 0 \quad \text{for all} \quad T_m \in \big\{ E_m, F_m , H_m \ | \ m \ge 1 \big\} \, .
\eeq
For level one, there is
only one affine ground state, the $\msl(2)$ singlet $\ket{-\brr}$. By contrast,
for $|\ell|\geq 2$ there will be infinitely many such affine ground 
states for all levels, of increasing depths. In all cases the action of the 
affine generators on such states is straightforward to evaluate by means of formulas 
\eqref{Hm} and~\eqref{EmFm}, with the result again being a DDF state.

In addition to the affine generators, the Sugawara generators can also be represented
in DDF form {\em for arbitrary levels} by plugging~\eqref{Hm} and~\eqref{EmFm} 
into~\eqref{Sugawara}~\cite{GKN,GN1}. 
This has been done explicitly in~\cite{GN1} for general hyperbolic Kac-Moody algebras. 
We specialize the result
given in (3.15) of~\cite{GN1} to $\mF$ by setting $d= 3$ and $h^\vee = 2$ to arrive at
\begin{align}\label{Sug2}
\begin{aligned}
{}^{[\ell]}\!\cL_m^\text{sug}(\ba) &\,=\, \frac{1}{2\ell} 
\sum_n \DD \, \Al_{\ell n} \Al_{\ell (m-n)} \DD 
+ \frac{1}{\ell( \ell +2)} \sum_{n \not= 0 \!\!\!\!\mod \ell} \DD \, \Al_{n} \Al_{\ell m-n} \DD 
+ g_\ell (\br_1\cdot \ba) \delta_{m,0} \\
&\quad - \frac{1}{2\ell(\ell+2)} \sum_{p=1}^{\ell-1} 
\frac{\zeta^{p \, \br_1 \cdot \ba}}{|\zeta^p - 1|^2} \oint \frac{\mathrm{d}z}{2\pi i} \ 
z^{\ell m-1} \, \DD 
\left( \exp\left[ \sum_{n \not= 0} \frac{\sqrt{2}}{n} \ \Al_n z^{-n} (1- \zeta^{-pn}) \right] 
- 1 \right) \DD \\
&\quad - \frac{1}{2\ell (\ell +2)} \sum_{p=1}^{\ell -1} 
\frac{\zeta^{-p \, \br_1 \cdot \ba}}{|\zeta^p - 1|^2} \oint \frac{\mathrm{d}z}{2\pi i} \ 
z^{\ell m-1} \, \DD \left( \exp\left[ - \sum_{n \not= 0} \frac{\sqrt{2}}{n} \ 
\Al_n z^{-n} (1- \zeta^{-pn}) \right] - 1 \right) \DD \, , 
\end{aligned}
\end{align}
where again $\ba$ is the tachyonic momentum of the state on which the 
Sugawara operator acts, and $\zeta$ is a primitive $\ell$-th root of unity.
Furthermore, we have introduced the periodic function
\beq
g_\ell (k) \,\coloneqq\, 
 \frac{\ell ^2-1}{12 \ell (\ell+2)} - \frac{f_\ell(k)}{\ell(\ell+2)} \\
\eeq
with~\cite{GKN}
\beq\label{fk}
f_\ell(k) \,\coloneqq\, \sum_{p=1}^{\ell-1} \frac{\zeta^{p k }}{|\zeta^p - 1|^2} 
\,=\, \frac{\ell^2-1}{12} - \frac{k(\ell-k)}{2} 
\qquad \text{for} \quad 0 \le k \le \ell-1 \, .
\eeq
This definition of $f_\ell(k)$ can be extended to all integer values of $k$ by 
means of the relation (which actually follows from the definition~\eqref{fk}
\beq
f_\ell(k) \,=\, f_\ell(k+n\ell) \,=\, f_\ell(-k) \, ,
\eeq
thus ensuring symmetry and periodicity of $g_\ell(k)$.
As before, the operator~\eqref{Sug2} is to be used and valid only
on the level-$\ell$ subspace $\mF^{(\ell)}$. As we can see,~\eqref{Sug2}, being 
non-polynomial, 
is no longer equivalent to a free field construction. Nevertheless, the
action of the Sugawara generators is again straightforward to evaluate
on any DDF state by means of~\eqref{Sug2}. 

\subsection{Ground states}\label{sec:MG}
As we explained in section~\ref{sec:2}, the crucial ingredients in analyzing the
higher level sectors of $\mF$ are the affine representation theory and 
the level-$\ell$ coset Virasoro algebra. Their combination motivates
the following definition valid for all levels $|\ell|\geq 2$.
\begin{definition}\label{def1}~\\
A {\bf Virasoro ground state (multiplet)} $\Psi^{(\ell)}\in \mF^{(1)}\otimes \mF^{(\ell -1)}$ at 
level $\ell\geq 2$ is an affine ground state multiplet, which in addition 
to (\ref{affg}) obeys the conditions
\beq\label{VirasoroGS}
{}^{[\ell]}\!\mL_m^{\rm{coset}}\,\Psi^{(\ell)} \,=\, 
0 \quad \text{for all} \quad m \ge 1 \, .
\eeq
\end{definition}
These product states are non-vanishing elements of the tensor product, but may
vanish as elements of the Lie algebra, that is, after the conversion of the tensor 
product into a Lie algebra commutator by means of the prescription~\eqref{StateOperator}.
In that case we refer to $\Psi^{(\ell)}$ as a {\em virtual ground state}. By definition
virtual ground states thus belong to the kernel of $\cI^{(\ell)}$ defined in~\eqref{Iell}. 

We can distinguish two level-$\ell$ Virasoro ground state multiplets of the same type
by their depth and ${}^{[\ell]}\!\mL_0^\text{coset}$ eigenvalue. On level 2 there are finitely 
many Virasoro ground states and on levels $\ell \ge 3$ there are infinitely many.
This motivates the additional definition.
\begin{definition}~\\
Let $\Psi^{(\ell)}\in \mF^{(1)}\otimes \mF^{(\ell -1)}$ be a Virasoro ground state at level
$\ell\geq 2$ and depth $\md$ with ${}^{[\ell]}\!\mL_0^\text{coset}$ eigenvalue 
$h_{r,s}^{(\ell)}$. Then $\Psi^{(\ell)}$ is said to be a {\bf maximal (Virasoro) ground
state (multiplet)} if there is no level-$\ell$ Virasoro ground state with
the same ${}^{[\ell]}\!\mL_0^\text{coset}$ eigenvalue and depth $< \md$.
\end{definition}
The number of maximal ground states on any level $\ell > 2$ is finite, and essentially in
one-to-one correspondence with the admissible level-$\ell$ weights (\ref{Lambdaell}).
At level 2 all Virasoro ground states are maximal. Moreover, all maximal ground states at
levels 2 and 3 are virtual. On level $\ell =4$ not all maximal ground states are virtual.

The main observation is now that one can generate from the collection of all Virasoro
ground states from Defintion~\ref{def1} {\em all} states in $\mF^{(1)}\otimes \mF^{(\ell-1)}$
by the combined action of the affine and 
coset Virasoro generators. Because $\cI^{(\ell)}$ is surjective, we obtain in this way
{\em all} elements of $\mF^{(\ell)}$ as DDF states after conversion of the tensor 
product to a Lie algebra commutator by means of~\eqref{StateOperator}, that is under 
the map $\cI^{(\ell)}$. The main open problem is then the determination of 
${\rm Ker}\,\cI^{(\ell)}$.
In section~\ref{sec:L3GS} we show at $\ell=3$ that we can obtain all Virasoro ground 
states from the finite set of maximal ground states with yet another operator. We
furthermore argue that there exists a generalization of this operator to all higher levels.
Thus, it would be enough to know the finite set of maximal ground states.

In practical terms the evaluation of the tensor product $\mF^{(1)} \otimes \mF^{(\ell -1)}$
boils down to the evaluation of a finite number of affine tensor products, with the previously
generated Virasoro representations for $\mF^{(\ell -1)}$ as `spectators'. 
Schematically, the general structure of such a tensor product is 
\beq\label{TP}
L(\bL_0 +2\bd) \otimes L(\bL^{(\ell -1)}) \,=\,
\bigoplus_a \, \Vir(c_\ell, h_{a}^{(\ell)}) \otimes L(\bL_a^{(\ell)} + n_a\bd)
\eeq
where $\bL^{(\ell-1)}$ is any admissible level-$(\ell -1)$ weight from
\eqref{Lambdaell}, that is with $p_0 +2p_1 = \ell -1$.
Likewise, $\bL_a^{(\ell)}$ is an admissible level-$\ell$ weight
from~\eqref{Lambdaell}, while the parameter $h_a^{(\ell)}$ is from the 
list~\eqref{hrs} of allowed level-$\ell$ eigenvalues of ${}^{[\ell]}\mL_0^{\text{coset}}$. 
In this schematic form, not all values of $h_a^{(\ell)} = h_{r,s}^{(\ell)}$ from~\eqref{hrs} 
are meant to occur, and which ones do depends on the specific weights on the left-hand
side. The formula~\eqref{TP} illustrates the {\em infinite reducibility} of
such tensor products, with the coset Virasoro algebra as an additional
ingredient to handle and distinguish an infinite number of identical
copies of the same affine representation, with appropriately shifted
(integer) coefficients $p_{-1}$ in~\eqref{Lambdaell}. This feature characterizes 
all tensor products for higher levels. 

The coset Virasoro module $ \Vir(c_\ell, h_a^{(\ell)})$ associated with the tensor 
product decomposition thus records the infinitely many isomorphic repetitions of the same
affine module with highest weight $\bL^{(\ell)}$, that occur shifted by multiples of $\bd$. 
Our convention here is that the coefficient $n_a$ is such that $\bL_a^{(\ell)} + n_a\bd$
is the first (highest) instance of the infinite repetition of affine highest weight states. 
In particular, there is always the case with $h_a^{(\ell)}=0$ (corresponding to $r=s=1$
in~\eqref{hrs} and $\bL_a^{(\ell)} + n_a \bd = \bL_0+2\bd + \bL^{(\ell-1)}$, corresponding
to the tensor product of the two affine highest weight states on the left-hand side 
of~\eqref{TP}. In Figures~\ref{fig:L2RootLattice}--\ref{fig:L4RootLattice}, we show the 
root diagrams of $\mF^{(\ell)}$ for $2 \leq \ell \leq 4$ and in those diagrams this highest
weight vector corresponds to the right-most red diamond.

When we pass to the characters corresponding to the above product
the notation is further refined by writing the character of a term
on the r.h.s. of~\eqref{TP} as 
\beq
\Ch \Vir\big(c_\ell, h_a^{(\ell)}\big)(q) \,\cdot\, 
\Ch L\big(\bL_a + (n_a +h_a^{(\ell)})\bd\big) \, .
\eeq
That is, we assign a fractional multiple of $-h_a^{(\ell)}\bd$ of the null root to the Virasoro 
character, in accord with the fact that we define the (minimal) Virasoro characters by 
${\rm Tr} (q^{L_0}) = q^h + \cdots$, see appendix~\ref{app:character}, and recall that 
$q=e^{-\bd}$. This shift has to be compensated 
for in the affine character, which explains the extra shift shown in the formula. 
Matters are further complicated by the fact that each affine representation 
$L(\bL^{(\ell -1)})$ in $\mF^{(\ell -1)}$ comes with its own baggage of factors 
of `porous' coset Virasoro representations from previous representation products, whose 
$\bd$-shifts must also be taken into account in the final formulas. This accounts for
increasingly more complicated patterns of fractional powers of $q$ in the final character
formulas.

\section{Level 2}

Generally speaking, the level-2 sector 
is spanned by all commutators of all level-one elements. It is thus contained 
in the antisymmetric product of two basic representations~\cite{FF}, which
results in~\eqref{L0L0}. The crucial point is that not all 
elements of this tensor product belong to the Lie algebra because some of 
them vanish on account of the Serre relations (and at higher levels 
also the Jacobi identities) once the tensor product is converted to a 
Lie algebra commutator. More precisely, we have~\cite{FF}
\beq
{\rm Ker}\, \cI^{(2)} \,=\, \cJ_2 \,=\, L\big(2\bL_1 + 3\bd\big) \, .
\eeq

\subsection{Maximal ground states for \texorpdfstring{$\ell =2$}{l=2}}
From~\eqref{Level2} we can read off that the dominant state of the virtual maximal ground
state multiplet has momentum $2\bL_1 + 3 \bd = -2\br_{-1} - \bd + \br_1$. 
So it belongs to a triplet and sits at depth $1$, as can also be seen from the position of the
right-most red diamond in Figure~\ref{fig:L2RootLattice}.
This triplet belongs to a coset Virasoro representation with eigenvalue $h= \frac12$. We
shall see below that this is enough information to uniquely characterize them.
Specifically, for level 2 there are no Virasoro ground state multiplets besides the maximal
ground state multiplet. The multiplet is an $\msl(2)$ triplet in 
$\mF^{(1)} \we \mF^{(1)}$ which consists of the three product states built out of
level-one DDF states
\beq\label{VTrip}
\Psi_{1,\bD,\pm 1}^{(2)} \,=\, \ket{\ba^{(1)}_{\pm 1}} \wedge \ket{\ba^{(1)}_0} \quad \text{and}
\quad \Psi_{1,\bD,0}^{(2)} \,=\, - \frac{1}{\sqrt{2}} \ {}^{[1]}\!A_{-1} \ket{\ba^{(1)}_0} 
\wedge \ket{\ba^{(1)}_0 } \, .
\eeq
The labeling on $\Psi$ is as follows:
\begin{itemize}\setlength\itemsep{0mm}
\item the first subscript gives 
the depth,
\item the second the $\msl(2)$ representation through its dimension,
\item and
the third entry is half the $h_1$ weight of the corresponding state in the given
$\msl(2)$ representation, 
\item while the superscript indicates the level.
\end{itemize}
These elements are perfectly well-defined (up to normalization) and non-vanishing as
elements of the tensor product $\mF^{(1)} \wedge \mF^{(1)}$, but when we convert 
the wedge products to actual commutators with~\eqref{StateOperator} they vanish
\beq
\big[ \ket{\ba^{(1)}_{\pm 1}} \,,\, \ket{\ba^{(1)}_0} \big] \,=\,
\big[ A_{-1} \ket{\ba^{(1)}_0} \,,\,\ket{\ba^{(1)}_0} \big] \,=\, 0
\eeq
because the associated momenta obey $(-2\brr - \bd \pm\br_1)^2 = 6 >2$ 
and $(-2\brr -\bd)^2 = 4 > 2$. In terms of multi-commutators of Chevalley-Serre
generators, the first of these states corresponds to $[f_{-1},[f_{-1},f_0]]$ which
vanishes by the Serre relation. However, here this vanishing does 
not need to be imposed `by hand' but rather follows directly from 
the formula~\eqref{StateOperator}. The ${}^{[2]}\!\mL_{0}^\text{coset}$
eigenvalue of any of the triplet states~\eqref{VTrip} is $\frac12$ in agreement 
with~\eqref{hrs}.

Acting with ${}^{[2]}\!\mL_{-1}^\text{coset}$ on the virtual triplet states 
we obtain three descendant states in $\mF^{(1)} \wedge \mF^{(1)}$
\beq\label{VTrip1}
{}^{[2]}\!\mL_{-1}^\text{coset} \Psi_{1,\bD,\pm 1}^{(2)} \,=\, 
\sqrt{2} \ket{\ba^{(1)}_{\pm 1}} \wedge A_{-1} \ket{\ba^{(1)}_0} \quad \text{and}
\quad {}^{[2]}\!\mL_{-1}^\text{coset} \Psi_{1,\bD,0}^{(2)} \,=\, 
- \frac{1}{\sqrt{2}} \ {}^{[1]}\!A_{-2} \ket{\ba^{(1)}_0} \wedge \ket{\ba^{(1)}_0 } \, , 
\eeq
which now carry momenta 
$2\bL_1 + 2\bd = -2\brr - 2\bd + \br_1$, $2\bL_1 + 2\bd-2\br_1 = -2\brr - 2\bd - \br_1$ 
and $2\bL_0 + 2\bd = -2\brr - 2\bd$ which square to
2, 2, and 0, respectively. Hence, the commutators no longer vanish but give 
honest non-vanishing elements of $\mF^{(2)}$. Explicitly, we obtain the following
$\msl(2)$ triplet states after the evaluation of the commutators by means
of~\eqref{StateOperator} (with the root labels in the left column)
\begin{align}\label{VTrip2}
\begin{aligned}
&-(2,2,3): &&\qquad\quad \psi_{2,\bD,-1}^{(2)} &&\hspace*{-.3cm}=\, 
\sqrt{2} \left[ \ket{\ba_{-1}^{(1)}} , \Ae_{-1} \ket{\ba_0^{(1)}} \right] 
&&\hspace*{-.3cm}=\, 2 \, \ket{\ba_{-1}^{(2)}} \, , \\
&-(2,2,2): &&\qquad\quad \psi_{2,\bD,0}^{(2)} 
&&\hspace*{-.3cm}=\, - \left[ \ket{\ba_{-1}^{(1)}} , \ket{\ba_1^{(1)}} \right] 
+ L_{-1}(\ldots) &&\hspace*{-.3cm}=\, - \sqrt{2} \ \Az_{-1} \ket{\ba_{0}^{(2)}} \, ,\\
&-(2,2,1): &&\qquad\quad \psi_{2,\bD,1}^{(2)} 
&&\hspace*{-.3cm}=\, - \sqrt{2} \left[ \ket{\ba_{1}^{(1)}} , \Ae_{-1} \ket{\ba_0^{(1)}} \right]
&&\hspace*{-.3cm}=\, 2 \, \ket{\ba_{1}^{(2)}} \, .
\end{aligned}
\end{align}
These three states form an affine ground state triplet at depth 2,
hence we denote them by $\psi^{(\ell)}$. In the following, descendant states which are 
not affine ground states will be denoted by $\phi^{(\ell)}$ with appropriate indices, as 
in~\eqref{VTrip2}. In this way we distinguish the elements of $\mF^{(\ell)}$ from 
the (virtual) ground states $\Psi^{(\ell)}$ that live in $\mF^{(1)} \otimes \mF^{(\ell-1)}$.
In general, there will always be $L_{-1}(\ldots)$ contributions to the 
DDF commutators, which must be dropped by~\eqref{Lm1}~\cite{GN}. Notice that 
there are no cocycle factors appearing anywhere in this section with our 
conventions in Appendix~\ref{app:Cocycle}. 

Let us also note that an analogous triplet exists {\em for all levels} 
with the three level-$\ell$ states
\beq\label{Triplet} 
2 \, \big| \ba^{(\ell)}_{\pm 1} \big\rangle \quad \text{and} \quad
- \sqrt{2} \, \Al_{-1} \big| \ba^{(\ell)}_0\big\rangle
\eeq
at depth $\ell$, and a virtual triplet analogous to~\eqref{VTrip} at depth $(\ell-1)$.
For the computation we evaluate the action of any coset
Virasoro generator on a tensor product state by means of formula~\eqref{mLm}, where
we use the standard expression for the Sugawara generators on the separate factors,
and the original expression~\eqref{Sugawara} on the third term in~\eqref{mLm},
a procedure that will also work at all higher levels. As we will see 
we can generate the full level-2 sector by repeated application of the
affine and coset Virasoro raising operators to the virtual ground state multiplet.

At level $\ell=2$ and depth 3 we have a total of seven DDF states that form two triplets and a singlet. 
These states are (without the commutators that generate them)
\begin{align}\label{someDDF}
\begin{aligned}
&\color{myblue}-(2,3,4) :&&\qquad\ \color{myblue} \psi_{3,\bD,-1}^{(2)}\,=\,
\left[ \Az_{-1} \Az_{-1} + \frac{3}{2} \ \Bz_{-2} \right] \ket{\ba_{-1}^{(2)}} 
\color{black} \, , \\[2ex]
&\color{myblue}-(2,3,3): &&\qquad \ \color{myblue} \psi_{3,\bD,0}^{(2)}\,=\, 
\left[ - \frac{7}{6\sqrt{2}} \ \Az_{-3} 
+ \frac{2\sqrt{2}}{3} \ \Az_{-1} \Az_{-1} \Az_{-1} 
- \frac{3}{2\sqrt{2}} \ \Az_{-1} \Bz_{-2} \right] \ket{\ba_0^{(2)}} \color{black} \, , \\[2ex]
&\color{myblue}-(2,3,2): &&\qquad\ \color{myblue} \psi_{3,\bD,1}^{(2)}\,=\,
\left[ \Az_{-1} \Az_{-1} + \frac{3}{2} \ \Bz_{-2} \right] \ket{\ba_1^{(2)}} \color{black} \, , 
\\[2ex]
&\color{myred}-(2,3,3): &&\qquad\ \color{myred} \phi_{3,\bE,0}^{(2)}\,=\,
- 4 \ \Az_{-2} \Az_{-1} \ket{\ba_0^{(2)}} \color{black} \, , \\[2ex]
&\color{mygreen}-(2,3,4): &&\qquad\ \color{mygreen} \phi_{3,\bD,-1}^{(2)}\,=\,
\sqrt{2} \ \Az_{-2} \ket{\ba_1^{(2)}} \color{black} \, , \\[2ex]
&\color{mygreen}-(2,3,3): &&\qquad\ \color{mygreen} \phi_{3,\bD,0}^{(2)}\,=\,
\left[ - \frac{\sqrt{2}}{3} \ \Az_{-3} - \frac{\sqrt{2}}{3} \ \Az_{-1} \Az_{-1} \Az_{-1} \right]
 \ket{\ba_0^{(2)}} \color{black} \, , \\[2ex]
&\color{mygreen}-(2,3,2): 
&&\qquad\ \color{mygreen} \phi_{3,\bD,1}^{(2)}\,=\,- \sqrt{2} \ \Az_{-2} \ket{\ba_{-1}^{(2)}}
 \color{black} \, .
\end{aligned}
\end{align}
The {\color{myblue}blue} states form an affine ground state triplet. The
{\color{myred}red} state is the singlet, and the {\color{mygreen}green} states form the
remaining triplet. The latter two are affine descendants of~\eqref{VTrip2}, as can also be
recognized from the fact that there appear no longitudinal states. By contrast, the blue
states are the result of the action of the coset Virasoro raising operators; the longitudinal
states result from the evaluation of the commutators. 

Continuing in this way, and starting from the dominant root of the maximal ground state
triplet, we construct the level-2 part of the root lattice shown in Fig.~\ref{fig:L2RootLattice}.
Expressing the first seven rows of Fig.~\ref{fig:L2RootLattice} in a 
table, we obtain Table~\ref{tab:lev7}.
\begin{table}
\begin{tabular}{|c|c|c|c|c|}
\hline
\ level \ & \ depth \ & $-(a_{-1}, a_0, a_1)$ & dim $\mathfrak{sl}(2)$ rep. 
& \ outer multiplicity \ \\
\hline
2 & 2 & (2,2,3) (2,2,2) (2,2,1) & 3 & 1 \\
\hline
2 & 3 & (2,3,4) (2,3,3) (2,3,2) & 3 & 2 \\
2 & 3 & (2,3,3) & 1 & 1 \\
\hline
2 & 4 & (2,4,6) \ldots (2,4,2) & 5 & 1 \\
2 & 4 & (2,4,5) (2,4,4) (2,4,3) & 3 & 4 \\
2 & 4 & (2,4,4) & 1 & 2 \\
\hline
2 & 5 & (2,5,7) \ldots (2,5,3) & 5 & 3 \\
2 & 5 & (2,5,6) (2,5,5) (2,5,4) & 3 & 8 \\
2 & 5 & (2,5,5) & 1 & 4 \\
\hline
2 & 6 & (2,6,9) \ldots (2,6,3) & 7 & 1 \\
2 & 6 & (2,6,8) \ldots (2,6,4) & 5 & 6 \\
2 & 6 & (2,6,7) (2,6,6) (2,6,5) & 3 & 15 \\
2 & 6 & (2,6,6) & 1 & 8 \\
\hline
2 & 7 & (2,7,10) \ldots (2,7,4) & 7 & 2 \\
2 & 7 & (2,7,9) \ldots (2,7,5) & 5 & 13 \\
2 & 7 & (2,7,8) (2,7,7) (2,7,6) & 3 & 27 \\
2 & 7 & (2,7,7) & 1 & 14 \\
\hline
\end{tabular}
\caption{\label{tab:lev7}Specific information about the $\mathfrak{sl}(2)$ multiplets at level-2 of 
$\mF^{(2)}$ up to depth 7.}
\end{table}
In appendix~\ref{app:mc}, we rewrite the states above in terms of multi-commutators 
of Chevalley-Serre generators.

\subsection{Coset Virasoro action}
As we already pointed out, there is no proper coset Virasoro representation 
on level-2 of $\mF$, because the Virasoro ground state triplet is absent from $\mF^{(2)}$. 
Acting with the coset Virasoro operators on the virtual triplet yields
\begin{align}
\begin{aligned}
{}^{[2]}\!\mL_{1}^{\text{coset}} \Psi_{1,\bD,w}^{(2)} \,&=\, 0 \, , \\
{}^{[2]}\!\mL_{0}^{\text{coset}} \Psi_{1,\bD,w}^{(2)} \,&=\, 
\frac{1}{2} \Psi_{1,\bD,w}^{(2)} \, , \\
\cI^{(2)} \left( {}^{[2]}\!\mL_{-1}^{\text{coset}} \Psi_{1,\bD,w}^{(2)} \right) \,&=\, 
\psi_{2,\bD,w}^{(2)} \, , \\
\cI^{(2)} \left({}^{[2]}\!\mL_{-2}^{\text{coset}} \Psi_{1,\bD,w}^{(2)} \right) \,&=\, 
\frac{3}{4} \psi_{3,\bD,w}^{(2)} \, .
\end{aligned}
\end{align}
The coset Virasoro eigenvalue of the virtual Virasoro ground states $\Psi_{1,\bD,w}^{(2)}$ 
is exactly the one we expect from~\eqref{Level2} and also $c_2 = \frac{1}{2}$. We see 
from the results above that we always obtain affine ground states when acting with 
${}^{[2]}\!\mL_{m}^{\text{coset}}$ for $m \le 0$ on $\Psi_{1,\bD,w}^{(2)}$. 
To show this we first use~\eqref{SugTmCom} which implies that the coset Virasoro
operator commutes with the affine generators
\beq\label{cosetAffine1}
[\, {}^{[\ell]}\!\mL_m^{\text{coset}} , T_n ] \, \Phi \,=\, 0 
\eeq
for any state $\Phi \in \mF^{(1)}\otimes \mF^{(\ell-1)}$. Taking $\Phi = \Psi_{1,\bD,w}^{(2)}$,
it is straightforward with the formulas~\eqref{EmFm} to check that 
\beq
E_1 \Psi_{1,\bD,-1}^{(2)} \,=\, H_1 \Psi_{1,\bD,-1}^{(2)} \,=\, E_1 \Psi_{1,\bD,0}^{(2)} \,=\, 
F_1 \Psi_{1,\bD,0}^{(2)} \,=\, F_1 \Psi_{1,\bD,1}^{(2)} \,=\, H_1 \Psi_{1,\bD,1}^{(2)} =0 
\eeq
and
\beq
F_1 \Psi_{1,\bD,-1}^{(2)} \,=\, H_1 \Psi_{1,\bD,0}^{(2)} \,=\, E_1 \Psi_{1,\bD,1}^{(2)} \,=\,
\ket{\ba_0^{(1)}} \otimes \ket{\ba_0^{(1)}} \, .
\eeq
Moreover,
\begin{align}\label{cosetAffine2}
\begin{aligned}
{}^{[2]}\!\mL_n^\text{coset} \left( \ket{\ba_0^{(1)}} \otimes \ket{\ba_0^{(1)}} \right) \,&=\,
 {}^{[1]}\!\mathcal{L}_n^\text{sug} \ket{\ba_0^{(1)}} \otimes \ket{\ba_0^{(1)}} 
+ \ket{\ba_0^{(1)}} \otimes {}^{[1]}\!\mathcal{L}_n^\text{sug} \ket{\ba_0^{(1)}}
- {}^{[2]}\!\mathcal{L}_n^\text{sug} \left( \ket{\ba_0^{(1)}} \otimes \ket{\ba_0^{(1)}} \right) \\
 \,&=\, L_{-1} ( \ldots ) 
\end{aligned}
\end{align}
because the argument is symmetric. Hence, the virtual triplet generates all affine ground
states on level-2 of $\mF$. 

There are two ways to reach the affine ground states $\psi_{3,\bD,w}^{(2)}$ at depth 3 
by applying either $ {}^{[2]}\!\mL_{-2}^{\text{coset}}$ or 
$\Big({}^{[2]}\!\mL_{-1}^{\text{coset}} \Big)^2$ to $\Psi_{1,\bD,w}$ and they must 
be related. This is expressed by the relation 
\beq
\left[ {}^{[2]}\!\mL_{-2}^{\text{coset}} - \frac34 
\Big({}^{[2]}\!\mL_{-1}^{\text{coset}} \Big)^2 \right] \Psi_{1,\bD,w}^{(2)} \,=\, 0
\eeq
in agreement with the first null vector from $\Vir(\ft,\ft)$~\cite{DiFrancesco}. 
For the next two coset Virasoro operators we find the relations
\beq
\left[ \frac{3}{4} (4n_1 + 3 n_2) {}^{[2]}\!\mL_{-3}^{\text{coset}} 
- n_2 \, {}^{[2]}\!\mL_{-1}^{\text{coset}} {}^{[2]}\!\mL_{-2}^{\text{coset}} 
- n_1 \Big({}^{[2]}\!\mL_{-1}^{\text{coset}} \Big)^3 \right] \Psi_{1,\bD,w} ^{(2)}\,=\, 0
\eeq
and
\begin{align}
\begin{aligned}
&\bigg[ (24m_1 + 18 m_2 + 8 m_3) {}^{[2]}\!\mL_{-4}^{\text{coset}} 
- \frac{4}{3} (12 m_1 + 9 m_2 + 4m_3) \Big( {}^{[2]}\!\mL_{-2}^{\text{coset}} \Big)^2 \\
&\quad + m_3 \, {}^{[2]}\!\mL_{-1}^{\text{coset}} {}^{[2]}\!\mL_{-3}^{\text{coset}} 
+ m_2 \Big( {}^{[2]}\!\mL_{-1}^{\text{coset}} \Big)^2 {}^{[2]}\!\mL_{-2}^{\text{coset}} 
+ m_1 \Big( {}^{[2]}\!\mL_{-1}^{\text{coset}} \Big)^4 \bigg] \Psi_{1,\bD,w}^{(2)} \,=\, 0 
\end{aligned}
\end{align}
from which all null vectors can be read off by choosing the coefficients 
$m_i, n_i \in \mathbb{C}$ appropriately.
Let us stress once more that the Virasoro null vectors are actually zero 
in our formalism. Altogether, the action of the coset Virasoro operator on the 
virtual maximal ground states thus yields two new multiplets at depth 5, 
in agreement with the general theory~\cite{DiFrancesco}. The dominant states 
in these multiplets are
\begin{align}
\begin{aligned}
\cI^{(2)} \left( {}^{[2]}\!\mL_{-4}^{\text{coset}} \Psi_{1,\bD,1}^{(2)} \right) \,=\, 
\bigg[ &\frac{51}{80} \ \Az_{-5} \Az_{-1} 
+ \frac{1}{144} \ \Az_{-3} \Az_{-3} 
+ \frac{11}{36} \ \Az_{-3} \Az_{-1} \Az_{-1} \Az_{-1} \\
& - \frac{23}{360} \ \Az_{-1} \Az_{-1} \Az_{-1} \Az_{-1} \Az_{-1} \Az_{-1} \\
& + \frac{3}{8} \ \Az_{-3} \Az_{-1} \Bz_{-2} 
+ \frac{1}{32} \ \Az_{-1} \Az_{-1} \Bz_{-2} \Bz_{-2} \\
& + \frac{1}{16} \ \Az_{-1} \Az_{-1} \Az_{-1} \Az_{-1} \Bz_{-2} 
+ \frac{11}{192} \ \Bz_{-2} \Bz_{-2} \Bz_{-2} \bigg] \ket{\ba_1^{(2)}} 
\end{aligned}
\end{align}
and
\begin{align}
\begin{aligned}
\cI^{(2)} \left({}^{[2]}\!\mL_{-3}^{\text{coset}} {}^{[2]}\!\mL_{-1}^{\text{coset}} 
\Psi_{1,\bD,1}^{(2)} \right) \,=\, \bigg[ &\frac{73}{80} \ \Az_{-5} \Az_{-1} 
+ \frac{11}{16} \ \Az_{-3} \Az_{-3} \\
&- \frac{1}{12} \ \Az_{-3} \Az_{-1} \Az_{-1} \Az_{-1} \\
& + \frac{17}{120} \ \Az_{-1} \Az_{-1} \Az_{-1} \Az_{-1} \Az_{-1} \Az_{-1} \\
& + \frac{1}{8} \ \Az_{-3} \Az_{-1} \Bz_{-2} 
+ \frac{3}{32} \ \Az_{-1} \Az_{-1} \Bz_{-2} \Bz_{-2} \\
& - \frac{1}{16} \ \Az_{-1} \Az_{-1} \Az_{-1} \Az_{-1} \Bz_{-2} \\
&+ \frac{13}{192} \ \Bz_{-2} \Bz_{-2} \Bz_{-2} \bigg] \ket{\ba_1^{(2)}} \\
& + {}^{[2]}\! H_{-1} \bigg[ \frac{5}{12} \ \Az_{-3} \Az_{-1} 
+ \frac{1}{24} \ \Az_{-1} \Az_{-1} \Az_{-1} \Az_{-1} \\
& \hphantom{+ {}^{[2]}\! H_{-1} \bigg[} + \frac{1}{8} \ \Az_{-1} \Az_{-1} \Bz_{-2} 
+ \frac{15}{96} \ \Bz_{-2} \Bz_{-2} \bigg] \ket{\ba_1^{(2)}}\, .
\end{aligned}
\end{align}
Notice that the result of the second calculation is not a pure affine ground state. 
This is no contradiction to~\eqref{cosetAffine1}--\eqref{cosetAffine2}.

We have already mentioned that $\mF$ does not decompose into proper representations
of the coset Virasoro algebra, as we will now make more explicit for $\mF^{(2)}$.
Hence, we shall work on the tensor product space and exchange tensor products for
commutators only after acting with the coset Virasoro operators. The tensor product space
forms a representation of the coset Virasoro algebra
\beq\label{VirAlgebra}
\left[ {}^{[\ell]}\!\mL_m^{\text{coset}} , {}^{[\ell]}\!\mL_n^{\text{coset}} \right] \,=\,
(m-n) \ {}^{[\ell]}\!\mL_{m+n}^{\text{coset}} 
+ \frac{c_\ell^{\text{coset}}}{12} (m^3 - m) \delta_{m+n,0} \, , 
\eeq
with central charge~\eqref{CentralCharge}.
The important point now is that on $\mF$ this algebra is {\em not} satisfied because we do
not have a complete Virasoro representation. This can be seen if one tries to apply
formula~\eqref{mLm} in the form
\beq\label{mLm1}
{}^{[\ell]}\!\mL_m^\text{coset} w \,\equiv\, {}^{[\ell]}\!\mL_m^\text{coset} 
\left( \left[ u \,,\, v \right]\right)
\,\stackrel{?}{=} \, \left[ {}^{[1]}\!\cL_m^\text{sug} u \,,\, v \right] \,+\, 
\left[ u \,,\, {}^{[\ell -1]}\!\cL_m^\text{sug} v \right] \, \,-\, {}^{[\ell]}\!\cL_m^\text{sug} w \, .
\eeq
Indeed, this erroneous application of the formula~\eqref{mLm} leads
to contradictions whenever the third term on the r.h.s. of~\eqref{mLm1} vanishes 
on account of the Serre relations. For example, we find for the dominant state of the
maximal ground state triplets
\beq
\left[ {}^{[2]}\!\mL_1^{\text{coset}} , {}^{[2]}\!\mL_{-1}^{\text{coset}} \right]
\psi_{2,\bD,1}^{(2)} \,=\, {}^{[2]}\!\mL_1^{\text{coset}} 
\left( - \psi_{3,\bD,1}^{(2)} + \phi_{3,\bD,1}^{(2)} \right) 
- {}^{[2]}\!\mL_{-1}^{\text{coset}} 0 \,=\, 4 \psi_{2,\bD,1}^{(2)} \, .
\eeq
By contrast, direct application of the result of the Virasoro commutator leads 
to a different answer:
\beq
2 \ {}^{[l]}\!\mL_0^{\text{coset}} \psi_{2,\bD,1}^{(2)} \,=\, 3 \psi_{2,\bD,1}^{(2)} 
\eeq
in contradiction with the Virasoro algebra.

\subsection{Affine characters at \texorpdfstring{$\ell =2$}{l=2}}
Of course, our findings can be rephrased using characters, the tool
mostly employed in the mathematical literature.
Recall that level 2 of $\mF$ is formally given by~\eqref{mg2}~\cite{FF}. 
In terms of characters this translates into the formula
\beq\label{Level2Ch} 
\Ch\, \mF^{(2)} \,=\,
q^{-7/2} \ \left( \chi_{2,1}^{4,3}(q) - q^{1/2} \right) \,\cdot \, \Ch\, L(2\bL_1) \, ,
\eeq
where we use the notation
\beq
\Ch\,\Vir \big(c^{p,p^\prime},h_{r,s}^{p,p^\prime}\big) (q) \, \equiv \, 
\chi_{r,s}^{p,p^\prime}(q)
= q^{h_{r,s}^{p,p^\prime}} + \ldots 
\eeq
for the Virasoro characters. The character of $L(2\bL_1)$ is given by~\cite{FF}
\beq\label{L2ModuleCh}
\Ch \, L(2\bL_1) \,=\,
\frac{q^{3/2}}{2}\frac{\varphi(q)}{\varphi(q^2)} \left[ \left( \frac{1}{\varphi(\sqrt{q})} 
- \frac{1}{\varphi(-\sqrt{q})} \right) 
\Theta_{2\boldsymbol{\Lambda}_0} 
+ \left( \frac{1}{\varphi(\sqrt{q})} + \frac{1}{\varphi(-\sqrt{q})} \right)
 \Theta_{2\boldsymbol{\Lambda}_1} \right]\,.
\eeq
If we combine these equations with the definitions of $\varphi(q)$, $\Theta_\bL$ and 
$\chi_{2,1}^{4,3}(q)$ given in Appendix~\ref{app:character}, we
obtain an expansion for $\Ch\, \mF^{(2)}$ similar to~\eqref{chLevel1}. 

At arbitrary level $\ell$ the character of $\mF^{(\ell)}$ 
can be computed recursively from the Weyl-Kac denominator 
formula~\cite{Kac} for $\mF$, or alternatively
with the program \texttt{SimpLie}~\cite{SimpLie} which offers an implementation of 
the Peterson recursion formula. 

\section{Level 3}
Ref.~\cite{BB} gives the following formula in eqn.(24) for the isomorphism as affine 
modules for the level-3 sector of the algebra
\beq\label{F3}
\mF^{(3)} \,\simeq \, \mF^{(1)}\otimes \mF^{(2)} / \wedge^3 \mF^{(1)}\,,
\eeq
which is equivalent to the statement that
\beq
{\rm Ker}\, \cI^{(3)} \,=\, \wedge^3 \mF^{(1)}
\eeq
and where $\wedge^3$ denotes the third exterior product of a representation.
Indeed, this subtraction eliminates the terms which vanish on account of
the Jacobi identity. In terms of characters this implies\footnote{At
higher level such a simple subtraction no longer works. For higher
rank algebras such as $E_{10}$, an extra subtraction is required
already at level $\ell =3$.}
\beq\label{CharF3}
\Ch \, \mF^{(3)} \,=\, \Ch\, \mF^{(1)} \cdot \Ch \, \mF^{(2)} \,-\, \Ch(\wedge^3 \mF^{(1)}) \, .
\eeq
We have verified this equation up to depth 30 in an explicit calculation.
In~\cite{BB} the authors also give an expression for the character of~\eqref{F3}, 
for which, however, we find a slightly different result, see below. To determine
$\mF^{(1)}\otimes \mF^{(2)}$ we first compute (for any $s$)
\begin{align}
\begin{aligned}
L(\bL_0+2\bd) \otimes L(2 \bL_1+s \bd )\,
&=\,\Vir(\tfrac{7}{10}, \tfrac{1}{10}) \otimes L(\bL_0 + 2 \bL_1 +(s+2) \bd) \\
&\quad \oplus \Vir(\tfrac{7}{10}, \tfrac{3}{2}) \otimes L(3\bL_0+(s+1)\bd) \, .
\end{aligned}
\end{align}
As the prefactor in~\eqref{Level2} encodes the infinite repetitions of representations
of this type according to 
\beq
\mF^{(2)} = \bigoplus_{n\geq 0 } a_n L(2\bL_1 + (2+n)\bd)\,,
\eeq
the (outer) multiplicities $a_n$ are the coefficients of the $q$-series
\beq\label{eq:c21}
\chi_{2,1}^{4,3} (q) -q^{1/2} = q^{3/2} \sum_{n\geq 0} a_n q^n\,.
\eeq
From this we deduce that, as a product of coset Virasoro and affine representations, we get 
\beq\label{chF1F2}
\mF^{(1)}\otimes \mF^{(2)} \, \simeq \, \ \bigoplus_{n\geq 0} a_n 
\Big( \Vir(\tfrac{7}{10}, \tfrac{1}{10}) \otimes L(\bL_0 + 2 \bL_1 + (4+n) \bd) 
\oplus \Vir(\tfrac{7}{10}, \tfrac{3}{2}) \otimes L(3\bL_0 + (3+n) \bd) \Big)\,.
\eeq
This is the level-3 analogue of~\eqref{L0L0}. Because there are now two 
affine representations
in this decomposition, we expect two kinds of Virasoro ground state multiplets on level 3. 
Namely, singlets and triplets.

In view of~\eqref{eq:c21} we can write the character of this tensor product also as 
\begin{align}
\label{eq:chF1F2}
\Ch (\mF^{(1)}\otimes \mF^{(2)}) = \big(\chi_{2,1}^{4,3} (q) -q^{1/2}\big) 
\Big( q^{-21/10}\chi_{3,3}^{5,4}(q)\, \Ch L(\bL_0 + 2 \bL_1 ) 
+ q^{-5/2} \chi_{3,1}^{5,4}(q) \, \Ch L(3\bL_0 ) \Big)\,.
\end{align}
Even though tempting, we do not write the vector space as a tensor product of 
truncated Virasoro modules.

Moreover, compared to~\eqref{L0L0}, the equation~\eqref{eq:chF1F2} exhibits a pile-up
of Virasoro characters which will give rise to infinitely many Virasoro ground states.
Recall that level 1 of $\mF$ is simply given by the affine module $L(\bL_0 + 2 \bd)$. 
Hence, there is one affine ground state (namely $\ket{\ba_0^{(1)}}$) from which we
obtain all DDF states on level 1 by the action of the affine generators~\eqref{Hm} and
\eqref{EmFm}. Then level 2 of $\mF$ is given by the tensor product of an affine module
and a (truncated) coset Virasoro representation ({\em cf.}~\eqref{Level2}). 
Hence, there are now infinitely many affine ground states that arise from the
action of the coset Virasoro raising operator on the Virasoro ground state multiplet
\eqref{VTrip}. In particular, there is only one Virasoro ground state multiplet.

Similarly on level 3 there are now infinitely many Virasoro
ground states which are generated from the action of the (truncated) coset Virasoro
representation $\Vir(\ft,\ft ) \ominus \mathbb{R} v_0$ on the two Virasoro ground states of
\beq
\Vir(\tfrac{7}{10}, \tfrac{1}{10}) \otimes L(\bL_0 + 2 \bL_1 + 5 \bd) \oplus 
\Vir(\tfrac{7}{10}, \tfrac{3}{2}) \otimes L(3\bL_0 + 4 \bd) \, .
\eeq
In the following, we disentangle this structure by first identifying the maximal 
ground states of $\mF^{(3)}$ and subsequently investigating how the action of
$\Vir(\ft,\ft ) \ominus \mathbb{R} v_0$ on these maximal ground states yields their
infinite duplication.

The discussion of the different kinds of ground states is summarized in the following table:
\begin{table}[H]
\begin{tabular}{c|ccc}
 & affine & Virasoro & maximal \\
\hline
level 1 & 1 & 0 & 0 \\
level 2 & $\infty$ & 1 & 1 \\
level 3 & $\infty$ & $\infty$ & 2 
\end{tabular}
\caption{Number of each kind of ground state for the levels 1 - 3. The number of maximal
ground states is finite for any level $\ell$.}
\end{table}

\subsection{Maximal ground states for \texorpdfstring{$\ell =3$}{l=3}}
Equation~\eqref{chF1F2} together with~\eqref{chF1F1F1} below tells us that we have
two virtual maximal ground state multiplets on level 3 of $\mF$. The first of which 
has dominant momentum $\bL_0 + 2 \bL_1 + 4 \bd = -3\br_{-1} - 2\bd + \br_1$.
\footnote{There is a difference of $+1\bd$ here compared to the argument of 
$L(\bL_0 + 2 \bL_1 + 5 \bd)$ because the $q$-series of 
$\Vir(\ft,\ft ) \ominus \mathbb{R} v_0$ is truncated and starts with $q^{3/2}$.}
Hence, this first multiplet is a triplet at depth 2. From the coset Virasoro prefactor
we can read off its eigenvalue $\frac{1}{10}$. Similarly, we can determine that the other
multiplet is a singlet at depth 3 with coset Virasoro eigenvalue $\frac{3}{2}$.
The explicit expressions of the dominant states in these 
multiplets in terms of DDF operators as elements of $\mF^{(1)}\otimes \mF^{(2)}$ are
\begin{align}
&\Psi^{(3)}_{2,\bD,1} \,=\, 2 \ket{\ba_0^{(1)}} \otimes \ket{\ba_{1}^{(2)}} \, , 
\label{L3VTrip} \\[2mm] 
&\begin{aligned}
\Psi^{(3)}_{3,\bE,0} \ \ \,= \ &2 \ket{\ba_{-1}^{(1)}} \otimes \ket{\ba_{1}^{(2)}} 
- 2 \ \Ae_{-1} \ket{\ba_0^{(1)}} \otimes \Az_{-1} \ket{\ba_0^{(2)}}
+ 2 \ket{\ba_1^{(1)}} \otimes \ket{\ba_{-1}^{(2)}} \\
&+ \ket{\ba_0^{(1)}} \otimes \Az_{-2} \Az_{-1} \ket{\ba_0^{(2)}} \, , \label{L3VSing}
\end{aligned}
\end{align}
with associated squared momenta $(-3 \br_{-1} - 2 \bd \pm \br_1)^2 = 8 > 2$ and 
$(-3 \br_{-1} - 2 \bd)^2 = 6 > 2$ 
for the triplet and $(-3 \br_{-1} - 3 \bd)^2 = 0$ for the singlet.
While the triplet thus vanishes by the Serre relation after conversion of
the tensor product into a commutator, the singlet state has allowed momentum. 
Hence, it must vanish for a different reason when replacing the tensor product with the
commutator
\begin{align}
\begin{aligned}
&2 \left[ \ket{\ba_{-1}^{(1)}} , \ket{\ba_{1}^{(2)}} \right] 
- 2 \left[ \Ae_{-1} \ket{\ba_0^{(1)}} , \Az_{-1} \ket{\ba_0^{(2)}} \right] 
+ 2 \left[ \ket{\ba_1^{(1)}} , \ket{\ba_{-1}^{(2)}} \right] 
+ \left[ \ket{\ba_0^{(1)}} , \Az_{-2} \Az_{-1} \ket{\ba_0^{(2)}} \right] \\
&\quad\,=\,
- \sqrt{2} \left[ \ket{\ba_{-1}^{(1)}} , 
\left[ \ket{\ba_{1}^{(1)}} , \Ae_{-1} \ket{\ba_0^{(1)}} \right] \right] 
- \sqrt{2} \left[ \Ae_{-1} \ket{\ba_0^{(1)}} , 
\left[ \ket{\ba_{-1}^{(1)}} , \ket{\ba_1^{(1)}} \right] \right] \\
&\quad \quad + \sqrt{2} \left[ \ket{\ba_1^{(1)}} , 
\left[ \ket{\ba_{-1}^{(1)}} , \Ae_{-1} \ket{\ba_0^{(1)}} \right] \right]
+ \left[ \ket{\ba_0^{(1)}} , 
\left[ \Ae_{-2} \Ae_{-1} \ket{\ba_0^{(1)}}, \ket{\ba_0^{(1)}} \right] \right] \, .
\end{aligned}
\end{align}
The first three terms on the right-hand side vanish with the Jacobi identity. The last term 
vanishes by itself, as it can be shown to be a null state. Thus, we see that
the states~\eqref{L3VSing} and~\eqref{L3VSing} are both in the kernel of $\cI^{(3)}$.
It is straightforward to check the affine and coset Virasoro vacuum conditions.
We can summarise the action of the coset Virasoro on the virtual states as follows.
\begin{align}
\begin{aligned}
{}^{[3]}\!\mL_{1}^{\text{coset}} \Psi_{2,\bD,w}^{(3)} \,&=\, 0 \, ,
&&\hphantom{\cI^{(3)} \quad } {}^{[3]}\!\mL_{1}^{\text{coset}} \Psi_{3,\bE,0}^{(3)} \,=\, 
0 \, , \\
{}^{[3]}\!\mL_{0}^{\text{coset}} \Psi_{2,\bD,w}^{(3)} \,&=\,
 \frac{1}{10} \Psi_{2,\bD,w}^{(3)} \, , 
&&\hphantom{\cI^{(3)} \quad } {}^{[3]}\!\mL_{0}^{\text{coset}} \Psi_{3,\bE,0}^{(3)} \,=\,
 \frac{3}{2} \Psi_{3,\bE,0}^{(3)} \, , \\
\cI^{(3)} \left( {}^{[3]}\!\mL_{-1}^{\text{coset}} \Psi_{2,\bD,w}^{(3)} \right) \,&=\, 
 \psi_{3,\bD,w}^{(3)} \, , 
\qquad \qquad
&&\cI^{(3)} \left( {}^{[3]}\!\mL_{-1}^{\text{coset}} \Psi_{3,\bE,0}^{(3)} \right) \,=\, 
\psi_{4,\bE,0}^{(3)} \, , \\
\cI^{(3)} \left( {}^{[3]}\!\mL_{-2}^{\text{coset}} \Psi_{2,\bD,w}^{(3)} \right) \,&=\, 
\psi_{4,\bD,w}^{(3)} \, ,
&&\cI^{(3)} \left( {}^{[3]}\!\mL_{-2}^{\text{coset}} \Psi_{3,\bE,0}^{(3)} \right) \,=\,
 \psi_{5,\bE,0}^{(3)} \, .
\end{aligned}
\end{align}
We use the notational convention explained below~\eqref{VTrip2} for distinguishing
Virasoro ground states from affine ground states.
In particular, we see that the coset Virasoro eigenvalues agree with the general 
prediction, and without shifts in $\bd$. The first descendant states, which are 
now elements of $\mF^{(3)}$, are given by
\beq\label{L3DDFGS}
\psi_{3,\bD,\pm 1}^{(3)} \,=\, - 2 \ket{\ba_{\pm 1}^{(3)}} \, , \quad
\psi_{3,\bD,0}^{(3)} \,=\, \sqrt{2}\ \Ad_{-1} \ket{\ba_0^{(3)}}
\eeq
and
\begin{align}
\begin{aligned}
\psi_{4,\bE,0}^{(3)} \,=\, \Big[&- \frac{4}{3} \ \Ad_{-2} \Ad_{-2} 
+ \ 3 \Ad_{-1} \Ad_{-1} \Ad_{-1} \Ad_{-1} - \frac{14}{3} \ \Ad_{-1} \Ad_{-1} \Bd_{-2} \\
&+ \frac{7}{18} \ \Bd_{-2} \Bd_{-2} - \frac{7}{18} \ \Bd_{-4} \Big] \ket{\ba_0^{(3)}} \, .
\end{aligned}
\end{align}
The states resulting from the application of ${}^{[3]}\!\mL_{-2}^{\text{coset}}$ 
on the maximal ground states and subsequent evaluation of the commutator are
\begin{align}
\begin{aligned}
\psi_{4,\bD,-1}^{(3)} \,=\, \bigg[- \frac{25}{54} \Bd_{-3} 
+ \frac{1}{36} \Ad_{-2} \Ad_{-1} + \frac{5 \sqrt{2}}{36} \Ad_{-1} \Bd_{-2}
 - \frac{4 \sqrt{2}}{27} \Ad_{-1} \Ad_{-1} \Ad_{-1} \bigg] \ket{\ba_{1}^{(3)}}
\end{aligned}
\end{align}
and
\begin{align}
\begin{aligned}
\psi_{5,\bE,0}^{(3)} \, =\, \bigg[&- \frac{385}{5832} \Bd_{-7} 
- \frac{15089}{9720} \Ad_{-5} \Ad_{-2} -\frac{385}{5832} \Bd_{-4} \Bd_{-3} 
- \frac{871}{1944} \Ad_{-4} \Ad_{-1} \Bd_{-2} \\
& - \frac{115}{486} \Ad_{-2} \Ad_{-2} \Bd_{-3} 
+ \frac{427}{1296} \Ad_{-2} \Ad_{-1} \Bd_{-4} 
+ \frac{385}{5832} \Bd_{-3} \Bd_{-2} \Bd_{-2} \\
& - \frac{379}{648} \Ad_{-4} \Ad_{-1} \Ad_{-1} \Ad_{-1} 
+ \frac{95}{108} \Ad_{-2} \Ad_{-2} \Ad_{-2} \Ad_{-1} \\
&- \frac{427}{1296} \Ad_{-2} \Ad_{-1} \Bd_{-2} \Bd_{-2} 
- \frac{196}{243} \Ad_{-1} \Ad_{-1} \Bd_{-3} \Bd_{-2} \\
&+ \frac{197}{216} \Ad_{-2} \Ad_{-1} \Ad_{-1} \Ad_{-1} \Bd_{-2} 
+ \frac{169}{324} \Ad_{-1} \Ad_{-1} \Ad_{-1} \Ad_{-1} \Bd_{-3} \\
& + \frac{3}{10} \Ad_{-2} \Ad_{-1} \Ad_{-1} \Ad_{-1} \Ad_{-1} \Ad_{-1} \bigg] 
\ket{\ba_{0}^{(3)}}\,.
\end{aligned}
\end{align}
All these states are affine ground states.

\subsection{Virasoro ground states for \texorpdfstring{$\ell =3$}{l=3}}\label{sec:L3GS}
The characters of the two terms in
\beq\label{partialCharacter}
\Vir(\tfrac{7}{10}, \tfrac{1}{10}) \otimes L(\bL_0 + 2 \bL_1 + 5 \bd) 
\oplus \Vir(\tfrac{7}{10}, \tfrac{3}{2}) \otimes L(3\bL_0 + 4 \bd) 
\eeq
are given by $q^{-21/10} \, \chi_{3,3}^{5,4}(q) \ \mathrm{ch} \, L(\boldsymbol{\Lambda}_0
+ 2\boldsymbol{\Lambda}_1 )$ and $q^{-5/2} \, \chi_{3,1}^{5,4}(q) \ \mathrm{ch} \,
L( 3 \boldsymbol{\Lambda}_0 )$ (see Appendix~\ref{app:character} for the relevant
 definitions). We can represent these characters by the following root systems:
\begin{figure}[H]
\vspace{.5cm}
\begin{tikzpicture}
\draw [fill] (-5,0) circle (.1);
\draw [fill] (-4,0) circle (.1);
\draw [fill] (-3,0) circle (.1);
\draw [fill] (-6,-1) circle (.1);
\draw [fill] (-5,-1) circle (.1);
\draw [fill] (-4,-1) circle (.1);
\draw [fill] (-3,-1) circle (.1);
\draw [fill] (-2,-1) circle (.1);
\draw [fill] (-6,-2) circle (.1);
\draw [fill] (-5,-2) circle (.1);
\draw [fill] (-4,-2) circle (.1);
\draw [fill] (-3,-2) circle (.1);
\draw [fill] (-2,-2) circle (.1);
\draw [fill] (-7,-3) circle (.1);
\draw [fill] (-6,-3) circle (.1);
\draw [fill] (-5,-3) circle (.1);
\draw [fill] (-4,-3) circle (.1);
\draw [fill] (-3,-3) circle (.1);
\draw [fill] (-2,-3) circle (.1);
\draw [fill] (-1,-3) circle (.1);
\node at (-4.6,0) {1};
\node at (-3.6,0) {1};
\node at (-2.6,0) {1};
\node at (-5.6,-1) {1};
\node at (-4.6,-1) {3};
\node at (-3.6,-1) {4};
\node at (-2.6,-1) {3};
\node at (-1.6,-1) {1};
\node at (-5.6,-2) {3};
\node at (-4.6,-2) {8};
\node at (-3.6,-2) {10};
\node at (-2.6,-2) {8};
\node at (-1.6,-2) {3};
\node at (-6.6,-3) {1};
\node at (-5.6,-3) {8};
\node at (-4.6,-3) {19};
\node at (-3.6,-3) {24};
\node at (-2.6,-3) {19};
\node at (-1.6,-3) {8};
\node at (-0.6,-3) {1};
\draw [fill] (3,-1) circle (.1);
\draw [fill] (2,-2) circle (.1);
\draw [fill] (3,-2) circle (.1);
\draw [fill] (4,-2) circle (.1);
\draw [fill] (1,-3) circle (.1);
\draw [fill] (2,-3) circle (.1);
\draw [fill] (3,-3) circle (.1);
\draw [fill] (4,-3) circle (.1);
\draw [fill] (5,-3) circle (.1);
\node at (3.4,-1) {1};
\node at (2.4,-2) {1};
\node at (3.4,-2) {2};
\node at (4.4,-2) {1};
\node at (1.4,-3) {1};
\node at (2.4,-3) {3};
\node at (3.4,-3) {6};
\node at (4.4,-3) {3};
\node at (5.4,-3) {1};
\end{tikzpicture}
\vspace{.5cm}
\caption{Partial visualization of $q^{-21/10} \chi_{3,3}^{5,4}(q) \ \mathrm{ch} \,
 L(\boldsymbol{\Lambda}_0 + 2\boldsymbol{\Lambda}_1 )$ (left) and $q^{-5/2} \chi_{3,1
 }^{5,4}(q) \ \mathrm{ch} \, L( 3 \boldsymbol{\Lambda}_0 )$ (right). The top right root of the
 left character is $(-3,-4,-3)$ (depth 4) and the top root of the right character 
is $(-3,-5,-5)$ (depth 5). The numbers in the figure indicate the multiplicities.}
\label{fig:F1F2ModulesCharacter}
\end{figure}
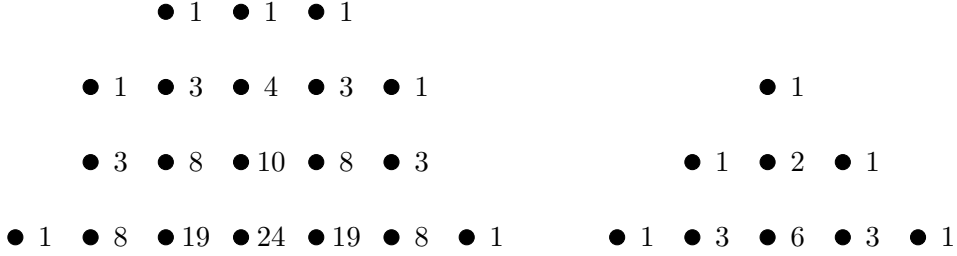
We can obtain all (DDF) states in these two diagrams from the action of the affine
generators $E_m$, $F_m$, $H_m$ and the coset Virasoro operator 
${}^{[3]}\!\mL_m^{\text{coset}}$ on the two maximal ground state 
multiplets~\eqref{L3VTrip} and~\eqref{L3VSing}. We illustrate this with the 
triplet (i.e. the left) character. Recall that (up to an at the moment unimportant 
$\boldsymbol{\delta}$ shift) its dominant maximal
ground state (i.e. the top right state) is given by 
\beq
\Psi_{2,\bD,1}^{(3)} \,=\, 2 \ket{\ba_0^{(1)}} \otimes \ket{\ba_1^{(2)}} \, .
\eeq
Acting with the affine generators and the coset Virasoro operator we obtain the three
states right below this state 
\begin{align}
\begin{aligned}\label{DescendantStates}
\Phi_{3,\bD,1}^{(3)} 
&\,=\, \left(H_{-1} + \frac{1}{2} F_0 E_{-1} \right) \Psi_{2,\bD ,1}^{(3)} \hspace*{-.3cm}
&&\,=\, \sqrt{2} \ \Ae_{-1} \ket{\ba_0^{(1)}} \otimes \ket{\ba_1^{(2)}} 
+ 2 \sqrt{2} \ket{\ba_0^{(1)}} \otimes \Az_{-2} \ket{\ba_1^{(2)}} \\
& &&\quad - \sqrt{2} \ket{\ba_1^{(1)}} \otimes \Az_{-1} \ket{\ba_0^{(2)}} \, , \\[2ex]
\Phi_{3,\mathbf{5},1}^{(3)} 
&\,=\, \frac{1}{2} F_0 E_{-1} \Psi_{2,\bD,1}^{(3)} 
&&\,=\, - \sqrt{2} \ \Ae_{-1} \ket{\ba_0^{(1)}} \otimes \ket{\ba_1^{(2)}} 
 - \sqrt{2} \ket{\ba_1^{(1)}} \otimes \Az_{-1} \ket{\ba_0^{(2)}} \, , \\[2ex]
\hat{\Phi}_{3,\bD,1}^{(3)} &\,=\, \mathfrak{L}_{-1}^{\mathrm{coset}} \Psi_{2,\bD,1}^{(3)} 
&&\,=\, - \frac{2\sqrt{2}}{5} \ \Ae_{-1} \ket{\ba_0^{(1)}} \otimes \ket{\ba_1^{(2)}} 
+ \frac{\sqrt{2}}{5} \ket{\ba_0^{(1)}} \otimes \Az_{-2} \ket{\ba_1^{(2)}} \\
& &&\quad + \frac{2 \sqrt{2}}{5} \ket{\ba_1^{(1)}} \otimes \Az_{-1} \ket{\ba_0^{(2)}} \, .
\end{aligned}
\end{align}
It can be checked that these three states are linearly independent and that any other action
of the affine generators on $\Psi_{2,\bD,1}^{(3)}$ yields a state in the linear span of these
three states.

To go from the characters described by Fig.~\ref{fig:F1F2ModulesCharacter} to the
character of $\mathfrak{F}^{(1)} \otimes \mathfrak{F}^{(2)}$ we can 
use~\eqref{eq:chF1F2}, according to which we must multiply the character 
of~\eqref{partialCharacter} by the $q$-series ({\em cf.}~\eqref{Level2Ch})
\begin{align}\label{PrefactorCharacter}
q^{-7/2} \left(\chi_{2,1}^{4,3}(q) - q^{1/2} \right) \,=\, 
q^{-2} \left( 1 + q + q^2 + 2 q^3 + 2 q^4 + 3 q^5 + 4 q^6 + 5 q^7 + 6 q^8 + \ldots \right)
 \, .
\end{align}
In terms of Fig.~\ref{fig:F1F2ModulesCharacter}, the multiplication with $m q^n$ amounts
to making $m$ copies of Fig. ~\ref{fig:F1F2ModulesCharacter} and shifting them by $n$
rows into the $-y$ direction. After the multiplication with~\eqref{PrefactorCharacter} the
two root systems in Fig.~\ref{fig:F1F2ModulesCharacter} become:
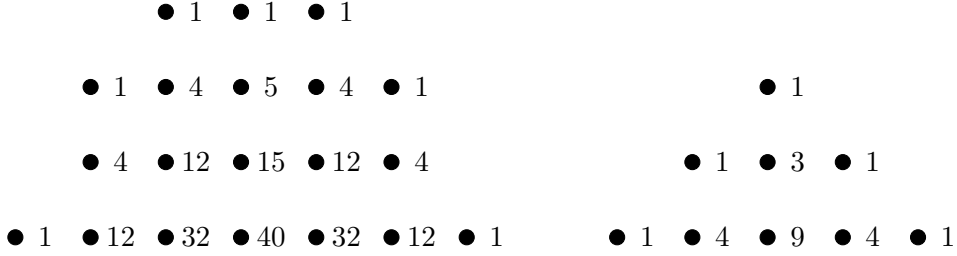
\begin{figure}[H]
\vspace{.5cm}
\begin{tikzpicture}
\draw [fill] (-5,0) circle (.1);
\draw [fill] (-4,0) circle (.1);
\draw [fill] (-3,0) circle (.1);
\draw [fill] (-6,-1) circle (.1);
\draw [fill] (-5,-1) circle (.1);
\draw [fill] (-4,-1) circle (.1);
\draw [fill] (-3,-1) circle (.1);
\draw [fill] (-2,-1) circle (.1);
\draw [fill] (-6,-2) circle (.1);
\draw [fill] (-5,-2) circle (.1);
\draw [fill] (-4,-2) circle (.1);
\draw [fill] (-3,-2) circle (.1);
\draw [fill] (-2,-2) circle (.1);
\draw [fill] (-7,-3) circle (.1);
\draw [fill] (-6,-3) circle (.1);
\draw [fill] (-5,-3) circle (.1);
\draw [fill] (-4,-3) circle (.1);
\draw [fill] (-3,-3) circle (.1);
\draw [fill] (-2,-3) circle (.1);
\draw [fill] (-1,-3) circle (.1);
\node at (-4.6,0) {1};
\node at (-3.6,0) {1};
\node at (-2.6,0) {1};
\node at (-5.6,-1) {1};
\node at (-4.6,-1) {4};
\node at (-3.6,-1) {5};
\node at (-2.6,-1) {4};
\node at (-1.6,-1) {1};
\node at (-5.6,-2) {4};
\node at (-4.6,-2) {12};
\node at (-3.6,-2) {15};
\node at (-2.6,-2) {12};
\node at (-1.6,-2) {4};
\node at (-6.6,-3) {1};
\node at (-5.6,-3) {12};
\node at (-4.6,-3) {32};
\node at (-3.6,-3) {40};
\node at (-2.6,-3) {32};
\node at (-1.6,-3) {12};
\node at (-0.6,-3) {1};
\draw [fill] (3,-1) circle (.1);
\draw [fill] (2,-2) circle (.1);
\draw [fill] (3,-2) circle (.1);
\draw [fill] (4,-2) circle (.1);
\draw [fill] (1,-3) circle (.1);
\draw [fill] (2,-3) circle (.1);
\draw [fill] (3,-3) circle (.1);
\draw [fill] (4,-3) circle (.1);
\draw [fill] (5,-3) circle (.1);
\node at (3.4,-1) {1};
\node at (2.4,-2) {1};
\node at (3.4,-2) {3};
\node at (4.4,-2) {1};
\node at (1.4,-3) {1};
\node at (2.4,-3) {4};
\node at (3.4,-3) {9};
\node at (4.4,-3) {4};
\node at (5.4,-3) {1};
\end{tikzpicture}
\vspace{.5cm}
\caption{Partial visualization of $q^{-28/5} (\chi_{2,1}^{4,3}(q) - q^{1/2} ) \, \chi_{3,3}^{5,4}
(q) \, \mathrm{ch} \, L(\boldsymbol{\Lambda}_0 + 2\boldsymbol{\Lambda}_1 )$ (left) and
$q^{-6} (\chi_{2,1}^{4,3}(q) - q^{1/2} ) \, \chi_{3,1}^{5,4}(q) \ \mathrm{ch} \, 
L( 3 \boldsymbol{\Lambda}_0 )$ (right). The top right root of the left character is 
$(-3,-2,-1)$ and the top root of the right character is $(-3,-3,-3)$. The numbers in 
the figure indicate the multiplicities.}
\label{fig:F1F2Character}
\end{figure}
Once again we study the left diagram. Below the maximal ground state 
$\Psi_{2,\bD,1}^{(3)}$ (sitting in the top right position) there are now four instead of the
three states that we had before. The fourth state is a Virasoro ground state. It cannot be
reached by the action of $E_m$, $F_m$, $H_m$ or ${}^{[3]}\!\mL_m^{\text{coset}}$ on
$\Psi_{2,\bD,1}^{(3)}$. Instead it arises from the prefactor associated with 
$\mathrm{Vir}(\tfrac{1}{2},\tfrac{1}{2})$ in front of $\mathrm{Vir}(\tfrac{7}{10}, 
\tfrac{1}{10}) \otimes L(\boldsymbol{\Lambda}_0 + 2\boldsymbol{\Lambda}_1 
+ 5 \boldsymbol{\delta})$. Because~\eqref{PrefactorCharacter} never terminates, there
are infinitely many such states. The action of $E_m$, $F_m$, $H_m$ and 
${}^{[3]}\!\mL_m^{\text{coset}}$ on each of these Virasoro ground states generates
one Virasoro and infinitely many affine multiplets. 

In the following, we explain how to derive all Virasoro ground states for $\ell = 3$ from 
the maximal ground states~\eqref{L3VTrip} and~\eqref{L3VSing}. Starting from the 
maximal ground state $\Psi_{2,\bD,1}^{(3)}$ we introduce the triple tensor product state
\beq\label{TensorState}
\Psi_{2,\bD,1}^{\otimes \, (3)} \,=\, \frac{1}{2\sqrt{2}} \ket{\ba_0^{(1)}} \otimes 
\left( \Ae_{-1} \ket{\ba_0^{(1)}} \wedge \ket{\ba_1^{(1)}} 
+ \Ae_{-1} \ket{\ba_1^{(1)}} \wedge \ket{\ba_0^{(1)}} \right) \, .
\eeq
This state is such that
\beq
(1 \otimes \mathcal{I}^{(2)}) \Psi_{2,\bD,1}^{\otimes \, (3)} \,=\, \Psi_{2,\bD,1}^{(3)}
\eeq
and that $\Psi_{2,\bD,1}^{\otimes \, (3)}$ is the affine ground state of a triplet multiplet.
This uniquely fixes $\Psi_{2,\bD,1}^\otimes$. Subsequently, we obtain the fourth state 
at $(-3,-3,-2)$ in the left root lattice of Fig.~\ref{fig:F1F2Character} via
\beq
\Psi_{3,\bD,1}^{(3)} \,=\, \Big(1 \otimes \mathcal{I}^{(2)} \circ 
{}^{[2]}\!\mL_{-1}^{\text{coset}} \Big) \Psi_{2,\bD,1}^{\otimes \, (3)} \,=\, 
2 \ket{\ba_0^{(1)}} \otimes \Az_{-1} \Az_{-1} \ket{\ba_1^{(2)}} 
+ 3 \ket{\ba_0^{(1)}} \otimes \Bz_{-2} \ket{\ba_1^{(2)}} \, .
\eeq
Besides being a Virasoro ground state, $\Psi_{3,\bD,1}^{(3)}$ is distinguished from 
$\hat{\Phi}_{3,\bD,1}^{(3)}$ by the fact that it has coset Virasoro eigenvalue 
$\frac{1}{10}$, while $\hat{\Psi}_{3,\bD,1}^{(3)}$ has coset Virasoro eigenvalue 
$\frac{1}{10} + 1 = \frac{11}{10}$. Moreover, let us note that under the action of 
$\cI^{(3)}$ both $\hat{\Phi}_{3,\bD,1}$ and $\Psi_{3,\bD,1}$ get mapped to the same 
DDF state, namely
\beq
\mathcal{I}^{(3)} \left( \hat{\Phi}_{3,\bD,1} \right) \,=\, 
\frac{1}{8} \, \mathcal{I}^{(3)} \left( \Psi_{3,\bD,1} \right) \,=\, - 2 \ket{\ba_1^{(3)}} \, .
\eeq
However, we do not expect this to happen for all Virasoro ground states and their
descendants.

Repeating this analysis for the singlet multiplet we find the following uniquely determined
affine singlet ground state (again a triple tensor product)
\begin{align}
\begin{aligned}
\Psi_{3,\bE,0}^{\otimes \, (3)} \,= \, & \frac{1}{2\sqrt{2}} \ket{\ba_{-1}^{(1)}} \otimes
\left( \Ae_{-1} \ket{\ba_1^{(1)}} \wedge \ket{\ba_0^{(1)}} 
+ \Ae_{-1} \ket{\ba_0^{(1)}} \wedge \ket{\ba_1^{(1)}} \right) \\
&- \frac{1}{4\sqrt{2}} \ \Ae_{-1} \ket{\ba_0^{(1)}} \otimes 
\left( \sqrt{2} \ \Ae_{-2} \ket{\ba_0^{(1)}} \wedge \ket{\ba_0^{(1)}} 
+ 2 \ \ket{\ba_{-1}^{(1)}} \wedge \ket{\ba_1^{(1)}} \right) \\
& - \frac{1}{2\sqrt{2}} \ket{\ba_1^{(1)}} \otimes 
\left( \Ae_{-1} \ket{\ba_{-1}^{(1)}} \wedge \ket{\ba_0^{(1)}} 
+ \Ae_{-1} \ket{\ba_0^{(1)}} \wedge \ket{\ba_{-1}^{(1)}} \right) \\
&+ \frac{1}{8\sqrt{2}} \ket{\ba_0^{(1)}} \otimes \
Big( \sqrt{2} \ \Ae_{-2} \ket{\ba_0^{(1)}} \wedge \Ae_{-1} \ket{\ba_0^{(1)}}
+ \sqrt{2} \ \Ae_{-2} \Ae_{-1} \ket{\ba_0^{(1)}} \wedge \ket{\ba_0^{(1)}} \\
&\hphantom{+ \frac{1}{8\sqrt{2}} \ket{\ba_0^{(1)}} \otimes 
\Big( } - 2 \ \Ae_{-1} \ket{\ba_1^{(1)}} \wedge \ket{\ba_{-1}^{(1)}}
+ 2 \ \Ae_{-1} \ket{\ba_{-1}^{(1)}} \wedge \ket{\ba_1^{(1)}} \Big) 
\end{aligned}
\end{align}
which satisfies
\beq
(1 \otimes \mathcal{I}^{(2)}) \Psi_{3,\bE,0}^{\otimes \, (3)} \,=\, \Psi_{3,\bE,0}^{(3)} \, .
\eeq
Again, we find the additional state below the top state in the singlet root system of
Fig.~\ref{fig:F1F2Character} compared to Fig.~\ref{fig:F1F2ModulesCharacter}.

Since the $q$ series of the prefactor can be generated using the coset Virasoro 
generators describing level $\ell=2$ we obtain that the infinite set of Virasoro 
ground states in $\mF^{(1)} \otimes \mF^{(2)}$ is given by
\beq\label{TensorCoset}
\Big( 1 \otimes \mathcal{I}^{(2)} \circ {}^{[2]}\!\mL_{-n_1}^{\text{coset}} \cdots 
{}^{[2]}\!\mL_{-n_N}^{\text{coset}} \Big) \quad \text{with} \quad 
n_1 \ge \ldots \ge n_N \quad \text{and} \quad N \ge 0 \,.
\eeq
We conjecture that a construction similar to~\eqref{TensorCoset} exists for all levels 
$\ell \ge 3$ and that for every level $\ell \ge 3$ the infinite set of Virasoro ground states 
can be obtained by the action of similar operators on the finite set of maximal ground
 states. For this conjecture we rely on Theorem~\ref{th:FF} and first write everything
 potential state as a tensor product state and then place coset Virasoro generators 
 followed by the maps $\mathcal{I}^{(k)}$ successively. We leave the derivation of 
 this operator and the further discussion of our conjecture for future work.

\subsection{Affine characters at \texorpdfstring{$\ell = 3$}{l=3}}
To get the character of equation~\eqref{chF1F2}, we need the character of the 
level-3 modules
\beq\label{L3Ch}
\Ch L(i \boldsymbol{\Lambda}_0 + 2j \boldsymbol{\Lambda}_1)\,=\, q^3
\sum_{\substack{(m,n) \in \{\mathbb{Z}_{\ge 0}, \mathbb{Z} \} \\ m+2n= 3}} 
C_{i,2j}^{m,2n} \, \Theta_{m \boldsymbol{\Lambda}_0 + 2n \boldsymbol{\Lambda}_1}
\eeq
with the string functions~\cite{BB}
\begin{align}\label{strings}
\begin{aligned}
C_{1,2}^{1,2} &= C_{1,2}^{5,-2}\,=\,\frac{q^{-2/5}}{\varphi(q)} \, \chi_{3,3}^{6,5}(q) \, , \\
C_{1,2}^{3,0} &= \frac{q^{-2/5}}{\varphi(q)} \, \left( \chi_{2,1}^{6,5}(q) 
+ \chi_{3,1}^{6,5} (q)\right) \, , \\
C_{3,0}^{1,2} &= C_{3,0}^{5,-2}\,=\,\frac{1}{\varphi(q)} \, \chi_{4,3}^{6,5}(q) \, , \\
C_{3,0}^{3,0} &= \frac{1}{\varphi(q)} \, \left( \chi_{1,1}^{6,5}(q) 
+ \chi_{4,1}^{6,5} (q) \right) \, , 
\end{aligned}
\end{align}
where the Virasoro characters $\chi_{r,s}^{p,p^\prime}(q)$ are defined in~\eqref{VirCh}.
Equation~\eqref{L3Ch} follows immediately from~\eqref{WeylW1} 
and~\eqref{ChStringTheta}. 
The fact that the string functions can be expressed in terms of (coset) Virasoro characters
for the central charges shown follows from the general property~\cite[Prop.~12.12]{Kac}
that they stem from a coset construction of $\msl(2)/\mathfrak{gl}(1)$ at level $\ell$,
where the $\mathfrak{gl}(1)$ is due to the Cartan subalgebra. The corresponding central
charge is in general $c= \frac{3\ell}{\ell+2} -1$, which for $\ell=3$ yields 
$c=\tfrac45 = c^{6,5}$.

For the character of the last term in~\eqref{F3} we find 
\begin{gather}\label{chF1F1F1}
\begin{gathered}
\Ch(\wedge^3 \mF^{(1)}) 
= \frac{ q^{-6}}{3} \bigg[- \chi_{1,1}^{5,4}(q) \chi_{1,1}^{4,3}(q) 
+ 2 \chi_{3,1}^{5,4}(q) \chi_{2,1}^{4,3}(q) 
+ q^{-1/15} \frac{\varphi(q)}{\varphi(q^3)} \chi_{3,3}^{6,5}(q) \bigg] \Ch \, L(3\bL_0) \\
+ \frac{q^{-28/5}}{3} \bigg[ - \chi_{3,2}^{5,4}(q) \chi_{1,1}^{4,3}(q) 
+ 2 \chi_{3,3}^{5,4}(q) \chi_{2,1}^{4,3}(q) 
- q^{-1/15} \frac{\varphi(q)}{\varphi(q^3)} \chi_{4,3}^{6,5}(q) \bigg] 
\Ch \, L(\bL_0+2\bL _1) \,.
\end{gathered}
\end{gather}
One important step in the derivation of~\eqref{chF1F1F1} is to write $\Theta_{3\bL_0}$ 
in terms of affine and Virasoro characters in the spirit of \S2 of~\cite{BB}, leading to 
\beq
\Theta_{3\bL_0} \,=\, \varphi(q) \left[ q^{-1/15} \, \chi_{3,3}^{6,5}(q) \, 
\Ch \, L(3\bL_0) - q^{1/3} \, \chi_{4,3}^{6,5}(q) \, \Ch\, L(\bL_0+2\bL_1)\right]\,.
\eeq
This together with the identity
\beq
\Ch(\mF^{(1)})^{*3} \,=\,\frac{q^{-3}}{\varphi(q^3)} \, \Theta_{3\Lambda_0}
\eeq
leads to~\eqref{chF1F1F1}. Combining these equations with formula~\eqref{CharF3} and
the known formulas for $\Ch \mF^{(1)}$ and $\Ch \mF^{(2)}$ gives the character of 
$\mF^{(3)}$
\begin{gather}\label{chF3}
\begin{gathered}
\Ch \, \mF^{(3)}
= q^{-6} \bigg[ \frac{1}{3} \chi_{1,1}^{5,4}(q) \chi_{1,1}^{4,3}(q) + \chi_{3,1}^{5,4}(q) 
\left( \frac{1}{3} \chi_{2,1}^{4,3}(q) - q^{1/2} \right) 
- \frac{q^{-1/15} }{3} \frac{\varphi(q)}{\varphi(q^3)} \chi_{3,3}^{6,5}(q) \bigg] 
\Ch \, L(3\bL_0) \\
+ q^{-28/5} \bigg[ \frac{1}{3} \chi_{3,2}^{5,4}(q) \chi_{1,1}^{4,3}(q) + \chi_{3,3}^{5,4}(q)
 \left(\frac{1}{3} \chi_{2,1}^{4,3}(q) - q^{1/2} \right) 
 + \frac{q^{-1/15} }{3} \frac{\varphi(q)}{\varphi(q^3)} \chi_{4,3}^{6,5}(q) \bigg]
 \Ch \, L(\bL_0+2\bL_1) \, .
\end{gathered}
\end{gather}
As a further confirmation of these formulae, we have checked that the root multiplicities
between the l.h.s. and the r.h.s. of~\eqref{F3} match up to depth 30. From~\eqref{chF3}
we can now in principle derive a closed form expression for the level-3 root multiplicities 
by substituting the formulas~\eqref{strings} for the affine characters. We refrain from
writing out the final result, as it is ``explicit, but ugly''~\cite{BB}, and as such not very
 illuminating. A further check would be to compare these results with those 
of~\cite{Kang1,Kang2}, a test which we leave to future work.

\section{Level 4}

At level four our formula~\eqref{FKer} reads
\beq\label{FKer4}
\mF^{(4)} \;\cong \, \big( \mF^{(1)} \otimes \mF^{(3)} \big) \,\big/ \, {\rm Ker} \, \cI^{(4)}\,.
\eeq
and we thus need to determine ${\rm Ker}\,\cI^{(4)}$. 
From Ref.~\cite{BB} and their equations (10), (13) and (22) we extract the 
following formula:
\beq\label{KerI4}
{\rm Ker}\, \cI^{(4)} \,=\, 
\mW^{(4)} \,-\, \big( \mF^{(1)}\otimes \cJ_3\big) \; \cap\; \mW^{(4)} 
\eeq
where
\beq
\cJ_3 \, \cong \, \mF^{(1)} \otimes \cJ_2 
\,\equiv \, \mF^{(1)} \otimes \big(\mF^{(1)} \wedge \mF^{(1)} - \mF^{(2)} \big)
\eeq
and
\beq
\mW^{(4)} \,=\, 
{\rm Weyl} \, \big( \mF^{(1)}\wedge \mF^{(1)} \big)^2 \,\coloneqq\,
{\rm Sym} \big( \mF^{(1)}\wedge \mF^{(1)} \big)^2 - \wedge^4 \mF^{(1)}
\eeq
projects out the Weyl tensor tableau. In principle, we should be able
to find expressions for the characters associated with Young-tableau-type
expressions or products of such Young-tableau-type expressions, 
but that is not so easy for the intersection term on the r.h.s. of~\eqref{KerI4},
as would be required for the derivation of an explicit formula for the
level-4 root multiplicities~\cite{BB}. Whereas the determination of such
intersections is straightforward for finite dimensional Lie groups and
their representations, this is not the case here, as one also needs to
match the Virasoro eigenvalues. We therefore leave the verification
of (\ref{KerI4}) to future work. A more general conjecture
would be that for all $\ell$, the kernel of $\cI^{(\ell)}$ can be expressed by such
Young-tableau-like combinations of lower $\mF^{(\ell)}$, in analogy with~\eqref{F3},
but in addition we expect there to be intersection terms which hamper more 
explicit calculations, because it is not clear how the permutation group
acts on them~\cite{BB}. 

For the product $\mF^{(1)}\otimes \mF^{(3)}$, the other ingredient
of~\eqref{FKer}, we can find an
explicit expression for the character by multiplication of the level 1 and level 3 modules.
In view of~\eqref{chF3}, there are two products to be considered, namely
\begin{align}\label{L0L3}
\begin{aligned}
L(\bL_0 + 2 \bd) \otimes L(3\bL_0 + 3 \bd) \, &\cong \, 
\mathrm{Vir}(\tfrac{4}{5},0) \otimes L(4\bL_0 + 5 \bd) 
\oplus \mathrm{Vir}(\tfrac{4}{5},\tfrac{2}{3}) \otimes L(2\bL_0 + 2\bL_1 + 4 \bd) \\
&\quad \oplus \mathrm{Vir}(\tfrac{4}{5},3) \otimes L(4 \bL_1 +\bd) \, 
\end{aligned}
\end{align}
and
\begin{align}\label{L0L0L1}
\begin{aligned}
L(\bL_0 + 2 \bd) \otimes L(\bL_0 + 2 \bL_1 + 4\bd) \,&\cong\, 
 \mathrm{Vir}(\tfrac{4}{5}, \tfrac{7}{5}) \otimes L(4\bL_0 +5 \bd) 
\oplus \mathrm{Vir}(\tfrac{4}{5}, \tfrac{1}{15}) \otimes L(2 \bL_0 + 2\bL_1 + 6 \bd ) \\
&\quad \oplus \mathrm{Vir}(\tfrac{4}{5}, \tfrac{2}{5}) \otimes L(4\bL_1 + 5\bd) \, ,
\end{aligned}
\end{align}
where again the $\bd$ shifts are in accord with our convention stated in 
section~\ref{sec:MG}.
Combining this with the level 3 Virasoro prefactors from~\eqref{chF3}
\beq
\mathcal{V}^{\, \text{singlet}} \,=\, q^{-6} \bigg[ \frac{1}{3} \chi_{1,1}^{5,4}(q) 
\chi_{1,1}^{4,3}(q) + \chi_{3,1}^{5,4}(q) \left( \frac{1}{3} \chi_{2,1}^{4,3}(q) - q^{1/2} \right) 
- \frac{q^{-1/15} }{3} \frac{\varphi(q)}{\varphi(q^3)} \chi_{3,3}^{6,5}(q) \bigg] 
\eeq
and
\beq
\mathcal{V}^{\, \text{triplet}} \,=\, q^{-28/5} \bigg[ \frac{1}{3} \chi_{3,2}^{5,4}(q)
 \chi_{1,1}^{4,3}(q) + \chi_{3,3}^{5,4}(q) \left( \frac{1}{3} \chi_{2,1}^{4,3}(q) 
 - q^{1/2} \right) + \frac{q^{-1/15} }{3} \frac{\varphi(q)}{\varphi(q^3)} \chi_{4,3}^{6,5}(q)
 \bigg]
\eeq
we obtain the character
\begin{align}
\begin{aligned}
\Ch(\mF^{(1)} \otimes \mF^{(3)}) \,=\, &\mathcal{V}^{\, \text{singlet}} \cdot \, 
\Ch( V(\bL_0 + 2 \bd) \otimes V(3\bL_0 + 3 \bd)) \\
&+ \mathcal{V}^{\, \text{triplet}} \cdot \, \Ch( V(\bL_0 + 2 \bd) \otimes 
V(2\bL_0 + \bL_1 + 4 \bd))\, .
\end{aligned}
\end{align}
Closed form expressions for the
characters of the level 4 modules $L(4\bL_0)$, $L(2\bL_0 + 2\bL_1)$ and $L(4\bL_1)$ 
can be deduced for example from the results in~\cite{KacPeterson,Mortenson} for the
string functions in terms of coset Virasoro modules (of central charge 
$\frac{3\ell}{\ell+2}-1$). 

From the tensor products, we can read off the ground state structure at level 4. To each
summand on the r.h.s. of~\eqref{L0L3} and~\eqref{L0L0L1} there is a maximal ground 
state, thus 
\begin{itemize}
\item
There are two singlets that both sit at depth 4. They have $\mL_0^{\text{coset}}$
eigenvalues $0$ and $\tfrac75$, respectively;
\item 
There are two triplets that sit at depths 5 and 3 with eigenvalues $\tfrac{2}{3}$ and 
$\tfrac{1}{15}$, respectively;
\item 
There are two fiveplets that sit at depths 4 and 8 with eigenvalues $\tfrac{2}{5}$ and 
$3$, respectively.
\end{itemize}
We next investigate what they look like as tensor products of DDF states and investigate
whether they are virtual or not. The tensor product form is obtained by making a general
ansatz in terms of DDF states and then imposing the conditions~\eqref{VirasoroGS}.
This we have succeeded in doing for all ground states but the final fiveplet at depth 7 
which is a deeply nested commutator of DDF states.

The dominant state of the triplet at depth 3 is given by
\beq
\Psi_{3,\bD,1}^{(4)} \,=\, - 2\ket{\ba_0^{(1)}} \otimes \ket{\ba_1^{(3)}} \, .
\eeq
The two singlets at depth 4 are
\begin{align}
\begin{aligned}
\Psi_{4,\bE,0}^{(4)} \,=\, &\ket{\ba_{0}^{(1)}} \otimes \Big[14 \Ad_{-3} \Ad_{-1} 
+ 3 \Ad_{-2} \Ad_{-2} + 2 \Ad_{-1} \Ad_{-1} \Ad_{-1} \Ad_{-1} \Big] \ket{\ba_{0}^{(3)}} \\
& + 42 \ket{\ba_{-1}^{(1)}} \otimes \ket{\ba_{1}^{(3)}} 
- 42 \ \Ad_{-1} \ket{\ba_{0}^{(1)}} \otimes \Ad_{-1} \ket{\ba_{0}^{(3)}} 
+ 42 \ket{\ba_{1}^{(1)}} \otimes \ket{\ba_{-1}^{(3)}}
\end{aligned}
\end{align}
and
\begin{align}
\begin{aligned}
\bar \Psi_{4,\bE,0}^{(4)} \,=\, &\ket{\ba_{0}^{(1)}} \otimes 
\Big[ -\frac{7}{54} \Bd_{-4} - \frac{4}{9} \Ad_{-2} \Ad_{-2} 
+ \frac{7}{54} \Bd_{-2} \Bd_{-2} - \frac{14}{9} \Ad_{-1} \Ad_{-1} \Bd_{-2} \\
&+ \Ad_{-1} \Ad_{-1} \Ad_{-1} \Ad_{-1} \Big] \ket{\ba_{0}^{(3)}} \, .
\end{aligned}
\end{align}
Moreover, the dominant state of the fiveplet at depth 4 is
\beq
\Psi_{4,\bF,2}^{(4)} \,=\, 24 \ket{\ba_{0}^{(1)}} \otimes \ket{\ba_{2}^{(3)}} 
+ 24 \ket{\ba_{1}^{(1)}} \otimes \ket{\ba_{1}^{(3)}} \, .
\eeq
All these states are virtual, i.e. they are in the kernel of the map $\cI^{(4)}$. For the triplet
and the fiveplet this is easy to see as their dominant states have momentum squared $>2$.
The two singlets have allowed momenta but their associated DDF states still vanish due to
the Serre relations and the Jacobi identity. 

{\allowdisplaybreaks
Subsequently, we investigate the triplet at depth 5. Its dominant state is given by
\begin{align}
\Psi_{5,\bD,-1}^{(4)} \,&=\, \ket{\ba_{0}^{(1)}} \nn\\
&\quad \quad \otimes \Bigg[ -\frac{22}{9} \ \Ad_{-5} \Ad_{-1} 
+ \frac{55}{36} \ \Ad_{-4} \Ad_{-2} - \frac{77}{18 \sqrt{2}} \ \Ad_{-4} \Bd_{-2} \nn\\
&\quad \quad \hphantom{ \otimes \Bigg[ } - \frac{7}{18 \sqrt{2}} \ \Ad_{-2} \Bd_{-4} 
+ \frac{11}{18 \sqrt{2}} \ \Ad_{-4} \Ad_{-1} \Ad_{-1} \nn\\
&\quad \quad \hphantom{ \otimes \Bigg[ } 
+ \frac{7}{36 \sqrt{2}} \ \Ad_{-2} \Ad_{-2} \Ad_{-2} 
- \frac{77}{36} \ \Ad_{-2} \Ad_{-2} \Bd_{-2} \nn\\
&\quad \quad \hphantom{ \otimes \Bigg[ } 
+ \frac{7}{18 \sqrt{2}} \ \Ad_{-2} \Bd_{-2} \Bd_{-2} 
- \frac{7}{18} \ \Ad_{-1} \Ad_{-1} \Bd_{-4} \nn\\
&\quad \quad \hphantom{ \otimes \Bigg[ } 
+ \frac{73}{36} \ \Ad_{-2} \Ad_{-2} \Ad_{-1} \Ad_{-1} 
+ \frac{35}{9 \sqrt{2}} \ \Ad_{-2} \Ad_{-1} \Ad_{-1} \Bd_{-2} \nn\\
&\quad \quad \hphantom{ \otimes \Bigg[ } 
+ \frac{7}{18} \ \Ad_{-1} \Ad_{-1} \Bd_{-2} \Bd_{-2} 
- \frac{23}{18 \sqrt{2}} \ \Ad_{-2} \Ad_{-1} \Ad_{-1} \Ad_{-1} \Ad_{-1} \nn\\
&\quad \quad \hphantom{ \otimes \Bigg[ } 
- \frac{7}{18} \ \Ad_{-1} \Ad_{-1} \Ad_{-1} \Ad_{-1} \Bd_{-2} \nn\\
&\quad \quad \hphantom{ \otimes \Bigg[ } 
- \frac{1}{18} \ \Ad_{-1} \Ad_{-1} \Ad_{-1} \Ad_{-1} \Ad_{-1} \Ad_{-1} \Bigg] 
\ket{\ba_{1}^{(3)}}  \\
& + \ket{\ba_{1}^{(1)}}\nn \\*
&\quad \quad \otimes \Bigg[\frac{7}{6} \ \Bd_{-4} 
+ 4 \Ad_{-2} \Ad_{-2} - \frac{7}{6} \ \Bd_{-2} \Bd_{-2} + 14 \Ad_{-1} \Ad_{-1} \Bd_{-2} \nn\\*
&\quad \quad \hphantom{ \otimes \Bigg[ } 
- 9 \ \Ad_{-1} \Ad_{-1} \Ad_{-1} \Ad_{-1} \Bigg] \ket{\ba_{0}^{(3)}} \, .\nn
\end{align}
This is the first maximal ground state which is not virtual. Instead we find
\begin{align}
\begin{aligned}
\cI^{(4)} \left( \Psi_{5,\bD,-1}^{(4)} \right) \,=\, \Bigg[&-\frac{45}{32} \ \Bv_{-4} 
- \frac{576}{32} \ \Av_{-3} \Av_{-1} - \frac{201}{32} \ \Av_{-2} \Av_{-2} 
- \frac{783}{32 \sqrt{2}} \ \Av_{-2} \Bv_{-2} \\
&- \frac{45}{8 \sqrt{2}} \ \Av_{-1} \Bd_{-3} + \frac{975}{256} \ \Bv_{-2} \Bv_{-2} 
+ \frac{717}{16 \sqrt{2}} \ \Av_{-2} \Av_{-1} \Av_{-1} \\
&- \frac{771}{64} \ \Av_{-1} \Av_{-1} \Bv_{-2} 
- \frac{915}{64} \ \Av_{-1} \Av_{-1} \Av_{-1} \Av_{-1} \Bigg] \ket{\ba_{1}^{(4)}} \, .
\end{aligned}
\end{align}
The sixth missing maximal fiveplet at depth 8 is currently out of reach for our 
computational machinery.}

The action of the coset Virasoro operator on the (virtual) maximal ground states confirms
the expected eigenvalues and gives rise to the descendant DDF states. For the triplet at
depth 3 we have
\begin{align}
\begin{aligned}
{}^{[4]}\!\mL_{1}^{\text{coset}} \Psi_{3,\bD,w}^{(4)} \,&=\, 0 \, , \\
{}^{[4]}\!\mL_{0}^{\text{coset}} \Psi_{3,\bD,w}^{(4)} \,&=\, 
\frac{1}{15} \Psi_{3,\bD,w}^{(4)} \, , \\
\cI^{(4)} \left( {}^{[4]}\!\mL_{-1}^{\text{coset}} \Psi_{3,\bD,w}^{(4)} \right) \,&=\, 
\psi_{4,\bD,w}^{(4)} 
\end{aligned}
\end{align}
with the affine ground states (as a special case of~\eqref{Triplet})
\beq
\psi_{4,\bD,\pm 1}^{(4)} \,=\, \frac{8}{5} \ket{\ba_{\pm 1}^{(4)}} \, , \quad 
\psi_{4,\bD,0}^{(4)} \,=\, -\frac{4}{5} \sqrt{2} \ \Av_{-1} \ket{\ba_{0}^{(4)}} \, .
\eeq
Similarly, we obtain for the singlet maximal ground states at depth 4
\begin{gather}
\begin{gathered}
\begin{aligned}
&\hphantom{\cI^{(4)} \quad} {}^{[4]}\!\mL_{1}^{\text{coset}} \Psi_{4,\bE,0}^{(4)} \,=\, 0 \, ,
&\hphantom{\cI^{(4)} \quad} {}^{[4]}\!\mL_{1}^{\text{coset}} \bar \Psi_{4,\bE,0}^{(4)} \,=\, 
0 \, , \\
&\hphantom{\cI^{(4)} \quad} {}^{[4]}\!\mL_{0}^{\text{coset}} \Psi_{4,\bE,0}^{(4)} \,=\,
 \frac{7}{5} \Psi_{4,\bE,0}^{(4)} \, , 
&\hphantom{\cI^{(4)} \quad} {}^{[4]}\!\mL_{0}^{\text{coset}} \bar \Psi_{4,\bE,0}^{(4)} \,=\,
 0 \, , \\
&\cI^{(4)} \left({}^{[4]}\!\mL_{-1}^{\text{coset}} \Psi_{4,\bE,0}^{(4)} \right) \,=\, 
\psi_{5,\bE,0}^{(4)} \, ,
&\quad \quad \cI^{(4)} \left({}^{[4]}\!\mL_{-1}^{\text{coset}} \bar \Psi_{4,\bE,0}^{(4)} \right)
 \,=\, 0 \, , \\
\end{aligned}
\end{gathered}
\end{gather}
where
\begin{align}
\begin{aligned}
\psi_{5,\bE,0}^{(4)}\,=\,\Bigg[ &-\frac{189}{40} \ \Bv_{-5} - 21 \ \Av_{-3} \Av_{-2} 
+ \frac{189}{40} \ \Bv_{-3} \Bv_{-2} - \frac{189}{5} \ \Av_{-2} \Av_{-1} \Bv_{-2} \\
&-\frac{189}{5} \ \Av_{-1} \Av_{-1} \Bv_{-3} 
+ \frac{336}{5} \ \Av_{-2} \Av_{-1} \Av_{-1} \Av_{-1} \Bigg] \ket{\ba_{0}^{(4)}} \, .
\end{aligned}
\end{align}
We observe that for the second singlet we actually have 
${}^{[4]}\!\mL_{-1}^{\text{coset}} \Psi_{4,\bF,w}^{(4)}$ without the application of 
$\cI^{(4)}$ since this state is actually a null state in the Virasoro module $\Vir(\frac45,0)$.

For the fiveplet we have
\begin{align}
\begin{aligned}
{}^{[4]}\!\mL_{1}^{\text{coset}} \Psi_{4,\bF,w}^{(4)} \,&=\, 0 \\
{}^{[4]}\!\mL_{0}^{\text{coset}} \Psi_{4,\bF,w}^{(4)} \,&=\, 
\frac{2}{5} \Psi_{4,\bF,w}^{(4)} \, , \\
\cI^{(4)} \left({}^{[4]}\!\mL_{-1}^{\text{coset}} \Psi_{4,\bF,w}^{(4)} \right) \,&=\, 0 \, .
\end{aligned}
\end{align}
Finally, we also confirm the eigenvalue of the triplet at depth 5 via
\begin{align}
\begin{aligned}
&{}^{[4]}\!\mL_{1}^{\text{coset}} \Psi_{5,\bD,w}^{(4)} \,=\, 0 \, , \\
&{}^{[4]}\!\mL_{0}^{\text{coset}} \Psi_{5,\bD,w}^{(4)} \,=\, 
\frac{2}{3} \Psi_{5,\bD,w}^{(4)} \, .
\end{aligned}
\end{align}

\section{Isotropy and anisotropy of root multiplicities}

It has been observed from the available data for root multiplicities that they can be
anisotropic. By isotropy we mean the the root multiplicities
depend only on the norm $\br^2$ of the root, and not on its 
orientation. From~\cite{FF} it follows that isotropy is respected at level-2.

In this section, we explain why the character of $\mF$ is isotropic only up to level 3 and
give a general criterion for determining whether a level is isotropic or not. This criterion is
based on the fact that roots at a given level belong to affine representations whose
characters in turn are expressed in terms of generalized theta functions and string
functions. The generalized theta functions relate elements that are related by (even) 
Weyl transformations and therefore all roots appearing in one theta function have the
same multiplicity and norm. 

The string functions correspond to adding multiples of $\bd$ to the elements from the 
$\Theta$-function. These elements all have the same multiplicity. Let $\ba$ be one of 
these elements on level $\ell$, meaning that $\ba= - \ell \br_{-1} -d \bd +n \br_1$, where 
$d$ is the depth of the root. The roots generated by the string function are of the form 
$\ba+m \bd$ and have norm $(\ba + m \bd)^2 = \ba^2 - 2 m\ell$ and so get more and 
more imaginary as one descends the string. 
Since $\bd$ is invariant under the affine Weyl group, the roots that are obtained by
descending $m$ steps from one top element have the same
multiplicities as those obtained from another top element of the same $\Theta$-function
and that is why the string functions and the $\Theta$-function factorize in the character. 

In general, the character of $\mF$ at a given level is a sum of $\Theta$-functions multiplied
by their string functions.\footnote{We here use the term string function more generally to
mean the full $q$-series that is produced from the affine character together with the
Virasoro characters.}
The question of isotropy or not of a given level can thus be determined, as a sufficient
condition, if the individual $\Theta$ functions appearing in the character of that level
generate disjoint sets of norms of roots. If they do, anisotropy is ruled out. We do not
know whether the converse is also true, but believe so based on available data.

Let us go through this analysis for the lowest levels. 
At level 1, we can express the character schematically as
\beq
\Ch \, \mF^{(1)}\,=\,f_1(q) \, \Theta_{\boldsymbol{\Lambda}_0} \, ,
\eeq
with a $q$-series $f_1(q)$. Hence, all root strings are connected to each other through
Weyl reflections. Since a Weyl reflection preserves the norm, the roots of $\mF$ on
level one are isotropic. 

On levels $\ell = 2$ and $\ell = 3$ we can express the character schematically as
\beq
\Ch \, \mF^{(\ell)}\,=\,f_{\ell, 1}(q) \, \Theta_{\ell, 1} + f_{\ell, 2}(q) \, \Theta_{\ell, 2} \, , 
\eeq
with some $q$-series $f_{\ell, i}$ and 
$\Theta_{2, 1} = \Theta_{2\boldsymbol{\Lambda}_1}$, 
$\Theta_{2,2} = \Theta_{2\boldsymbol{\Lambda}_0} $ and 
$\Theta_{3, 1} = \Theta_{\boldsymbol{\Lambda}_0+2\boldsymbol{\Lambda}_0} 
+ \Theta_{5\boldsymbol{\Lambda}_0-2\boldsymbol{\Lambda}_0} $, 
$\Theta_{3,2} = \Theta_{3\boldsymbol{\Lambda}_0}$. The norms of the roots in the 
$q$-series $f_{\ell, 1}(q)$ are $\{ 2 - 2m \cdot \ell \ | \ m \in \mathbb{Z}_{\ge 0} \}$. 
The norms of the roots in the $q$-series $f_{\ell, 2}(q)$ are 
$\{- 2m \cdot \ell \ | \ m \in \mathbb{Z}_{\ge 0} \}$. These two sets of norms are 
disjoint and thus two roots of the same norm belong to the same Weyl orbit and 
hence must have the same multiplicity. Thus the roots of $\mF$ on level 2 and 3 
are isotropic. 

On level 4 we do not have isotropy anymore. We can express the character schematically as
\beq\label{SchematicCharacter}
\Ch \, \mF^{(4)}\,=\,f_{4, 1}(q) \, \Theta_{4\boldsymbol{\Lambda}_0} 
+ f_{4, 2}(q) \, \left( \Theta_{2\boldsymbol{\Lambda}_0+2\boldsymbol{\Lambda}_1} 
+ \Theta_{6\boldsymbol{\Lambda}_0-2\boldsymbol{\Lambda}_1} \right) 
+ f_{4, 3}(q) \, \left( \Theta_{4\boldsymbol{\Lambda}_1} 
+ \Theta_{8\boldsymbol{\Lambda}_0-4\boldsymbol{\Lambda}_1} \right) \, .
\eeq
The roots in the corresponding $q$-series $f_{4, i}(q)$ have norms
\beq\label{Norms}
\{- 2 \cdot 4 m \ | \ m \in \mathbb{Z}_{\ge 0} \} \, , \quad 
\{ 2 - 2 \cdot 4 m \ | \ m \in \mathbb{Z}_{\ge 0} \} \quad \text{and} \quad 
\{ - 2 \cdot 4 m \ | \ m \in \mathbb{Z}_{\ge 0} \} \, .
\eeq
Thus the roots in the singlet and 5-plet $q$-series have the same norms. However, the 
$q$-series themselves do not agree. We find the first mismatch for the roots 
$(-4, -12, -12)$ and $(-4, -13, -11)$ which both have norm $-64$ but multiplicity 
$10107$ respectively $10108$. 

For the general case we make the following conjecture. The roots on a given level 
$\ell$ are isotropic if and only if $\ell$ is prime or $\ell = 2p$ with $p>2$ and $p$ prime. 
We have tested this conjecture up to level 25 explicitly. We arrive at this conjecture by
considering the possible level-$\ell$ modules and their associated norms similar 
to~\eqref{SchematicCharacter} and~\eqref{Norms}.

\section{Outlook}

In this paper we have initiated a new approach to studying the hyperbolic KMA $\mF$.
This approach relies on the methods developed in~\cite{GN,GN1}, and we have shown that 
these methods have the potential to reach beyond the low level sectors of $\mF$ 
studied so far. At the very least, they offer a much more concrete realization of $\mF$.
In contradistinction to more conventional approaches relying on the division by 
`Serre ideals' 
here the main open problem, besides working out products of affine representations,
is in determining and understanding the kernel of the map $\cI^{(\ell)}$ introduced 
in~\eqref{Iell}. A further complication as one moves up in level $\ell$, is that each lower 
level factor comes with its own coset Virasoro representations, so that the products of
such representations increase without bound as $\ell\rightarrow\infty$. We have presented
partial evidence that, beyond the affine and coset Virasoro operators, there exist
operators also for the `spectator Virasoro representations'. These should eventually enable
us to generate the full level-$\ell$ sectors from a finite set of maximal ground states
for each level. Importantly, our approach is not so much aimed at obtaining 
multiplicity formulas (a main focus in the mathematical literature), but rather at
understanding the Lie algebra structure itself. Since $\mF$ is a subalgebra of the 
vertex operator algebra which is realized here with DDF operators, we also have a 
very concrete realization of the Lie bracket of $\mF$ with (\ref{StateOperator}) that 
can be used for calculation. 

It remains to be see whether or not our approach can give insight into 
all-level properties of $\mF$. Nevertheless, implications for fundamental physics
(which we mentioned only briefly in the introduction) are of considerable interest,
especially with regard to quantum cosmology and the physics of the Big Bang.
Their exploration will be a fascinating topic for further study. 

\subsubsection*{Acknowledgements}
We would like to thank Denis Bernard, Ilka Brunner, Stefan Fredenhagen, and especially
Alex Feingold for illuminating discussions. 
HN thanks Luis Alvarez-Gaum\'e and the Simons Center for Geometry and Physics 
for hospitality and support during the final redaction of this paper. SC would like to thank the Humboldt-Universität of Berlin and the Erasmus+ program, respectively for hospitality and support during part of the work. AK, HM and HN would also like to thank the Mathematisches Forschungsinstitut Oberwolfach, where 
these results were first presented, for hospitality.  This work was supported in part by 
the European Research Council (ERC) under the European Union’s Horizon 2020 
research and innovation programme (grant agreement No 740209).

\appendix
\section{Proof of Theorem 1}\label{app:pf}

In this appendix we provide a proof of Theorem~\ref{th:FF}. We start by noting the
standard fact that $\mF$ has a basis of in terms of standard multicommutators, {\em i.e.}
\beq
\mF^{(\ell)} \,=\, \mathrm{span} \{ f_{i_1 \ldots i_n} \ | \ \text{with } 
\ell \text{ generators} \ f_{-1} \ \text{and} \ n \ge \ell \} \, ,
\eeq
which implies that 
the inclusion $[\mF^{(1)}, \mF^{(\ell-1)}] \,\subseteq \, \mF^{(\ell)}$ is obvious. 
We use the notation of appendix~\ref{app:mc} for 
multi-commutators and the set $\{f_{i_1 \ldots i_n}\}$ whose span is taken is not a basis. 

For the reverse inclusion $\mF^{(\ell)} \subseteq [\mF^{(1)}, \mF^{(\ell-1)}]$ 
let $f_{i_1 \ldots i_n} \in \mF^{(\ell)}$. We proceed by induction on the first place $k$
for which $i_k=-1$.
Therefore assume that $i_k = -1$ and $-1 \not \in \{i_1, \ldots , i_{k-1}\}$ 
and we will show that we can write $f_{i_1 \ldots i_n}$ as a linear combination of
 commutators of level 1 and level $\ell-1$ elements.

\subsubsection*{Base Case: $\mathbf{k=1}$}
For $k=1$ we simply have
\beq
f_{i_1 \ldots i_n} \,=\, \left[ f_{-1}, f_{i_2 \ldots i_n} \right] \, .
\eeq
Clearly $f_{-1} \in \mF^{(1)}$ and $f_{i_2 \ldots i_n} \in \mF^{(\ell - 1)}$ because 
$f_{i_2 \ldots i_n}$ contains $\ell -1$ elements $f_{-1}$. 

\subsection*{Induction Step: $k \to k+1$}
Consider the multi-commutator
\beq
f_{i_1 \ldots i_n} \,=\, f_{i_1 \ldots i_k, -1 , i_{k+1}, \ldots i_n}\,=\,
\left[ f_{i_1}, f_{i_2, \ldots , i_k, -1, i_{k+1}, \ldots i_n} \right] \, .
\eeq
Since $f_{i_1} \not= f_{-1}$ we can write $f_{i_2, \ldots , i_k, -1, i_{k+1}, \ldots i_n}$ 
by the induction hypothesis 
as a linear combination of multi-commutators of level 1 and level $\ell -1$. Schematically
 we have a finite sum
\beq
f_{i_2, \ldots , i_k, -1, i_{k+1}, \ldots i_n} \,= \sum_a \left[x_a^{(1)} , x_a^{(\ell-1)} \right]\,,
\eeq
where $x^{(1)}_a \in \mF^{(1)}$ and $x^{(\ell-1)}_a \in \mF^{(\ell-1)}$. We then use 
the Jacobi identity to get
\begin{align}
\begin{aligned}
f_{i_1,i_2, \ldots , i_k, -1, i_{k+1}, \ldots i_n} \,&=\, 
\left[ f_{i_1}, \sum_a \left[x_a^{(1)} , x_a^{(\ell-1)} \right] \right]\\
&=\, \sum_i \left[ \left[f_{i_1} , x_a^{(1)}\right], x^{(\ell-1)}_a \right] 
+\sum_i \left[ x_a^{(1)}, \left[f_{i_1} , x^{(\ell-1)}_a \right] \right] \,,
\end{aligned}
\end{align}
which, since $f_{i_1}\neq f_{-1}$ does not change the level, is clearly a linear 
combination of commutators in $[\mF^{(1)}, \mF^{(\ell-1)}]$ as claimed.
\qed

\section{Character formulas}\label{app:character}

Here, we collect all our conventions and important formulae for computing the character 
of $\mathfrak{F}^{(\ell)}$.

Let $P = \mathbb{C} \bL_{-1} \oplus \mathbb{Z} \bL_0 \oplus \mathbb{Z}\bL_1$ be the
weight lattice. Moreover let $P(\bL)$ be the set of all weights in the integrable
highest-weight module $L(\boldsymbol{\Lambda})$ and $P^\ell$ the set of all weights
of level $\ell$. Then the formal character of $L(\boldsymbol{\Lambda})$ is given by
\beq\label{Cha}
\Ch \, L(\boldsymbol{\Lambda}) \,=\, \sum_{\bla \in P(\bL)} \mathrm{mult}(\bla) \, e^\bla \, , 
\eeq
where $\mathrm{mult}(\bla)$ is the multiplicity of $\bla \in P(\bL)$. The formal
characters $\Ch \, \mF^{(\ell)}$ are defined analogously, except that they are
(in general infinite) sums of affine characters of type~\eqref{Cha}.
The characters~\eqref{Cha} are functions of three variables in general, and we will 
make particular use of the variable $q=e^{\bL_{-1}}= e^{-\bd}$. All characters can
be determined from the Weyl-Kac character formula~\cite{Kac}
\beq
\Ch \, L(\bL) \,=\, R^{-1} \sum_{w\in\cW} e^{w(\brho + \bL) -\brho} \, , 
\eeq
where the sum is over the Weyl group 
$\cW = \mathbb{Z}\rtimes \mathbb{Z}_2 \equiv \mT\rtimes \mathbb{Z}_2$,
with the Weyl vector
\beq
\brho \,=\, \bL_{-1} + \bL_0 + \bL_1 \,=\, -2\brr - 5\bd + \frac12 \br_1 \;,
\eeq 
and the denominator for affine $\mathfrak{sl}_2$ is
\beq 
R\,\coloneqq \, \prod_{n\geq 0} \big(1 - e^{\br_0 + n\bd}\big)
\prod_{n> 0} \big(1 - e^{ n\bd}\big)\prod_{n\geq 0} \big(1 - e^{\br_1 + n\bd}\big) \, .
\eeq
The abelian translation subgroup of the Weyl group is 
$\mT=\{\mt^n \,|\, n\in \mathbb{Z}\}$ where $\mt=w_1w_0$ in terms of simple 
reflections and $\mt^n$ acts on a weight 
$\bla= p_{-1} \bL_{-1} + p_0 \bL_0 + p_1 \bL_1 \in P$ by
\beq
\mt^n(\bla) \,=\, \bla + \big[ p_0 n^2 + p_1 n(n-1) \big] \bL_{-1} +2n ( p_0 + p_1 )\bL_0 
- 2n( p_0 + p_1 ) \bL_1 \,.
\eeq
In particular, the coefficient $p_{-1}$ of $\bL_{-1}=-\bd$ does not enter in the orbit
under translations. The so-called generalized theta function of a weight $\bla \in P^\ell$ 
is defined by
\beq\label{genth}
\Theta_{\bla} \,=\, e^{-\frac{|\bla|^2}{2\ell} \bd} \ \sum_{\mt \in \mT} e^{\mt(\bla)}\,=\,
e^{-\frac{|\bla|^2}{2\ell} \bd} \ \sum_{n\in\mathbb{Z}} e^{\mt^n(\bla)} \, .
\eeq
We note that the definition is invariant under shifts of $\bla$ by multiples of $\bd$, 
i.e.\ $\Theta_{\bla+ s \bd}\,=\,\Theta_{\bla}$ for any $s\in\mathbb{C}$.
The full Weyl group is given by the semi-direct product $W\,=\,\{1,w_1\} \ltimes \mT$, where
\beq\label{WeylW1}
w_1(p_{-1} \bL_{-1} + p_0 \bL_0 + p_1 \bL_1 ) \,=\, 
p_{-1} \bL_{-1} + (p_0+2p_1) \bL_0 - p_1 \bL_1 \, 
\eeq 
and $w_1 \mt = \mt^{-1} w_1$. 

Since any Weyl transformation preserves the multiplicity of a root, we may organize the
character along orbits of the translation group using the generalized theta 
function~\eqref{genth}. Since the orbits for weights shifted by $\bla$ look similar, 
we additionally need to keep track of the multiplicities of weights along a string of weights
shifted by $\bla$. To this end we define the set of \textit{maximal} weights 
$\text{max}(\bL)$ associated with a highest weight representation $L(\bL)$ to be the 
set of $\bla$ such that $\bla+\bd\notin P(\bL)$. Then we can write the
character of $L(\bL)$ for $\bL$ a highest weight of level $\ell$ as
\beq\label{ChStringTheta}
\Ch \, L(\bL) \,=\, \sum_{\substack{ \bla \in \text{max}(\bL) \\ \bla \!\!\!\! \mod \mT}}
 C_\bL^\bla \ \Theta_\bla \, ,
\eeq
where $C_\bL^\bla$ is a so-called string function. It describes the multiplicities of a string
of weights extending in the $-\bd$ direction. Note that with the definition~\eqref{genth},
the string function 
\beq
C_\bL^\bla \,= \,e^{\frac{|\bL|^2}{2\ell}\bd} + \ldots \,=\, q^{-\frac{|\bL|^2}{2\ell}} + \ldots
\eeq
We also make use of the inverse of the generating series 
of partitions with $\vp(q) = \prod\limits_{n\ge 1}^\infty (1-q^n)$, such that 
\beq\label{phi}
\frac1{\varphi(q)} \,=\, \sum_{n\geq 0} p(n) q^n
\eeq
with $p(n)$ the classical number of partitions of $n$ into positive integers 
(with repetitions). 

As an example, we give the character of the basic representation on level $\ell=1$ in 
this form: 
\beq
\Ch \, L(\bL_0+n \bd) \,=\,q^{-n} \, \Ch \, L(\bL_0)\,=\,
q^{-n} \, C^{\bL_0}_{\bL_0} \Theta_{\Lambda_0}\,=\, 
\frac{q^{1-n}}{\varphi(q)} \, \Theta_{\Lambda_0}
\eeq
since there is only one maximal weight up to the action by $\mT$ and the multiplicities
in the basic module are simply given by the classical integer partitions~\cite{Kac}.

In the main text, we have already introduced the Sugawara central charge and allowed
eigenvalues of unitary minimal models. They are a special case of the $(p, p^\prime)$
minimal model with central charge and allowed eigenvalues (with $p>p^\prime$ without
loss of generality)
\begin{align}
\begin{aligned}
&c^{p,p^\prime} \,=\, 1- 6 \frac{(p-p')^2}{pp'} \, , \\
&h_{r,s}^{p, p^\prime} \,=\, \frac{(pr-p's)^2-(p-p')^2}{4pp'} \, , 
\end{aligned}
\end{align}
with $1\leq r \leq p'-1$ and $1\leq s\leq p-1$ as well as $p^\prime s \le p r$. 
These are only unitary for $|p-p'|= 1$ and then give the same formulas
as~\eqref{CentralCharge} and~\eqref{hrs}, but with a different parametrization where
\beq
\ell \,=\, \frac12 \frac{p+p'}{p-p'} - \frac32\,.
\eeq
Finally, we define the Virasoro character of the minimal model 
$\Vir(c^{p,p^\prime},h_{r,s}^{p, p^\prime})$ by~\cite{DiFrancesco}
\beq\label{VirCh}
\chi_{r,s}^{p,p^\prime}(q) \,=\, \frac{1}{\varphi(q)} \left[ q^{h_{r,s}^{p,p^\prime}} 
+ \sum_{k=1}^\infty (-1)^k \left( q^{h_1(k)}+q^{h_2(k)}\right)\right]
\,=\, q^{h_{r,s}^{p,p^\prime}} + \ldots \,.
\eeq
with 
\beq
h_1(k)\,=\,h_{r+kp',(-1)^k s+(1-(-1)^k) p/2}^{p,p^\prime}\,,\quad\quad
h_2(k)\,=\,h_{r,kp + (-1)^k s+(1-(-1)^k) p/2}^{p,p^\prime}\,.
\eeq

\section{Cocycle factors}\label{app:Cocycle}

In this appendix, we discuss the so-called cocycle factors and fix the related convention
used throughout this paper. Cocycle factors were first introduced 
in~\cite{FK,Segal:1981ap,Goddard:1988fw} to ensure the anti-symmetry of the
commutator~\eqref{StateOperator}. For this work we adopt the cocycle conventions 
of~\cite{Kac}. We begin by repeating some important notation.

Recall that any DDF state is of the form
\beq
\varphi_\br \,=\, \prod_{i=1}^M \Al_{-m_i} \prod_{j=1}^N \Bl_{-n_j} \ket{\ba_n^{(l)}} 
\quad \quad M, N \ge 0
\eeq
with total momentum
\beq
\br \,=\, \ba_n^{(l)} + \left( \sum_{i=1}^M m_i + \sum_{j=1}^N n_j \right) \bk_l \, .
\eeq
If the DDF state $\varphi_\br$ is an element of the Lie algebra $\mF$ then the total
momentum $\br$ is an element of the root lattice $Q$, i.e.
\beq
\br \,=\, a_{-1} \br_{-1} + k_{-1} \bd + a_1 \br_1 \quad \text{with} \quad 
a_{-1}, k_{-1}, a_1 \in \mathbb{Z} \, .
\eeq
In~\cite{GN} it is explained in detail how the commutator of any two DDF states is defined
via the state-operator correspondence. However, it is well known that this definition does
not make the commutator anti-symmetric for all DDF states.
Hence~\cite{FK,Segal:1981ap,Goddard:1988fw} introduced the cocycle factors 
$c_\br$ which commute with all DDF states $\varphi_\bs$ ($\br, \bs \in Q$) and satisfy 
\beq
c_\br c_\bs \,=\, \epsilon(\br, \bs) c_{\br+\bs} \, , 
\eeq
where $\epsilon: Q \times Q \mapsto \{\pm 1\}$ is a 2-cocycle. We then define 
$\varphi \coloneqq \varphi_\br c_\br$ and $\psi \coloneqq \psi_\bs c_\bs$ such that the
 commutator~\eqref{StateOperator} becomes
\beq
[ \vp, \psi ] \,=\, \oint \frac{\rd z}{2\pi i} \ \cV( \vp_r c_\br;z) \, \psi_s c_\bs 
\,=\, \oint \frac{\rd z}{2\pi i} \ \cV( \vp_r ;z) \, c_\br \psi_s c_\bs 
\,=\, \epsilon(\br, \bs) \oint \frac{\rd z}{2\pi i} \ \cV( \vp_r ;z) \, \psi_s c_{\br+\bs} \, ,
\eeq
where $\oint \frac{\rd z}{2\pi i} \ \cV( \vp_r ;z) \, \psi_s$ is a new DDF state with total
momentum $\br + \bs$. This commutator is now anti-symmetric for all DDF 
states~\cite{FK,Segal:1981ap,Goddard:1988fw}.

For the 2-cocycle $\epsilon$ we impose the following conditions~\cite{Kac}
\begin{align}\label{bimult}
\begin{aligned}
&\epsilon(\br + \br^\prime , \bs ) \,=\, \epsilon(\br , \bs) \epsilon(\br^\prime , \bs) \, , \\
&\epsilon(\br , \bs + \bs^\prime ) \,=\, \epsilon(\br , \bs) \epsilon(\br , \bs^\prime) 
\end{aligned}
\end{align}
and
\beq\label{cocycle}
\epsilon(\br, \br) \,=\, (-1)^{\frac{1}{2} (\br \cdot \br)}
\eeq
for all $\br, \bs \in Q$. Moreover, we normalize the cocycle factor to
\beq\label{normalization}
\epsilon(0, 0) \,=\, 1 \, .
\eeq
By replacing $\br$ with $\br + \bs$ in~\eqref{cocycle} we obtain
\beq
 \epsilon(\br, \bs) \epsilon(\bs, \br) \,=\, (-1)^{\br \cdot \bs} \, .
\eeq
The bi-multiplicativity~\eqref{bimult} and the normalization~\eqref{normalization} imply
\beq
\epsilon(m \br , n \br) \,=\, \epsilon(\br, \br)^{mn} \,=\, (-1)^{\frac{1}{2} mn (\br \cdot \br)} \, .
 \qquad (m,n \in \mathbb{Z})
\eeq
These conditions are sufficient to specify
\begin{align}
\begin{aligned}
&\epsilon(\br_{-1} , \br_{-1}) \,=\, -1 \, , \\
&\epsilon(\bd, \bd) \,=\, 1 \, , \\
&\epsilon(\br_1 , \br_1) \,=\, - 1 \, \\
\end{aligned}
\end{align}
and
\begin{align}
\begin{aligned}
&\epsilon(\br_{-1} , \bd ) \,=\, - \epsilon( \bd , \br_{-1} ) \, , \\
&\epsilon(\br_{-1} , \br_1) \,=\, \epsilon(\br_1 , \br_{-1}) \, , \\
& \epsilon( \bd , \br_1) \,=\, \epsilon(\br_1 , \bd ) \, . 
\end{aligned}
\end{align}
Subsequently, we choose
\beq
\epsilon(\br_{-1} , \bd ) \,=\, -1 \quad \text{and} \quad \epsilon(\br_{-1} , \br_1) \,=\, 
\epsilon( \bd, \br_1 ) \,=\, 1 
\eeq
in agreement with the conventions given in~\cite{Kac}. Fixing also these last three cocycle
factors explicitly has the advantage that now the cocycle factor of any two momenta 
$\br, \bs \in Q$ is completely determined which considerably simplifies the results of all
computations in the main text involving commutators and affine generators. 

The affine generators ${}^{[\ell]}\!E_m$ and ${}^{[\ell]}\!F_m$ defined in~\eqref{EmFm} 
also come with a cocycle factor because they change the total momentum of the state 
on which they act. Let ${}^{[\ell]}\!E_m \equiv {}^{[\ell]}\!\hat{E}_m c_{\br_1}$ and 
${}^{[\ell]}\!F_m \equiv {}^{[\ell]}\!\hat{F}_m c_{-\br_1}$, so we find
\begin{align}
\begin{aligned}
&{}^{[\ell]}\!E_m \varphi \,=\, {}^{[\ell]}\!\hat{E}_m c_{\br_1} \varphi_\bs c_\bs \,=\, 
\epsilon( \br_1, \bs) {}^{[\ell]}\!\hat{E}_m \varphi_\bs c_\mathbf{s + \br_1} \, , \\
&{}^{[\ell]}\!F_m \varphi \,=\, {}^{[\ell]}\!\hat{F}_m c_{-\br_1} \varphi_\bs c_\bs \,=\, 
\epsilon(- \br_1, \bs) {}^{[\ell]}\!\hat{F}_m \varphi_\bs c_\mathbf{s - \br_1} \, .
\end{aligned}
\end{align}
Finally, we remark that the cocycle factors do not commute with the affine generators 
${}^{[\ell]}\!\hat{E}_m$ and ${}^{[\ell]}\!\hat{F}_m$ since 
$c_\br e^{i \bs \cdot \mathbf{Q}} = \epsilon(\br, \bs) e^{i \bs \cdot \mathbf{Q}} c_\br$.

\section{Translation of DDF states to multi-commutators at level 2}\label{app:mc}

For completeness we spell out the relation between these DDF 
states~\eqref{someDDF} and multi-commutators of the Chevalley-Serre generators more
explicitly. To this end we introduce the shorthand notation for words $w= l w'$ where 
$l$ is a letter in the non-commutative alphabet $\{-1,0,1\}$ recursively by
\beq\label{elw}
f_{w}\,=\,f_{lw'} \coloneqq [f_l, f_{w'}]
\eeq
with $f_w= f_{l w'}= f_l$ for words $w$ of length one (i.e. $w'=\emptyset$).

For low depths and level 2 the translation from DDF states to multi-commutators can 
be done relatively easily. For the above three ground states at depth 2, we find
\begin{align}
\begin{aligned}
& \psi_{2,\bD,-1}^{(2)} \,=\, \frac{1}{2} \ f_{-1,-1,0,1,0} \, , \\
& \psi_{2,\bD,0}^{(2)} \,=\, - \frac{1}{4} \ f_{-1,-1,0,1,0,1} \, , \\
& \psi_{2,\bD,1}^{(2)} \,=\, - \frac{1}{4} \ f_{-1,-1,1,0,1,0,1} \, .
\end{aligned}
\end{align}
with the notation~\eqref{elw}. At depth 3 the relation between the seven states and the 
seven linearly independent multi-commutators is a little more complicated. The 
multi-commutators are
\begin{align}
\begin{aligned}
&-(2,3,4): &&\chi_{3,6}^{(2)} \,=\, f_{-1, -1, 1, 0, 1, 0, 1, 0, 1} 
&&\chi_{3,7}^{(2)} \,=\, f_{-1, 1, 0, -1, 1, 0, 1, 0, 1} \\[2ex]
&-(2,3,3): && \chi_{3,3}^{(2)} \,=\, f_{-1, -1, 0, 1, 0, 1, 0, 1} 
&&\chi_{3,4}^{(2)} \,=\, f_{-1, 0, -1, 1, 0, 1, 0, 1} 
&&\chi_{3,5}^{(2)} \,=\, f_{-1, 1, 0, -1, 0, 1, 0, 1} \\[2ex]
&-(2,3,2): &&\chi_{3,1}^{(2)} \,=\, f_{-1, -1, 0, 0, 1, 0, 1} 
&& \chi_{3,2}^{(2)} \,=\, f_{-1, 0, -1, 0, 1, 0, 1} \, .
\end{aligned}
\end{align}
Then we find that
\beq
\begin{pmatrix} 
\psi_{3,\bD,-1}^{(2)} \\ \psi_{3,\bD,0}^{(2)} \\ \psi_{3,\bD,1}^{(2)} \\ 
\phi_{3,\bE,0}^{(2)} \\ \phi_{3,\bD,-1}^{(2)} 
\\ \phi_{3,\bD,0}^{(2)} \\ \phi_{3,\bD,1}^{(2)} 
\end{pmatrix} 
\,=\, 
\begin{pmatrix} 
1 & -1 & 0 & 0 & 0 & 0 & 0 \\
0 & 0 & \frac{1}{2} & 0 & -\frac{1}{2} & 0 & 0 \\
0 & 0 & 0 & 0 & 0 & \frac{1}{2} & - \frac{1}{2} \\
0 & 0 & 0 & 1 & -1 & 0 & 0 \\
- \frac{1}{4} & \frac{1}{2} & 0 & 0 & 0 & 0 & 0 \\
0 & 0 & -\frac{1}{8} & 0 &\frac{1}{4} & 0 & 0 \\
0 & 0 & 0 & 0 & 0 & - \frac{1}{8} & \frac{1}{4}
\end{pmatrix}
\begin{pmatrix}
\chi_{\bD,1}^{(2)} \\ \chi_{\bD,2}^{(2)} \\ \chi_{\bD,3}^{(2)} \\ 
\chi_{\bD,4}^{(2)} \\ \chi_{\bD,5}^{(2)} \\ \chi_{\bD,6}^{(2)} \\ 
\chi_{\bD,7}^{(2)}
\end{pmatrix} \, .
\eeq
It is not hard to see that this transformation is invertible. Similar relations exist for higher depths.

\newpage
\section{Root systems}\label{app:Figures}

We provide figures of the root system of $\mathfrak{F}$ at levels $\ell= 1,2,3, 4$.
For an interactive 2D and 3D version of these root systems see~\cite{VisualLie}.
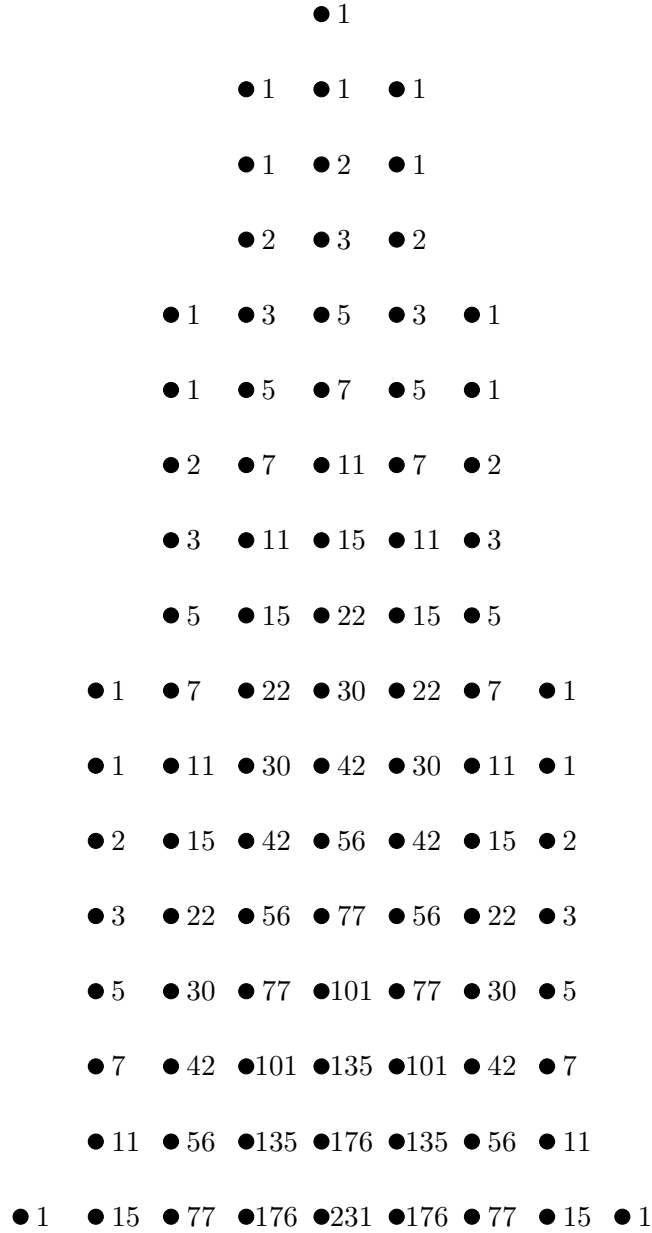
\begin{figure}[H]
\vspace{.5cm}
\begin{tikzpicture}
\draw [fill] (0,0) circle (.1);
\draw [fill] (-1,-1) circle (.1);
\draw [fill] (0,-1) circle (.1);
\draw [fill] (1,-1) circle (.1);
\draw [fill] (-1,-2) circle (.1);
\draw [fill] (0,-2) circle (.1);
\draw [fill] (1,-2) circle (.1);
\draw [fill] (-1,-3) circle (.1);
\draw [fill] (0,-3) circle (.1);
\draw [fill] (1,-3) circle (.1);
\draw [fill] (-2,-4) circle (.1);
\draw [fill] (-1,-4) circle (.1);
\draw [fill] (0,-4) circle (.1);
\draw [fill] (1,-4) circle (.1);
\draw [fill] (2,-4) circle (.1);
\draw [fill] (-2,-5) circle (.1);
\draw [fill] (-1,-5) circle (.1);
\draw [fill] (0,-5) circle (.1);
\draw [fill] (1,-5) circle (.1);
\draw [fill] (2,-5) circle (.1);
\draw [fill] (-2,-6) circle (.1);
\draw [fill] (-1,-6) circle (.1);
\draw [fill] (0,-6) circle (.1);
\draw [fill] (1,-6) circle (.1);
\draw [fill] (2,-6) circle (.1);
\draw [fill] (-2,-7) circle (.1);
\draw [fill] (-1,-7) circle (.1);
\draw [fill] (0,-7) circle (.1);
\draw [fill] (1,-7) circle (.1);
\draw [fill] (2,-7) circle (.1);
\draw [fill] (-2,-8) circle (.1);
\draw [fill] (-1,-8) circle (.1);
\draw [fill] (0,-8) circle (.1);
\draw [fill] (1,-8) circle (.1);
\draw [fill] (2,-8) circle (.1);
\draw [fill] (-3,-9) circle (.1);
\draw [fill] (-2,-9) circle (.1);
\draw [fill] (-1,-9) circle (.1);
\draw [fill] (0,-9) circle (.1);
\draw [fill] (1,-9) circle (.1);
\draw [fill] (2,-9) circle (.1);
\draw [fill] (3,-9) circle (.1);
\draw [fill] (-3,-10) circle (.1);
\draw [fill] (-2,-10) circle (.1);
\draw [fill] (-1,-10) circle (.1);
\draw [fill] (0,-10) circle (.1);
\draw [fill] (1,-10) circle (.1);
\draw [fill] (2,-10) circle (.1);
\draw [fill] (3,-10) circle (.1);
\draw [fill] (-3,-11) circle (.1);
\draw [fill] (-2,-11) circle (.1);
\draw [fill] (-1,-11) circle (.1);
\draw [fill] (0,-11) circle (.1);
\draw [fill] (1,-11) circle (.1);
\draw [fill] (2,-11) circle (.1);
\draw [fill] (3,-11) circle (.1);
\draw [fill] (-3,-12) circle (.1);
\draw [fill] (-2,-12) circle (.1);
\draw [fill] (-1,-12) circle (.1);
\draw [fill] (0,-12) circle (.1);
\draw [fill] (1,-12) circle (.1);
\draw [fill] (2,-12) circle (.1);
\draw [fill] (3,-12) circle (.1);
\draw [fill] (-3,-13) circle (.1);
\draw [fill] (-2,-13) circle (.1);
\draw [fill] (-1,-13) circle (.1);
\draw [fill] (0,-13) circle (.1);
\draw [fill] (1,-13) circle (.1);
\draw [fill] (2,-13) circle (.1);
\draw [fill] (3,-13) circle (.1);
\draw [fill] (-3,-14) circle (.1);
\draw [fill] (-2,-14) circle (.1);
\draw [fill] (-1,-14) circle (.1);
\draw [fill] (0,-14) circle (.1);
\draw [fill] (1,-14) circle (.1);
\draw [fill] (2,-14) circle (.1);
\draw [fill] (3,-14) circle (.1);
\draw [fill] (-3,-15) circle (.1);
\draw [fill] (-2,-15) circle (.1);
\draw [fill] (-1,-15) circle (.1);
\draw [fill] (0,-15) circle (.1);
\draw [fill] (1,-15) circle (.1);
\draw [fill] (2,-15) circle (.1);
\draw [fill] (3,-15) circle (.1);
\draw [fill] (-4,-16) circle (.1);
\draw [fill] (-3,-16) circle (.1);
\draw [fill] (-2,-16) circle (.1);
\draw [fill] (-1,-16) circle (.1);
\draw [fill] (0,-16) circle (.1);
\draw [fill] (1,-16) circle (.1);
\draw [fill] (2,-16) circle (.1);
\draw [fill] (3,-16) circle (.1);
\draw [fill] (4,-16) circle (.1);
\node at (.3,0) {1};
\node at (-.7,-1) {1};
\node at (.3,-1) {1};
\node at (1.3,-1) {1};
\node at (-.7,-2) {1};
\node at (.3,-2) {2};
\node at (1.3,-2) {1};
\node at (-.7,-3) {2};
\node at (.3,-3) {3};
\node at (1.3,-3) {2};
\node at (-1.7,-4) {1};
\node at (-.7,-4) {3};
\node at (.3,-4) {5};
\node at (1.3,-4) {3};
\node at (2.3,-4) {1};
\node at (-1.7,-5) {1};
\node at (-.7,-5) {5};
\node at (.3,-5) {7};
\node at (1.3,-5) {5};
\node at (2.3,-5) {1};
\node at (-1.7,-6) {2};
\node at (-.7,-6) {7};
\node at (.4,-6) {11};
\node at (1.3,-6) {7};
\node at (2.3,-6) {2};
\node at (-1.7,-7) {3};
\node at (-.6,-7) {11};
\node at (.4,-7) {15};
\node at (1.4,-7) {11};
\node at (2.3,-7) {3};
\node at (-1.7,-8) {5};
\node at (-.6,-8) {15};
\node at (.4,-8) {22};
\node at (1.4,-8) {15};
\node at (2.3,-8) {5};
\node at (-2.7,-9) {1};
\node at (-1.7,-9) {7};
\node at (-.6,-9) {22};
\node at (.4,-9) {30};
\node at (1.4,-9) {22};
\node at (2.3,-9) {7};
\node at (3.3,-9) {1};
\node at (-2.7,-10) {1};
\node at (-1.6,-10) {11};
\node at (-.6,-10) {30};
\node at (.4,-10) {42};
\node at (1.4,-10) {30};
\node at (2.4,-10) {11};
\node at (3.3,-10) {1};
\node at (-2.7,-11) {2};
\node at (-1.6,-11) {15};
\node at (-.6,-11) {42};
\node at (.4,-11) {56};
\node at (1.4,-11) {42};
\node at (2.4,-11) {15};
\node at (3.3,-11) {2};
\node at (-2.7,-12) {3};
\node at (-1.6,-12) {22};
\node at (-.6,-12) {56};
\node at (.4,-12) {77};
\node at (1.4,-12) {56};
\node at (2.4,-12) {22};
\node at (3.3,-12) {3};
\node at (-2.7,-13) {5};
\node at (-1.6,-13) {30};
\node at (-.6,-13) {77};
\node at (.4,-13) {101};
\node at (1.4,-13) {77};
\node at (2.4,-13) {30};
\node at (3.3,-13) {5};
\node at (-2.7,-14) {7};
\node at (-1.6,-14) {42};
\node at (-.6,-14) {101};
\node at (.4,-14) {135};
\node at (1.4,-14) {101};
\node at (2.4,-14) {42};
\node at (3.3,-14) {7};
\node at (-2.6,-15) {11};
\node at (-1.6,-15) {56};
\node at (-.6,-15) {135};
\node at (.4,-15) {176};
\node at (1.4,-15) {135};
\node at (2.4,-15) {56};
\node at (3.4,-15) {11};
\node at (-3.7,-16) {1};
\node at (-2.6,-16) {15};
\node at (-1.6,-16) {77};
\node at (-.6,-16) {176};
\node at (.4,-16) {231};
\node at (1.4,-16) {176};
\node at (2.4,-16) {77};
\node at (3.4,-16) {15};
\node at (4.3,-16) {1};
\end{tikzpicture}
\vspace{.5cm}
\caption{The partial level-1 root system associated to $\mF^{(1)}$. The numbers 
specify the number of DDF states at each point
and coincide with the multiplicities of the associated roots.
They are in one-to-one correspondence with~\eqref{chLevel1} and~\eqref{SL2Rep}.
The uppermost root is $(-1,0,0)$ the triplet below is $(-1,-1,-2)$, $(-1,-1,-1)$ and 
$(-1,-1,0)$ from the left to right. 
The corresponding DDF states are $\ket{\ba_0}$ and 
$\ket{\ba_{- 1}}$, $A_{-1} \ket{\ba_0}$, $\ket{\ba_{+1}}$.
As explained in the main text the first number indicates the level, the second number
indicates the depth and the third number is the sum of the depth and the weight. 
Thus, the states in the center column are $(-1,-\md, -\md)$. A state $n$ points left of 
the center column is $(-1,-\md, -\md - n)$. This correspondence holds at all levels.}
\label{fig:L1RootLattice}
\end{figure}

\begin{figure}[H]
\vspace{.5cm}
\begin{tikzpicture}
\node [fill, myred, diamond, draw, scale=0.7] at (-1,1) {};
\node [fill, myred, diamond, draw, scale=0.7] at (0,1) {};
\node [fill, myred, diamond, draw, scale=0.7] at (1,1) {};
\draw [fill] (-1,0) circle (.1);
\draw [fill] (0,0) circle (.1);
\draw [fill] (1,0) circle (.1);
\draw [fill] (-1,-1) circle (.1);
\draw [fill] (0,-1) circle (.1);
\draw [fill] (1,-1) circle (.1);
\draw [fill] (-2,-2) circle (.1);
\draw [fill] (-1,-2) circle (.1);
\draw [fill] (0,-2) circle (.1);
\draw [fill] (1,-2) circle (.1);
\draw [fill] (2,-2) circle (.1);
\draw [fill] (-2,-3) circle (.1);
\draw [fill] (-1,-3) circle (.1);
\draw [fill] (0,-3) circle (.1);
\draw [fill] (1,-3) circle (.1);
\draw [fill] (2,-3) circle (.1);
\draw [fill] (-3,-4) circle (.1);
\draw [fill] (-2,-4) circle (.1);
\draw [fill] (-1,-4) circle (.1);
\draw [fill] (0,-4) circle (.1);
\draw [fill] (1,-4) circle (.1);
\draw [fill] (2,-4) circle (.1);
\draw [fill] (3,-4) circle (.1);
\draw [fill] (-3,-5) circle (.1);
\draw [fill] (-2,-5) circle (.1);
\draw [fill] (-1,-5) circle (.1);
\draw [fill] (0,-5) circle (.1);
\draw [fill] (1,-5) circle (.1);
\draw [fill] (2,-5) circle (.1);
\draw [fill] (3,-5) circle (.1);
\draw [fill] (-3,-6) circle (.1);
\draw [fill] (-2,-6) circle (.1);
\draw [fill] (-1,-6) circle (.1);
\draw [fill] (0,-6) circle (.1);
\draw [fill] (1,-6) circle (.1);
\draw [fill] (2,-6) circle (.1);
\draw [fill] (3,-6) circle (.1);
\draw [fill] (-3,-7) circle (.1);
\draw [fill] (-2,-7) circle (.1);
\draw [fill] (-1,-7) circle (.1);
\draw [fill] (0,-7) circle (.1);
\draw [fill] (1,-7) circle (.1);
\draw [fill] (2,-7) circle (.1);
\draw [fill] (3,-7) circle (.1);
\draw [fill] (-4,-8) circle (.1);
\draw [fill] (-3,-8) circle (.1);
\draw [fill] (-2,-8) circle (.1);
\draw [fill] (-1,-8) circle (.1);
\draw [fill] (0,-8) circle (.1);
\draw [fill] (1,-8) circle (.1);
\draw [fill] (2,-8) circle (.1);
\draw [fill] (3,-8) circle (.1);
\draw [fill] (4,-8) circle (.1);
\draw [fill] (-4,-9) circle (.1);
\draw [fill] (-3,-9) circle (.1);
\draw [fill] (-2,-9) circle (.1);
\draw [fill] (-1,-9) circle (.1);
\draw [fill] (0,-9) circle (.1);
\draw [fill] (1,-9) circle (.1);
\draw [fill] (2,-9) circle (.1);
\draw [fill] (3,-9) circle (.1);
\draw [fill] (4,-9) circle (.1);
\draw [fill] (-4,-10) circle (.1);
\draw [fill] (-3,-10) circle (.1);
\draw [fill] (-2,-10) circle (.1);
\draw [fill] (-1,-10) circle (.1);
\draw [fill] (0,-10) circle (.1);
\draw [fill] (1,-10) circle (.1);
\draw [fill] (2,-10) circle (.1);
\draw [fill] (3,-10) circle (.1);
\draw [fill] (4,-10) circle (.1);
\draw [fill] (-4,-11) circle (.1);
\draw [fill] (-3,-11) circle (.1);
\draw [fill] (-2,-11) circle (.1);
\draw [fill] (-1,-11) circle (.1);
\draw [fill] (0,-11) circle (.1);
\draw [fill] (1,-11) circle (.1);
\draw [fill] (2,-11) circle (.1);
\draw [fill] (3,-11) circle (.1);
\draw [fill] (4,-11) circle (.1);
\draw [fill] (-5,-12) circle (.1);
\draw [fill] (-4,-12) circle (.1);
\draw [fill] (-3,-12) circle (.1);
\draw [fill] (-2,-12) circle (.1);
\draw [fill] (-1,-12) circle (.1);
\draw [fill] (0,-12) circle (.1);
\draw [fill] (1,-12) circle (.1);
\draw [fill] (2,-12) circle (.1);
\draw [fill] (3,-12) circle (.1);
\draw [fill] (4,-12) circle (.1);
\draw [fill] (5,-12) circle (.1);
\draw [fill] (-5,-13) circle (.1);
\draw [fill] (-4,-13) circle (.1);
\draw [fill] (-3,-13) circle (.1);
\draw [fill] (-2,-13) circle (.1);
\draw [fill] (-1,-13) circle (.1);
\draw [fill] (0,-13) circle (.1);
\draw [fill] (1,-13) circle (.1);
\draw [fill] (2,-13) circle (.1);
\draw [fill] (3,-13) circle (.1);
\draw [fill] (4,-13) circle (.1);
\draw [fill] (5,-13) circle (.1);
\draw [fill] (-5,-14) circle (.1);
\draw [fill] (-4,-14) circle (.1);
\draw [fill] (-3,-14) circle (.1);
\draw [fill] (-2,-14) circle (.1);
\draw [fill] (-1,-14) circle (.1);
\draw [fill] (0,-14) circle (.1);
\draw [fill] (1,-14) circle (.1);
\draw [fill] (2,-14) circle (.1);
\draw [fill] (3,-14) circle (.1);
\draw [fill] (4,-14) circle (.1);
\draw [fill] (5,-14) circle (.1);
\draw [fill] (-5,-15) circle (.1);
\draw [fill] (-4,-15) circle (.1);
\draw [fill] (-3,-15) circle (.1);
\draw [fill] (-2,-15) circle (.1);
\draw [fill] (-1,-15) circle (.1);
\draw [fill] (0,-15) circle (.1);
\draw [fill] (1,-15) circle (.1);
\draw [fill] (2,-15) circle (.1);
\draw [fill] (3,-15) circle (.1);
\draw [fill] (4,-15) circle (.1);
\draw [fill] (5,-15) circle (.1);
\node at (-.7,1) {{\color{myred}1}};
\node at (.3,1) {{\color{myred}1}};
\node at (1.3,1) {{\color{myred}1}};
\node at (-.7,0) {1};
\node at (.3,0) {1};
\node at (1.3,0) {1};
\node at (-.7,-1) {2};
\node at (.3,-1) {3};
\node at (1.3,-1) {2};
\node at (-1.7,-2) {1};
\node at (-.7,-2) {5};
\node at (.3,-2) {7};
\node at (1.3,-2) {5};
\node at (2.3,-2) {1};
\node at (-1.7,-3) {3};
\node at (-.6,-3) {11};
\node at (.4,-3) {15};
\node at (1.4,-3) {11};
\node at (2.3,-3) {3};
\node at (-2.7,-4) {1};
\node at (-1.7,-4) {7};
\node at (-.6,-4) {22};
\node at (.4,-4) {30};
\node at (1.4,-4) {22};
\node at (2.3,-4) {7};
\node at (3.3,-4) {1};
\node at (-2.7,-5) {2};
\node at (-1.6,-5) {15};
\node at (-.6,-5) {42};
\node at (.4,-5) {56};
\node at (1.4,-5) {42};
\node at (2.4,-5) {15};
\node at (3.3,-5) {2};
\node at (-2.7,-6) {5};
\node at (-1.6,-6) {30};
\node at (-.6,-6) {77};
\node at (.4,-6) {101};
\node at (1.4,-6) {77};
\node at (2.4,-6) {30};
\node at (3.3,-6) {5};
\node at (-2.6,-7) {11};
\node at (-1.6,-7) {56};
\node at (-.6,-7) {135};
\node at (.4,-7) {176};
\node at (1.4,-7) {135};
\node at (2.4,-7) {56};
\node at (3.4,-7) {11};
\node at (-3.7,-8) {1};
\node at (-2.6,-8) {22};
\node at (-1.6,-8) {101};
\node at (-.6,-8) {231};
\node at (.4,-8) {297};
\node at (1.4,-8) {231};
\node at (2.4,-8) {101};
\node at (3.4,-8) {22};
\node at (4.3,-8) {1};
\node at (-3.7,-9) {3};
\node at (-2.6,-9) {42};
\node at (-1.6,-9) {176};
\node at (-.6,-9) {385};
\node at (.4,-9) {490};
\node at (1.4,-9) {385};
\node at (2.4,-9) {176};
\node at (3.4,-9) {42};
\node at (4.3,-9) {3};
\node at (-3.7,-10) {7};
\node at (-2.6,-10) {77};
\node at (-1.6,-10) {297};
\node at (-.6,-10) {626};
\node at (.4,-10) {791};
\node at (1.4,-10) {626};
\node at (2.4,-10) {297};
\node at (3.4,-10) {77};
\node at (4.3,-10) {7};
\node at (-3.6,-11) {15};
\node at (-2.6,-11) {135};
\node at (-1.6,-11) {490};
\node at (-.55,-11) {\footnotesize{1001}};
\node at (.45,-11) {\footnotesize{1253}};
\node at (1.45,-11) {\footnotesize{1001}};
\node at (2.4,-11) {490};
\node at (3.4,-11) {135};
\node at (4.4,-11) {15};
\node at (-4.7,-12) {1};
\node at (-3.6,-12) {30};
\node at (-2.6,-12) {231};
\node at (-1.6,-12) {791};
\node at (-.55,-12) {\footnotesize{1571}};
\node at (.45,-12) {\footnotesize{1953}};
\node at (1.45,-12) {\footnotesize{1571}};
\node at (2.4,-12) {791};
\node at (3.4,-12) {231};
\node at (4.4,-12) {30};
\node at (5.3,-12) {1};
\node at (-4.7,-13) {2};
\node at (-3.6,-13) {56};
\node at (-2.6,-13) {385};
\node at (-1.55,-13) {\footnotesize{1253}};
\node at (-.55,-13) {\footnotesize{2429}};
\node at (.45,-13) {\footnotesize{3000}};
\node at (1.45,-13) {\footnotesize{2429}};
\node at (2.45,-13) {\footnotesize{1253}};
\node at (3.4,-13) {385};
\node at (4.4,-13) {56};
\node at (5.3,-13) {2};
\node at (-4.7,-14) {5};
\node at (-3.6,-14) {101};
\node at (-2.6,-14) {626};
\node at (-1.55,-14) {\footnotesize{1953}};
\node at (-.55,-14) {\footnotesize{3702}};
\node at (.45,-14) {\footnotesize{4544}};
\node at (1.45,-14) {\footnotesize{3702}};
\node at (2.45,-14) {\footnotesize{1953}};
\node at (3.4,-14) {626};
\node at (4.4,-14) {101};
\node at (5.3,-14) {5};
\node at (-4.6,-15) {11};
\node at (-3.6,-15) {176};
\node at (-2.55,-15) {\footnotesize{1001}};
\node at (-1.55,-15) {\footnotesize{3000}};
\node at (-.55,-15) {\footnotesize{5576}};
\node at (.45,-15) {\footnotesize{6804}};
\node at (1.45,-15) {\footnotesize{5576}};
\node at (2.45,-15) {\footnotesize{3000}};
\node at (3.45,-15) {\footnotesize{1001}};
\node at (4.4,-15) {176};
\node at (5.4,-15) {11};
\end{tikzpicture}
\vspace{.5cm}
\caption{The partial level-2 root system associated to $\mF^{(2)}$. The numbers 
specify the number of states at each point,
and are equal to the multiplicities of the associated roots.
The {\color{red}red} diamonds denote the virtual triplet~\eqref{VTrip} at depth one
that is not part of $\mF^{(2)}$ but only $\mF^{(1)} \wedge \mF^{(1)}$.
The top left root is $(-2,-1,-2)$. Not drawn in this picture (nor included in the
multiplicities) is the affine tower erected over the red diamonds as none of its 
states belong to $\mF^{(2)}$.} 
\label{fig:L2RootLattice}
\end{figure}
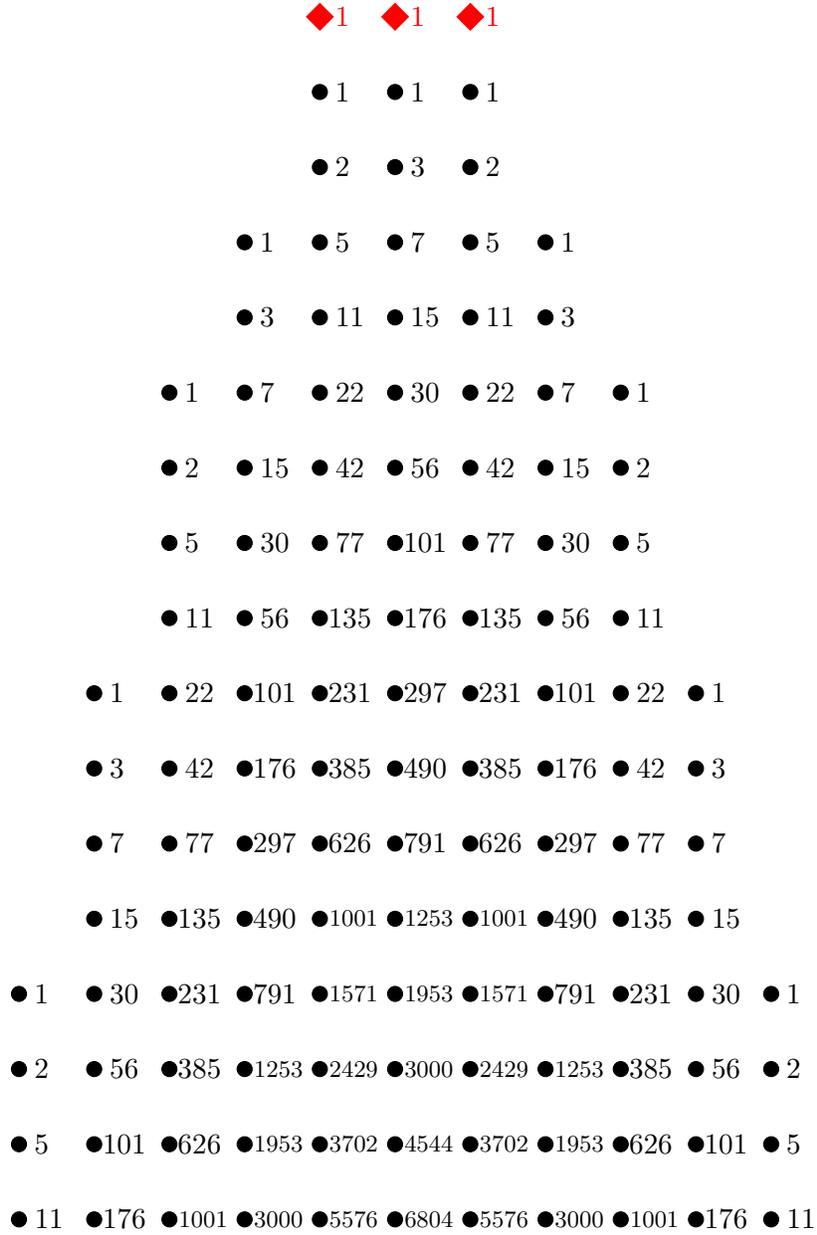

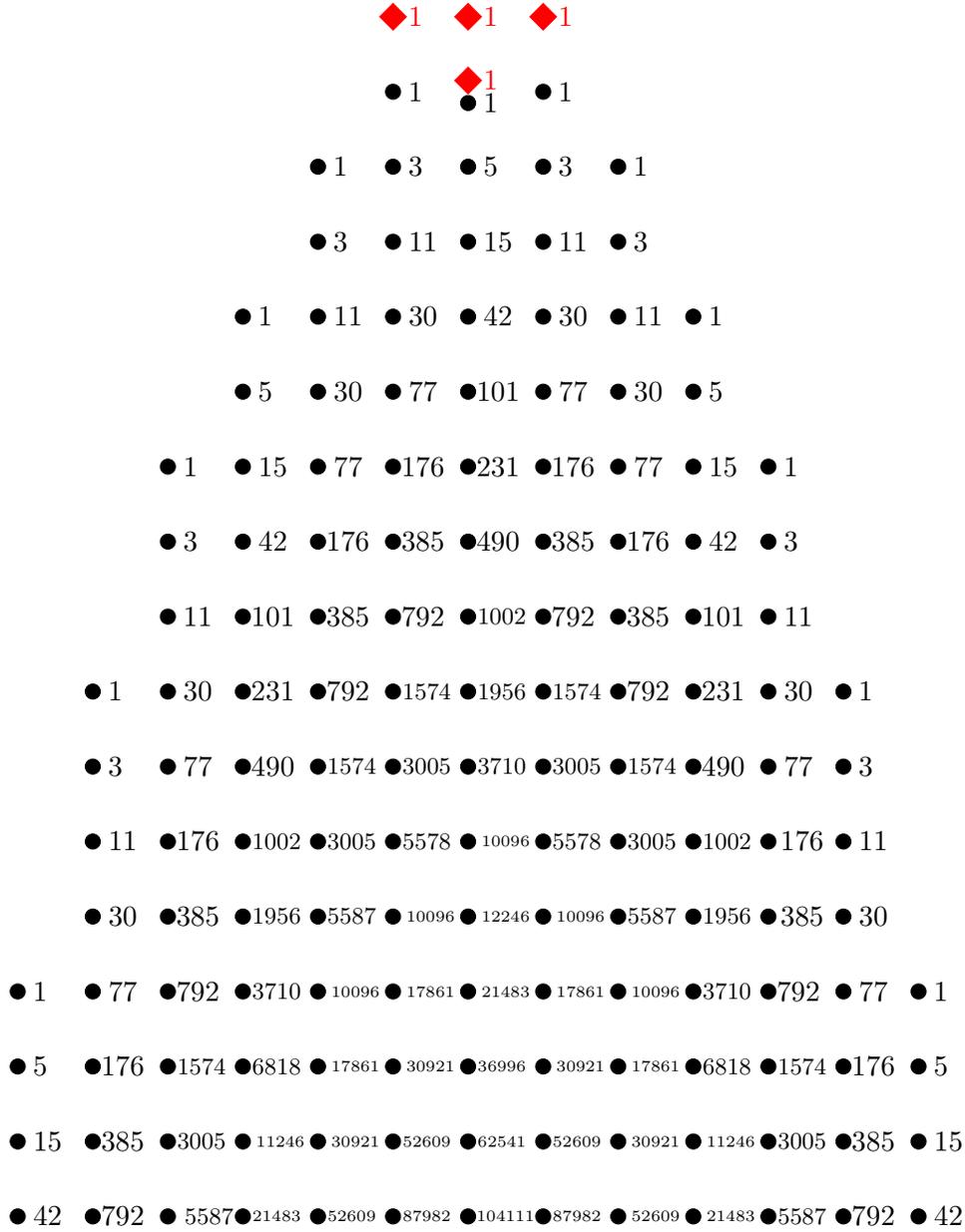
\begin{figure}[H]
\vspace{.5cm}
\begin{tikzpicture}
\node [fill, myred, diamond, draw, scale=0.7] at (-1,1) {};
\node [fill, myred, diamond, draw, scale=0.7] at (0,1) {};
\node [fill, myred, diamond, draw, scale=0.7] at (1,1) {};
\node [fill, myred, diamond, draw, scale=0.7] at (0,.15) {};
\draw [fill] (-1,0) circle (.1);
\draw [fill] (0,-.15) circle (.1);
\draw [fill] (1,0) circle (.1);
\draw [fill] (-2,-1) circle (.1);
\draw [fill] (-1,-1) circle (.1);
\draw [fill] (0,-1) circle (.1);
\draw [fill] (1,-1) circle (.1);
\draw [fill] (2,-1) circle (.1);
\draw [fill] (-2,-2) circle (.1);
\draw [fill] (-1,-2) circle (.1);
\draw [fill] (0,-2) circle (.1);
\draw [fill] (1,-2) circle (.1);
\draw [fill] (2,-2) circle (.1);
\draw [fill] (-3,-3) circle (.1);
\draw [fill] (-2,-3) circle (.1);
\draw [fill] (-1,-3) circle (.1);
\draw [fill] (0,-3) circle (.1);
\draw [fill] (1,-3) circle (.1);
\draw [fill] (2,-3) circle (.1);
\draw [fill] (3,-3) circle (.1);
\draw [fill] (-3,-4) circle (.1);
\draw [fill] (-2,-4) circle (.1);
\draw [fill] (-1,-4) circle (.1);
\draw [fill] (0,-4) circle (.1);
\draw [fill] (1,-4) circle (.1);
\draw [fill] (2,-4) circle (.1);
\draw [fill] (3,-4) circle (.1);
\draw [fill] (-4,-5) circle (.1);
\draw [fill] (-3,-5) circle (.1);
\draw [fill] (-2,-5) circle (.1);
\draw [fill] (-1,-5) circle (.1);
\draw [fill] (0,-5) circle (.1);
\draw [fill] (1,-5) circle (.1);
\draw [fill] (2,-5) circle (.1);
\draw [fill] (3,-5) circle (.1);
\draw [fill] (4,-5) circle (.1);
\draw [fill] (-4,-6) circle (.1);
\draw [fill] (-3,-6) circle (.1);
\draw [fill] (-2,-6) circle (.1);
\draw [fill] (-1,-6) circle (.1);
\draw [fill] (0,-6) circle (.1);
\draw [fill] (1,-6) circle (.1);
\draw [fill] (2,-6) circle (.1);
\draw [fill] (3,-6) circle (.1);
\draw [fill] (4,-6) circle (.1);
\draw [fill] (-4,-7) circle (.1);
\draw [fill] (-3,-7) circle (.1);
\draw [fill] (-2,-7) circle (.1);
\draw [fill] (-1,-7) circle (.1);
\draw [fill] (0,-7) circle (.1);
\draw [fill] (1,-7) circle (.1);
\draw [fill] (2,-7) circle (.1);
\draw [fill] (3,-7) circle (.1);
\draw [fill] (4,-7) circle (.1);
\draw [fill] (-5,-8) circle (.1);
\draw [fill] (-4,-8) circle (.1);
\draw [fill] (-3,-8) circle (.1);
\draw [fill] (-2,-8) circle (.1);
\draw [fill] (-1,-8) circle (.1);
\draw [fill] (0,-8) circle (.1);
\draw [fill] (1,-8) circle (.1);
\draw [fill] (2,-8) circle (.1);
\draw [fill] (3,-8) circle (.1);
\draw [fill] (4,-8) circle (.1);
\draw [fill] (5,-8) circle (.1);
\draw [fill] (-5,-9) circle (.1);
\draw [fill] (-4,-9) circle (.1);
\draw [fill] (-3,-9) circle (.1);
\draw [fill] (-2,-9) circle (.1);
\draw [fill] (-1,-9) circle (.1);
\draw [fill] (0,-9) circle (.1);
\draw [fill] (1,-9) circle (.1);
\draw [fill] (2,-9) circle (.1);
\draw [fill] (3,-9) circle (.1);
\draw [fill] (4,-9) circle (.1);
\draw [fill] (5,-9) circle (.1);
\draw [fill] (-5,-10) circle (.1);
\draw [fill] (-4,-10) circle (.1);
\draw [fill] (-3,-10) circle (.1);
\draw [fill] (-2,-10) circle (.1);
\draw [fill] (-1,-10) circle (.1);
\draw [fill] (0,-10) circle (.1);
\draw [fill] (1,-10) circle (.1);
\draw [fill] (2,-10) circle (.1);
\draw [fill] (3,-10) circle (.1);
\draw [fill] (4,-10) circle (.1);
\draw [fill] (5,-10) circle (.1);
\draw [fill] (-5,-11) circle (.1);
\draw [fill] (-4,-11) circle (.1);
\draw [fill] (-3,-11) circle (.1);
\draw [fill] (-2,-11) circle (.1);
\draw [fill] (-1,-11) circle (.1);
\draw [fill] (0,-11) circle (.1);
\draw [fill] (1,-11) circle (.1);
\draw [fill] (2,-11) circle (.1);
\draw [fill] (3,-11) circle (.1);
\draw [fill] (4,-11) circle (.1);
\draw [fill] (5,-11) circle (.1);
\draw [fill] (-6,-12) circle (.1);
\draw [fill] (-5,-12) circle (.1);
\draw [fill] (-4,-12) circle (.1);
\draw [fill] (-3,-12) circle (.1);
\draw [fill] (-2,-12) circle (.1);
\draw [fill] (-1,-12) circle (.1);
\draw [fill] (0,-12) circle (.1);
\draw [fill] (1,-12) circle (.1);
\draw [fill] (2,-12) circle (.1);
\draw [fill] (3,-12) circle (.1);
\draw [fill] (4,-12) circle (.1);
\draw [fill] (5,-12) circle (.1);
\draw [fill] (6,-12) circle (.1);
\draw [fill] (-6,-13) circle (.1);
\draw [fill] (-5,-13) circle (.1);
\draw [fill] (-4,-13) circle (.1);
\draw [fill] (-3,-13) circle (.1);
\draw [fill] (-2,-13) circle (.1);
\draw [fill] (-1,-13) circle (.1);
\draw [fill] (0,-13) circle (.1);
\draw [fill] (1,-13) circle (.1);
\draw [fill] (2,-13) circle (.1);
\draw [fill] (3,-13) circle (.1);
\draw [fill] (4,-13) circle (.1);
\draw [fill] (5,-13) circle (.1);
\draw [fill] (6,-13) circle (.1);
\draw [fill] (-6,-14) circle (.1);
\draw [fill] (-5,-14) circle (.1);
\draw [fill] (-4,-14) circle (.1);
\draw [fill] (-3,-14) circle (.1);
\draw [fill] (-2,-14) circle (.1);
\draw [fill] (-1,-14) circle (.1);
\draw [fill] (0,-14) circle (.1);
\draw [fill] (1,-14) circle (.1);
\draw [fill] (2,-14) circle (.1);
\draw [fill] (3,-14) circle (.1);
\draw [fill] (4,-14) circle (.1);
\draw [fill] (5,-14) circle (.1);
\draw [fill] (6,-14) circle (.1);
\draw [fill] (-6,-15) circle (.1);
\draw [fill] (-5,-15) circle (.1);
\draw [fill] (-4,-15) circle (.1);
\draw [fill] (-3,-15) circle (.1);
\draw [fill] (-2,-15) circle (.1);
\draw [fill] (-1,-15) circle (.1);
\draw [fill] (0,-15) circle (.1);
\draw [fill] (1,-15) circle (.1);
\draw [fill] (2,-15) circle (.1);
\draw [fill] (3,-15) circle (.1);
\draw [fill] (4,-15) circle (.1);
\draw [fill] (5,-15) circle (.1);
\draw [fill] (6,-15) circle (.1);
\node at (-.7,1) {{\color{myred}1}};
\node at (.3,1) {{\color{myred}1}};
\node at (1.3,1) {{\color{myred}1}};
\node at (.3,.15) {{\color{myred}1}};
\node at (-.7,0) {1};
\node at (.3,-.15) {1};
\node at (1.3,0) {1};
\node at (-1.7,-1) {1};
\node at (-.7,-1) {3};
\node at (.3,-1) {5};
\node at (1.3,-1) {3};
\node at (2.3,-1) {1};
\node at (-1.7,-2) {3};
\node at (-.6,-2) {11};
\node at (.4,-2) {15};
\node at (1.4,-2) {11};
\node at (2.3,-2) {3};
\node at (-2.7,-3) {1};
\node at (-1.6,-3) {11};
\node at (-.6,-3) {30};
\node at (.4,-3) {42};
\node at (1.4,-3) {30};
\node at (2.4,-3) {11};
\node at (3.3,-3) {1};
\node at (-2.7,-4) {5};
\node at (-1.6,-4) {30};
\node at (-.6,-4) {77};
\node at (.4,-4) {101};
\node at (1.4,-4) {77};
\node at (2.4,-4) {30};
\node at (3.3,-4) {5};
\node at (-3.7,-5) {1};
\node at (-2.6,-5) {15};
\node at (-1.6,-5) {77};
\node at (-.6,-5) {176};
\node at (.4,-5) {231};
\node at (1.4,-5) {176};
\node at (2.4,-5) {77};
\node at (3.4,-5) {15};
\node at (4.3,-5) {1};
\node at (-3.7,-6) {3};
\node at (-2.6,-6) {42};
\node at (-1.6,-6) {176};
\node at (-.6,-6) {385};
\node at (.4,-6) {490};
\node at (1.4,-6) {385};
\node at (2.4,-6) {176};
\node at (3.4,-6) {42};
\node at (4.3,-6) {3};
\node at (-3.6,-7) {11};
\node at (-2.6,-7) {101};
\node at (-1.6,-7) {385};
\node at (-.6,-7) {792};
\node at (.45,-7) {\footnotesize{1002}};
\node at (1.4,-7) {792};
\node at (2.4,-7) {385};
\node at (3.4,-7) {101};
\node at (4.4,-7) {11};
\node at (-4.7,-8) {1};
\node at (-3.6,-8) {30};
\node at (-2.6,-8) {231};
\node at (-1.6,-8) {792};
\node at (-.55,-8) {\footnotesize{1574}};
\node at (.45,-8) {\footnotesize{1956}};
\node at (1.45,-8) {\footnotesize{1574}};
\node at (2.4,-8) {792};
\node at (3.4,-8) {231};
\node at (4.4,-8) {30};
\node at (5.3,-8) {1};
\node at (-4.7,-9) {3};
\node at (-3.6,-9) {77};
\node at (-2.6,-9) {490};
\node at (-1.55,-9) {\footnotesize{1574}};
\node at (-.55,-9) {\footnotesize{3005}};
\node at (.45,-9) {\footnotesize{3710}};
\node at (1.45,-9) {\footnotesize{3005}};
\node at (2.45,-9) {\footnotesize{1574}};
\node at (3.4,-9) {490};
\node at (4.4,-9) {77};
\node at (5.3,-9) {3};
\node at (-4.6,-10) {11};
\node at (-3.6,-10) {176};
\node at (-2.55,-10) {\footnotesize{1002}};
\node at (-1.55,-10) {\footnotesize{3005}};
\node at (-.55,-10) {\footnotesize{5578}};
\node at (.5,-10) {\tiny{10096}};
\node at (1.45,-10) {\footnotesize{5578}};
\node at (2.45,-10) {\footnotesize{3005}};
\node at (3.45,-10) {\footnotesize{1002}};
\node at (4.45,-10) {176};
\node at (5.4,-10) {11};
\node at (-4.6,-11) {30};
\node at (-3.6,-11) {385};
\node at (-2.55,-11) {\footnotesize{1956}};
\node at (-1.55,-11) {\footnotesize{5587}};
\node at (-.5,-11) {\tiny{10096}};
\node at (.5,-11) {\tiny{12246}};
\node at (1.5,-11) {\tiny{10096}};
\node at (2.45,-11) {\footnotesize{5587}};
\node at (3.45,-11) {\footnotesize{1956}};
\node at (4.45,-11) {385};
\node at (5.4,-11) {30};
\node at (-5.7,-12) {1};
\node at (-4.6,-12) {77};
\node at (-3.6,-12) {792};
\node at (-2.55,-12) {\footnotesize{3710}};
\node at (-1.5,-12) {\tiny{10096}};
\node at (-.5,-12) {\tiny{17861}};
\node at (.5,-12) {\tiny{21483}};
\node at (1.5,-12) {\tiny{17861}};
\node at (2.5,-12) {\tiny{10096}};
\node at (3.45,-12) {\footnotesize{3710}};
\node at (4.4,-12) {792};
\node at (5.4,-12) {77};
\node at (6.3,-12) {1};
\node at (-5.7,-13) {5};
\node at (-4.6,-13) {176};
\node at (-3.55,-13) {\footnotesize{1574}};
\node at (-2.55,-13) {\footnotesize{6818}};
\node at (-1.5,-13) {\tiny{17861}};
\node at (-.5,-13) {\tiny{30921}};
\node at (.45,-13) {\tiny{36996}};
\node at (1.5,-13) {\tiny{30921}};
\node at (2.5,-13) {\tiny{17861}};
\node at (3.45,-13) {\footnotesize{6818}};
\node at (4.45,-13) {\footnotesize{1574}};
\node at (5.4,-13) {176};
\node at (6.3,-13) {5};
\node at (-5.6,-14) {15};
\node at (-4.6,-14) {385};
\node at (-3.55,-14) {\footnotesize{3005}};
\node at (-2.5,-14) {\tiny{11246}};
\node at (-1.5,-14) {\tiny{30921}};
\node at (-.55,-14) {\tiny{52609}};
\node at (.45,-14) {\tiny{62541}};
\node at (1.45,-14) {\tiny{52609}};
\node at (2.5,-14) {\tiny{30921}};
\node at (3.5,-14) {\tiny{11246}};
\node at (4.45,-14) {\footnotesize{3005}};
\node at (5.4,-14) {385};
\node at (6.4,-14) {15};
\node at (-5.6,-15) {42};
\node at (-4.6,-15) {792};
\node at (-3.45,-15) {\footnotesize{5587}};
\node at (-2.55,-15) {\tiny{21483}};
\node at (-1.55,-15) {\tiny{52609}};
\node at (-.55,-15) {\tiny{87982}};
\node at (.5,-15) {\tiny{104111}};
\node at (1.45,-15) {\tiny{87982}};
\node at (2.5,-15) {\tiny{52609}};
\node at (3.5,-15) {\tiny{21483}};
\node at (4.45,-15) {\footnotesize{5587}};
\node at (5.4,-15) {792};
\node at (6.4,-15) {42};
\end{tikzpicture}
\vspace{.5cm}
\caption{The partial level-3 root system associated to $\mF^{(3)}$. 
The numbers specify the number of 
states at each point, and are equal to the multiplicities of the associated roots.
The {\color{red}red} diamonds denote the virtual singlet~\eqref{L3VSing} at depth three 
and virtual triplet~\eqref{L3VTrip} at depth two that are not part of $\mF^{(3)}$. 
At depth 3 the virtual singlet sits among the regular states. Again we have
not depicted the affine towers built on the virtual red diamond states.}
\label{fig:L3RootLattice}
\end{figure}

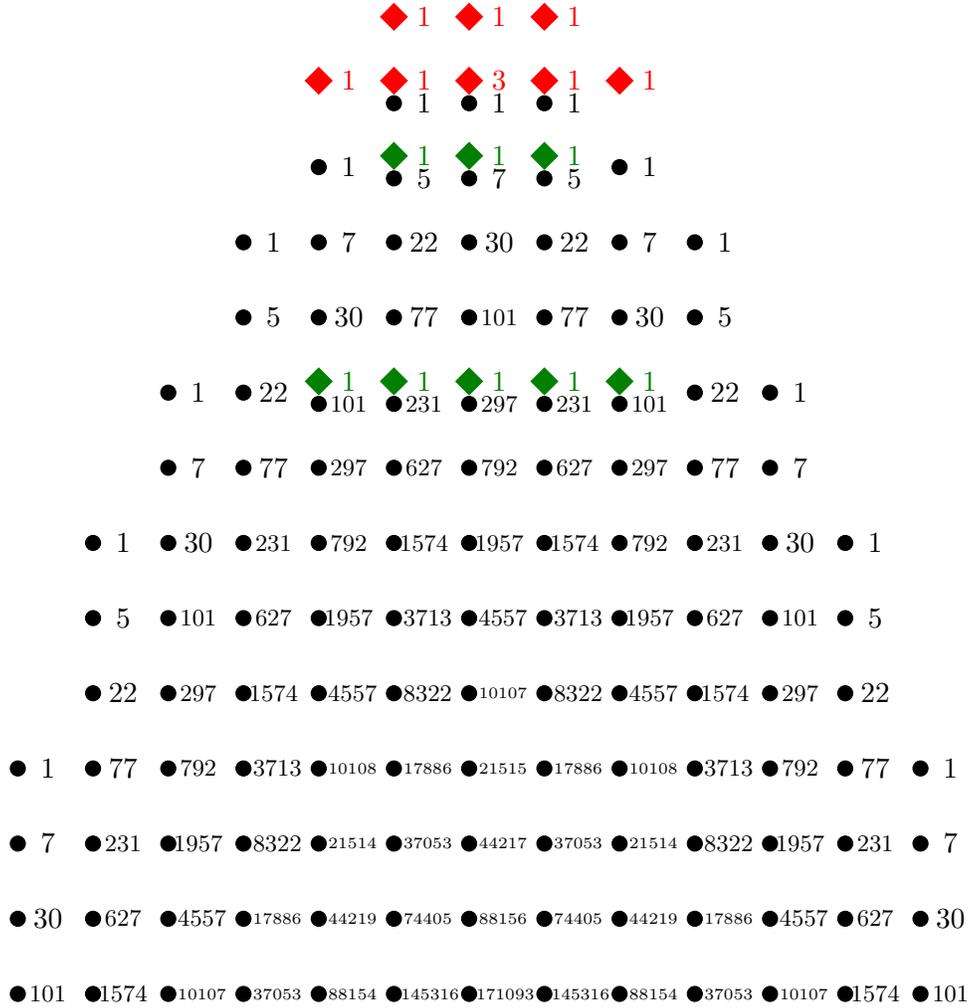
\begin{figure}[H]
\vspace{.5cm}
\begin{tikzpicture}
\node [fill, myred, diamond, draw, scale=0.7] at (-1,1) {};
\node [fill, myred, diamond, draw, scale=0.7] at (0,1) {};
\node [fill, myred, diamond, draw, scale=0.7] at (1,1) {};
\node [fill, myred, diamond, draw, scale=0.7] at (-2,.15) {};
\node [fill, myred, diamond, draw, scale=0.7] at (-1,.15) {};
\node [fill, myred, diamond, draw, scale=0.7] at (0,.15) {};
\node [fill, myred, diamond, draw, scale=0.7] at (1,.15) {};
\node [fill, myred, diamond, draw, scale=0.7] at (2,.15) {};
\draw [fill] (-1,-.15) circle (.1);
\draw [fill] (0,-.15) circle (.1);
\draw [fill] (1,-.15) circle (.1);
\node [fill, mygreen, diamond, draw, scale=0.7] at (-1,.-.85) {};
\node [fill, mygreen, diamond, draw, scale=0.7] at (0,.-.85) {};
\node [fill, mygreen, diamond, draw, scale=0.7] at (1,.-.85) {};
\draw [fill] (-2,-1) circle (.1);
\draw [fill] (-1,-1.15) circle (.1);
\draw [fill] (0,-1.15) circle (.1);
\draw [fill] (1,-1.15) circle (.1);
\draw [fill] (2,-1) circle (.1);
\draw [fill] (-3,-2) circle (.1);
\draw [fill] (-2,-2) circle (.1);
\draw [fill] (-1,-2) circle (.1);
\draw [fill] (0,-2) circle (.1);
\draw [fill] (1,-2) circle (.1);
\draw [fill] (2,-2) circle (.1);
\draw [fill] (3,-2) circle (.1);
\draw [fill] (-3,-3) circle (.1);
\draw [fill] (-2,-3) circle (.1);
\draw [fill] (-1,-3) circle (.1);
\draw [fill] (0,-3) circle (.1);
\draw [fill] (1,-3) circle (.1);
\draw [fill] (2,-3) circle (.1);
\draw [fill] (3,-3) circle (.1);
\node [fill, mygreen, diamond, draw, scale=0.7] at (-2,-3.85) {};
\node [fill, mygreen, diamond, draw, scale=0.7] at (-1,-3.85) {};
\node [fill, mygreen, diamond, draw, scale=0.7] at (0,-3.85) {};
\node [fill, mygreen, diamond, draw, scale=0.7] at (1,-3.85) {};
\node [fill, mygreen, diamond, draw, scale=0.7] at (2,-3.85) {};
\draw [fill] (-4,-4) circle (.1);
\draw [fill] (-3,-4) circle (.1);
\draw [fill] (-2,-4.15) circle (.1);
\draw [fill] (-1,-4.15) circle (.1);
\draw [fill] (0,-4.15) circle (.1);
\draw [fill] (1,-4.15) circle (.1);
\draw [fill] (2,-4.15) circle (.1);
\draw [fill] (3,-4) circle (.1);
\draw [fill] (4,-4) circle (.1);
\draw [fill] (-4,-5) circle (.1);
\draw [fill] (-3,-5) circle (.1);
\draw [fill] (-2,-5) circle (.1);
\draw [fill] (-1,-5) circle (.1);
\draw [fill] (0,-5) circle (.1);
\draw [fill] (1,-5) circle (.1);
\draw [fill] (2,-5) circle (.1);
\draw [fill] (3,-5) circle (.1);
\draw [fill] (4,-5) circle (.1);
\draw [fill] (-5,-6) circle (.1);
\draw [fill] (-4,-6) circle (.1);
\draw [fill] (-3,-6) circle (.1);
\draw [fill] (-2,-6) circle (.1);
\draw [fill] (-1,-6) circle (.1);
\draw [fill] (0,-6) circle (.1);
\draw [fill] (1,-6) circle (.1);
\draw [fill] (2,-6) circle (.1);
\draw [fill] (3,-6) circle (.1);
\draw [fill] (4,-6) circle (.1);
\draw [fill] (5,-6) circle (.1);
\draw [fill] (-5,-7) circle (.1);
\draw [fill] (-4,-7) circle (.1);
\draw [fill] (-3,-7) circle (.1);
\draw [fill] (-2,-7) circle (.1);
\draw [fill] (-1,-7) circle (.1);
\draw [fill] (0,-7) circle (.1);
\draw [fill] (1,-7) circle (.1);
\draw [fill] (2,-7) circle (.1);
\draw [fill] (3,-7) circle (.1);
\draw [fill] (4,-7) circle (.1);
\draw [fill] (5,-7) circle (.1);
\draw [fill] (-5,-8) circle (.1);
\draw [fill] (-4,-8) circle (.1);
\draw [fill] (-3,-8) circle (.1);
\draw [fill] (-2,-8) circle (.1);
\draw [fill] (-1,-8) circle (.1);
\draw [fill] (0,-8) circle (.1);
\draw [fill] (1,-8) circle (.1);
\draw [fill] (2,-8) circle (.1);
\draw [fill] (3,-8) circle (.1);
\draw [fill] (4,-8) circle (.1);
\draw [fill] (5,-8) circle (.1);
\draw [fill] (-6,-9) circle (.1);
\draw [fill] (-5,-9) circle (.1);
\draw [fill] (-4,-9) circle (.1);
\draw [fill] (-3,-9) circle (.1);
\draw [fill] (-2,-9) circle (.1);
\draw [fill] (-1,-9) circle (.1);
\draw [fill] (0,-9) circle (.1);
\draw [fill] (1,-9) circle (.1);
\draw [fill] (2,-9) circle (.1);
\draw [fill] (3,-9) circle (.1);
\draw [fill] (4,-9) circle (.1);
\draw [fill] (5,-9) circle (.1);
\draw [fill] (6,-9) circle (.1);
\draw [fill] (-6,-10) circle (.1);
\draw [fill] (-5,-10) circle (.1);
\draw [fill] (-4,-10) circle (.1);
\draw [fill] (-3,-10) circle (.1);
\draw [fill] (-2,-10) circle (.1);
\draw [fill] (-1,-10) circle (.1);
\draw [fill] (0,-10) circle (.1);
\draw [fill] (1,-10) circle (.1);
\draw [fill] (2,-10) circle (.1);
\draw [fill] (3,-10) circle (.1);
\draw [fill] (4,-10) circle (.1);
\draw [fill] (5,-10) circle (.1);
\draw [fill] (6,-10) circle (.1);
\draw [fill] (-6,-11) circle (.1);
\draw [fill] (-5,-11) circle (.1);
\draw [fill] (-4,-11) circle (.1);
\draw [fill] (-3,-11) circle (.1);
\draw [fill] (-2,-11) circle (.1);
\draw [fill] (-1,-11) circle (.1);
\draw [fill] (0,-11) circle (.1);
\draw [fill] (1,-11) circle (.1);
\draw [fill] (2,-11) circle (.1);
\draw [fill] (3,-11) circle (.1);
\draw [fill] (4,-11) circle (.1);
\draw [fill] (5,-11) circle (.1);
\draw [fill] (6,-11) circle (.1);
\draw [fill] (-6,-12) circle (.1);
\draw [fill] (-5,-12) circle (.1);
\draw [fill] (-4,-12) circle (.1);
\draw [fill] (-3,-12) circle (.1);
\draw [fill] (-2,-12) circle (.1);
\draw [fill] (-1,-12) circle (.1);
\draw [fill] (0,-12) circle (.1);
\draw [fill] (1,-12) circle (.1);
\draw [fill] (2,-12) circle (.1);
\draw [fill] (3,-12) circle (.1);
\draw [fill] (4,-12) circle (.1);
\draw [fill] (5,-12) circle (.1);
\draw [fill] (6,-12) circle (.1);
\node at (-.6,1) {{\color{myred}1}};
\node at (.4,1) {{\color{myred}1}};
\node at (1.4,1) {{\color{myred}1}};
\node at (-1.6,.15) {{\color{myred}1}};
\node at (-.6,.15) {{\color{myred}1}};
\node at (.4,.15) {{\color{myred}3}};
\node at (1.4,.15) {{\color{myred}1}};
\node at (2.4,.15) {{\color{myred}1}};
\node at (-.6,-.85) {{\color{mygreen}1}};
\node at (.4,-.85) {{\color{mygreen}1}};
\node at (1.4,-.85) {{\color{mygreen}1}};
\node at (-1.6,-3.85) {{\color{mygreen}1}};
\node at (-.6,-3.85) {{\color{mygreen}1}};
\node at (.4,-3.85) {{\color{mygreen}1}};
\node at (1.4,-3.85) {{\color{mygreen}1}};
\node at (2.4,-3.85) {{\color{mygreen}1}};
\node at (.4,-.15) {{1}};
\node at (.4,-1.15) {{7}};
\node at (.4,-2) {{30}};
\node at (.4,-3) {\footnotesize{101}};
\node at (.4,-4.15) {\footnotesize{297}};
\node at (.4,-5) {\footnotesize{792}};
\node at (.4,-6) {\footnotesize{1957}};
\node at (.45,-7) {\footnotesize{4557}};
\node at (.45,-8) {\tiny{10107}};
\node at (.45,-9) {\tiny{21515}};
\node at (.45,-10) {\tiny{44217}};
\node at (.45,-11) {\tiny{88156}};
\node at (.48,-12) {\tiny{171093}};
\node at (-.6,-.15) {{1}};
\node at (-.6,-1.15) {{5}};
\node at (-.6,-2) {{22}};
\node at (-.6,-3) {{77}};
\node at (-.6,-4.15) {\footnotesize{231}};
\node at (-.6,-5) {\footnotesize{627}};
\node at (-.6,-6) {\footnotesize{1574}};
\node at (-.55,-7) {\footnotesize{3713}};
\node at (-.55,-8) {\footnotesize{8322}};
\node at (-.55,-9) {\tiny{17886}};
\node at (-.55,-10) {\tiny{37053}};
\node at (-.55,-11) {\tiny{74405}};
\node at (-.52,-12) {\tiny{145316}};
\node at (1.4,-.15) {{1}};
\node at (1.4,-1.15) {{5}};
\node at (1.4,-2) {{22}};
\node at (1.4,-3) {{77}};
\node at (1.4,-4.15) {\footnotesize{231}};
\node at (1.4,-5) {\footnotesize{627}};
\node at (1.4,-6) {\footnotesize{1574}};
\node at (1.45,-7) {\footnotesize{3713}};
\node at (1.45,-8) {\footnotesize{8322}};
\node at (1.45,-9) {\tiny{17886}};
\node at (1.45,-10) {\tiny{37053}};
\node at (1.45,-11) {\tiny{74405}};
\node at (1.48,-12) {\tiny{145316}};
\node at (3.4,-2) {{1}};
\node at (3.4,-3) {{5}};
\node at (3.4,-4) {{22}};
\node at (3.4,-5) {{77}};
\node at (3.4,-6) {\footnotesize{231}};
\node at (3.4,-7) {\footnotesize{627}};
\node at (3.4,-8) {\footnotesize{1574}};
\node at (3.45,-9) {\footnotesize{3713}};
\node at (3.45,-10) {\footnotesize{8322}};
\node at (3.45,-11) {\tiny{17886}};
\node at (3.45,-12) {\tiny{37053}};
\node at (-2.6,-2) {{1}};
\node at (-2.6,-3) {{5}};
\node at (-2.6,-4) {{22}};
\node at (-2.6,-5) {{77}};
\node at (-2.6,-6) {\footnotesize{231}};
\node at (-2.6,-7) {\footnotesize{627}};
\node at (-2.6,-8) {\footnotesize{1574}};
\node at (-2.55,-9) {\footnotesize{3713}};
\node at (-2.55,-10) {\footnotesize{8322}};
\node at (-2.55,-11) {\tiny{17886}};
\node at (-2.55,-12) {\tiny{37053}};
\node at (-1.6,-1) {{1}};
\node at (-1.6,-2) {{7}};
\node at (-1.6,-3) {{30}};
\node at (-1.6,-4.15) {\footnotesize{101}};
\node at (-1.6,-5) {\footnotesize{297}};
\node at (-1.6,-6) {\footnotesize{792}};
\node at (-1.6,-7) {\footnotesize{1957}};
\node at (-1.55,-8) {\footnotesize{4557}};
\node at (-1.55,-9) {\tiny{10108}};
\node at (-1.55,-10) {\tiny{21514}};
\node at (-1.55,-11) {\tiny{44219}};
\node at (-1.55,-12) {\tiny{88154}};
\node at (2.4,-1) {{1}};
\node at (2.4,-2) {{7}};
\node at (2.4,-3) {{30}};
\node at (2.4,-4.15) {\footnotesize{101}};
\node at (2.4,-5) {\footnotesize{297}};
\node at (2.4,-6) {\footnotesize{792}};
\node at (2.4,-7) {\footnotesize{1957}};
\node at (2.45,-8) {\footnotesize{4557}};
\node at (2.45,-9) {\tiny{10108}};
\node at (2.45,-10) {\tiny{21514}};
\node at (2.45,-11) {\tiny{44219}};
\node at (2.45,-12) {\tiny{88154}};
\node at (4.4,-4) {{1}};
\node at (4.4,-5) {{7}};
\node at (4.4,-6) {{30}};
\node at (4.4,-7) {\footnotesize{101}};
\node at (4.4,-8) {\footnotesize{297}};
\node at (4.4,-9) {\footnotesize{792}};
\node at (4.4,-10) {\footnotesize{1957}};
\node at (4.45,-11) {\footnotesize{4557}};
\node at (4.45,-12) {\tiny{10107}};
\node at (-3.6,-4) {{1}};
\node at (-3.6,-5) {{7}};
\node at (-3.6,-6) {{30}};
\node at (-3.6,-7) {\footnotesize{101}};
\node at (-3.6,-8) {\footnotesize{297}};
\node at (-3.6,-9) {\footnotesize{792}};
\node at (-3.6,-10) {\footnotesize{1957}};
\node at (-3.55,-11) {\footnotesize{4557}};
\node at (-3.55,-12) {\tiny{10107}};
\node at (-4.6,-6) {{1}};
\node at (-4.6,-7) {{5}};
\node at (-4.6,-8) {{22}};
\node at (-4.6,-9) {{77}};
\node at (-4.6,-10) {\footnotesize{231}};
\node at (-4.6,-11) {\footnotesize{627}};
\node at (-4.6,-12) {\footnotesize{1574}};
\node at (5.4,-6) {{1}};
\node at (5.4,-7) {{5}};
\node at (5.4,-8) {{22}};
\node at (5.4,-9) {{77}};
\node at (5.4,-10) {\footnotesize{231}};
\node at (5.4,-11) {\footnotesize{627}};
\node at (5.4,-12) {\footnotesize{1574}};
\node at (-5.6,-9) {{1}};
\node at (-5.6,-10) {{7}};
\node at (-5.6,-11) {{30}};
\node at (-5.6,-12) {\footnotesize{101}};
\node at (6.4,-9) {{1}};
\node at (6.4,-10) {{7}};
\node at (6.4,-11) {{30}};
\node at (6.4,-12) {\footnotesize{101}};
\end{tikzpicture}
\vspace{.5cm}
\caption{The partial level-4 root system associated to the level-4 part 
$\mathfrak{F}^{(4)}$ indicated in black,
again with root multiplicities indicated. The highest root on this level is $(-4,-4,-3)$ 
and is the top right black dot. The red diamonds indicate the virtual maximal ground 
states described in~\eqref{L0L3} and~\eqref{L0L0L1} and on which affine and coset
Virasoro modules can be built that contain all level-4 elements of the hyperbolic algebra.
Similarly the green diamonds denote the maximal ground states which also belong to 
$\mF^{(4)}$. For the fiveplet at depth 8 we do not yet have an expression in terms of 
DDF states. Hence we cannot yet tell whether it is virtual or not.}
\label{fig:L4RootLattice}
\end{figure}


\newpage
\begin{thebibliography}{99}
\bibitem{FF}
A.~J.~Feingold and I.~B.~Frenkel,
\textit{A hyperbolic Kac-Moody algebra and the theory of Siegel modular forms of genus 2},
\href{https://doi.org/10.1007/BF01457086}%
{Math. Ann. \textbf{263}, 87–144 (1983)}. 
%
\bibitem{Kac}
V.~G.~Kac,
\textit{Infinite-Dimensional Lie algebras},
\href{https://doi.org/10.1017/CBO9780511626234}%
{3rd ed. Cambridge University Press 1990}.
%
\bibitem{KMW}
V.~G.~ Kac, R.~V.~Moody and M.~Wakimoto,
\textit{On $E_{10}$}, 
in \href{https://doi.org/10.1007/978-94-015-7809-7}%
{eds. K.~Bleuler and M.~Werner 
\textit{Differential Geometrical Methods in Theoretical Physics}, Springer Dordrecht 1988}.
%
\bibitem{Kang1}
S.~J.~Kang,
\textit{Root Multiplicities of the Hyperbolic Kac-Moody Lie Algebra $HA_1^{(1)}$}, 
\href{https://doi.org/10.1006/jabr.1993.1198}%
{Journal of Algebra \textbf{160}, 492-523 (1993)}.
\bibitem{Kang2}
S.~J.~Kang,
\textit{Root Multiplicities of Kac-Moody Algebras},
\href{https://doi.org/10.1215/S0012-7094-94-07423-1}%
{Duke Math. J. \textbf{74}, 635-666 (1994)}.
%
\bibitem{BB}
M.~Bauer and D.~Bernard, 
\textit{On root multiplicities of some hyperbolic Kac-Moody algebras}, 
\href{https://doi.org/10.1023/A:1007317602691}%
{Lett. Math. Phys. \textbf{42} (1997), 153-166},
\href{https://arxiv.org/abs/hep-th/9612210}{\texttt{arXiv:hep-th/9612210}}.
%
\bibitem{KW1}
V.~G.~Kac and M.~Wakimoto,
\textit{Unitarizable highest weight representations of the Virasoro, Neveu-Schwarz and
Ramond algebras},
\href{https://math.mit.edu/~kac/not-easily-available/unitarizable.pdf}%
{Lecture Notes in Physics 261 (1986), 345-372}.
%
\bibitem{GKO}P.~Goddard, A.~Kent and D.~I.~Olive, 
\textit{Virasoro Algebras and Coset Space Models}, 
\href{https://doi.org/10.1016/0370-2693(85)91145-1}%
{Phys. Lett. B \textbf{152} (1985), 88-92}.
%
\bibitem{VisualLie}
H.~Malcha,
\textit{VisualLie}, (2024),
\href{https://hmalcha.github.io/VisualLie/}{\texttt{https://hmalcha.github.io/VisualLie/}}.
%
\bibitem{GN}
R.~W.~Gebert and H.~Nicolai,
\textit{On E(10) and the DDF construction},
\href{https://doi.org/10.1007/BF02101809}%
{Commun. Math. Phys. \textbf{172} (1995), 571-622},
\href{https://arxiv.org/abs/hep-th/9406175}{\texttt{arXiv:hep-th/9406175}}.
%
\bibitem{GN1}
R.~W.~Gebert and H.~Nicolai,
\textit{An affine string vertex operator construction at arbitrary level},
\href{https://doi.org/10.1063/1.532135}{J. Math. Phys. \textbf{38} (1997), 4435-4450}
\href{https://arxiv.org/abs/hep-th/9608014}{\texttt{arXiv:hep-th/9608014}}.
%
\bibitem{Borcherds}
R.~E.~Borcherds
\textit{Vertex algebras, Kac-Moody algebras, and the Monster},
\href{https://doi.org/10.1073/pnas.83.10.3068}%
{Proceedings of the National Academy of Sciences \textbf{83}, 10 (1986), 3068-3071}. 
%
\bibitem{IF}
I.~B.~Frenkel, 
\textit{Representations of Kac-Moody algebras and dual resonance models},
in 
\textit{Applications of Group Theory in Physics and Mathematical Physics}
(Chicago, 1982), 325-353, Lectures in Appl. Math. {\bf 21}, Amer. Math. Soc., 
Providence, RI, 1985.
%
\bibitem{FLM}
I.~B.~Frenkel, J.~Lepowsky and A.~Meurman, 
\textit{Vertex Operator Algebras and the Monster},
Pure and Applied Mathematics Vol. 134, San Diego, CA: Academic Press, 1988.
%
\bibitem{DDF}
E.~Del Giudice, P.~Di Vecchia and S.~Fubini,
\textit{General properties of the dual resonance model},
\href{https://doi.org/10.1016/0003-4916(72)90272-2}%
{Annals Phys. \textbf{70} (1972), 378-398}.
%
\bibitem{Julia}
B.~Julia,
\textit{Infinite Lie algebras in physics}, in 
\href{https://lib-extopc.kek.jp/preprints/PDF/1981/8112/8112035.pdf}
{Johns Hopkins Workshop on Current Problems in
Particle Physics: Unified theories and Beyond, Johns Hopkins University, Baltimore (1981)}.
%
\bibitem{BM}
P.~Breitenlohner and D.~Maison,
\textit{On the Geroch group}, 
\href{https://lib-extopc.kek.jp/preprints/PDF/1987/8702/8702388.pdf}{
Ann. Inst. H. Poincar\'e. Phys. Th\'eor. {\bf 46} (1987) 215}.
%
\bibitem{Nic0}
H.~Nicolai,
\textit{A Hyperbolic Lie algebra from supergravity},
\href{https://doi.org/10.1016/0370-2693(92)90328-2}%
{Phys. Lett. B \textbf{276} (1992), 333-340}.
%
\bibitem{Penna:2021apa}
R.~F.~Penna,
\textit{The Geroch Group in One Dimension}, 
\href{https://arxiv.org/abs/2112.05661}{\texttt{arXiv:2112.05661 [hep-th]}}.
%
\bibitem{DHN}
T~ Damour, M.~Henneaux and H.~Nicolai,
\textit{Cosmological Billiards},
\href{https://doi.org/10.1007/0-387-24992-3_5}%
{Class.Quant.Grav. {\bf 20} (2003) R145-R200}.
%
\bibitem{BKL} V.~A.~Belinskii, I.~M.~Khalatnikov and E.~M.~Lifshitz, 
\textit{Oscillatory approach to a singular point in the relativistic cosmology}, 
Adv. Phys. {\bf 19} (1970) 525.
%
\bibitem{Nic}
H.~Nicolai,
\textit{Complexity and the Big Bang},
\href{https://doi.org/10.1088/1361-6382/ac1b07}%
{Class. Quant. Grav. \textbf{38} (2021) no.18, 187001},
\href{https://arxiv.org/abs/2104.09626}{\texttt{arXiv:2104.09626 [gr-qc]}}.
%
\bibitem{GO}
P.~Goddard and D.~I.~Olive,
\textit{Kac-Moody and Virasoro Algebras in Relation to Quantum Physics},
\href{https://doi.org/10.1142/S0217751X86000149}%
{Int. J. Mod. Phys. A \textbf{1} (1986), 303}.
%
\bibitem{DiFrancesco}
P.~Di Francesco, P.~Mathieu and D.~Senechal,
\textit{Conformal Field Theory},
\href{https://doi.org/10.1007/978-1-4612-2256-9}{Springer-Verlag, 1997}.
%
\bibitem{Brower:1972wj}
R.~C.~Brower,
\textit{Spectrum generating algebra and no ghost theorem for the dual model},
\href{https://doi.org/10.1103/PhysRevD.6.1655}{Phys. Rev. D \textbf{6} (1972), 1655-1662}.
%
\bibitem{FK}
I.~B.~Frenkel and V.~G.~Kac, 
\textit{Basic Representations of Affine Lie Algebras and Dual Resonance Models},
\href{https://doi.org/10.1007/BF01391662}{Invent Math 62, 23–66 (1980)}. 
%
\bibitem{Segal:1981ap}
G.~Segal,
\textit{Unitarity Representations of Some Infinite Dimensional Groups},
\href{https://doi.org/10.1007/BF01208274}%
{Commun. Math. Phys. \textbf{80} (1981), 301-342}.
%
\bibitem{GO0}
P.~Goddard and D.~I. Olive, 
in \textit{Vertex operators in Mathematics and physics},
\href{https://doi.org/10.1007/978-1-4613-9550-8}%
{eds. J. Lepowsky, S. Mandelstam and I.M. Singer, MSRI Publications, Springer Verlag
(1983)}.
%
\bibitem{GKN}
R.~W.~Gebert, K.~Koepsell and H.~Nicolai,
\textit{The Sugawara generators at arbitrary level},
\href{https://doi.org/10.1007/s002200050055}%
{Commun. Math. Phys. \textbf{184} (1997), 119-141},
\href{https://arxiv.org/abs/hep-th/9604155}{\texttt{arXiv:hep-th/9604155}}.
%
\bibitem{SimpLie}
T.~Nutma,
\textit{SimpLie}, (2015), 
\href{https://github.com/teake/simplie}{\texttt{https://github.com/teake/simplie}}.
%
\bibitem{KacPeterson}
V.~G.~Kac and D.~H.~Peterson,
\textit{Infinite-dimensional Lie algebras, theta functions and modular forms},
\href{https://doi.org/10.1016/0001-8708(84)90032-X}{Adv. Math. 53, 125-264 (1984)}.
%
\bibitem{Mortenson}
E.~T.~Mortenson
\textit{On Hecke-type double-sums and general string functions for the affine
Lie algebra $A_{1}^{(1)}$},
\href{https://doi.org/10.1007/s11139-023-00737-x}%
{Ramanujan J. \textbf{63}, No. 3, 553-582 (2024)},
\href{https://arxiv.org/abs/2110.02615}{\texttt{arXiv:2110.02615 [math.NT]}}.
%
\bibitem{Goddard:1988fw}
P.~Goddard and D.~I.~Olive,
\textit{Kac-Moody and Virasoro Algebras: A Preprint Volume for Physicists},
\href{https://doi.org/10.1142/0485}{Adv. Ser. Math. Phys. \textbf{3} (1988)}.
\end{thebibliography}
\end{document}